\documentclass[final]{aa}  

\usepackage{txfonts}
\usepackage{colortbl}
\usepackage[table]{xcolor}	

\RequirePackage{bbm}
\RequirePackage{rotating}
\RequirePackage{subfig,color,pifont}
\RequirePackage{graphicx,array,hyperref} 
\RequirePackage{blkarray,hhline,multirow}

\definecolor{dkgreen2}  	{rgb}{0.10,0.40,0.10}
\definecolor{lightGreen}  	{rgb}{0.75,0.95,0.75}
\definecolor{lightGreen2}  {rgb}{0.85,0.98,0.85}
\definecolor{lightRed}  	{rgb}{0.95,0.75,0.75}
\definecolor{lightRed1} 	{rgb}{0.95,0.94,0.94}
\definecolor{lightRed2} 	{rgb}{0.98,0.90,0.90}
\definecolor{lightYellow}    {rgb}{0.99,0.99,0.92}
\definecolor{lightGray} 	{rgb}{0.95,0.95,0.95}
\definecolor{dkblue}          {rgb}{0.05,0.05,0.70}
\definecolor{dkgray}          {rgb}{0.40,0.40,0.40}
\definecolor{dkgray2}   	{rgb}{0.20,0.20,0.20}
\definecolor{dkmauve}  	{rgb}{0.70,0.01,0.50}
\definecolor{orng}              {rgb}{0.80,0.40,0.10}
\definecolor{prColo}           {rgb}{0.60,0.20,0.40}
\definecolor{trColo}            {rgb}{0.10,0.40,0.10}
\definecolor{trColo2}          {rgb}{0.20,0.70,0.20}

\renewcommand{\arraystretch}{1.80}
\newcommand{\indep}{\rotatebox[origin=c]{90}{$\models$}}
\newcommand{\mynewformat}{\fontsize{8.5pt}{8pt}\selectfont}
\RequirePackage[labelfont=bf]{caption}  %
\DeclareCaptionFormat{myformat}{\fontsize{9pt}{8pt}\selectfont#1#2#3}
\hypersetup{ bookmarks=true, unicode=true, pdftoolbar=true, pdfmenubar=true, pdffitwindow=true, pdfstartview={FitH}, pdfauthor={SJ}, 
colorlinks=true, linkcolor=dkblue, citecolor=dkblue, urlcolor=dkmauve, breaklinks }
 
 \let\oldenumerate\enumerate \renewcommand{\enumerate}{
	\oldenumerate
	\setlength{\itemsep}{2.0pt}
	\setlength{\parskip}{0.0pt}
	\setlength{\parsep}{0.0pt}
}
\let\olditemize\itemize \renewcommand{\itemize}{
	\olditemize
	\setlength{\itemsep}{2.0pt}
	\setlength{\parskip}{0.0pt}
	\setlength{\parsep}{0.0pt}
}
 
\usepackage[squaren, Gray, cdot]{SIunits}
 
\begin{document}
{

   \title{\Large{Automated reliability assessment for spectroscopic redshift measurements}}
   
   \author{
   S.~Jamal                           \inst{\ref{inst1}}
   \and  V.~Le Brun               \inst{\ref{inst1}}
   \and  O.~Le F\`{e}vre        \inst{\ref{inst1}}
   \and  D.~Vibert                  \inst{\ref{inst1}}
   \and  A.~Schmitt                \inst{\ref{inst1}}
   \and  C.~Surace                \inst{\ref{inst1}}
    \and  Y.~Copin                  \inst{\ref{inst2}}
   \and  B.~Garilli                   \inst{\ref{inst3}}
   \and  M.~Moresco              \inst{\ref{inst4},\ref{inst5}}
   \and  L.~Pozzetti                \inst{\ref{inst5}}
   %
     }

   \institute{ 
        {Aix Marseille Univ., CNRS, LAM, Laboratoire d'Astrophysique de Marseille, Marseille, France\\
        \email{
                \href{mailto:sara.jamal@lam.fr}{\textrm{sara.jamal@lam.fr};
                \href{mailto:vincent.lebrun@lam.fr}{\textrm{vincent.lebrun@lam.fr}
        }}}
        \label{inst1}
        }
  \and 
        {Universit\'{e} Lyon, F-69622, Lyon, France; Universit\'{e} Lyon 1, Villeurbanne; CNRS/IN2P3, Institut de Physique Nucl\'{e}aire de Lyon}
        \label{inst2}
  \and 
        {INAF - Istituto di Astrofisica Spaziale e Fisica Cosmica Milano, via Bassini 15, 20133 Milano, Italy}
        \label{inst3}
  \and 
        {Dipartimento di Fisica e Astronomia, Universit\`a di Bologna, Via Gobetti 93/2, I-40129, Bologna, Italy}          
        \label{inst4}
  \and
        {INAF - Osservatorio Astronomico di Bologna, Via Gobetti 93/3, I-40129, Bologna, Italy}                                   
        \label{inst5}
    }
    
   \date{Received 2 June 2017/ Accepted 9 September 2017}		

   \abstract
{
Future large-scale surveys, such as the ESA Euclid mission, will produce a large set of galaxy redshifts ($\geq10^{6}$) that will require fully automated data-processing pipelines to analyze the data, extract crucial information and ensure that all requirements are met.\\
A fundamental element in these pipelines is to associate to each galaxy redshift measurement a quality, or reliability, estimate.
}
{
In this work, we introduce a new approach to automate the spectroscopic redshift reliability assessment based on machine learning (ML) and characteristics of the redshift probability density function.
}
{
We propose to rephrase the spectroscopic redshift estimation into a Bayesian framework, in order to incorporate all sources of information and uncertainties related to the redshift estimation process and produce a redshift posterior probability density function (PDF).\\
To automate the assessment of a reliability flag, we exploit key features in the redshift posterior PDF and machine learning algorithms.
}
{
As a working example, public data from the VIMOS VLT Deep Survey is exploited to present and test this new methodology.
We first tried to reproduce the existing reliability flags using supervised classification in order to describe different types of redshift PDFs, but due to the subjective definition of these flags (classification accuracy $\sim$58\%), we soon opted for a new homogeneous partitioning of the data into distinct clusters via unsupervised classification.
After assessing the accuracy of the new clusters via resubstitution and test predictions (classification accuracy $\sim$98\%), we projected unlabeled data from preliminary mock simulations for the Euclid space mission into this mapping to predict their redshift reliability labels.
} 
{
Through the development of a methodology in which a system can build its own experience to assess the quality of a parameter, we are able to set a preliminary basis of an automated reliability assessment for spectroscopic redshift measurements.\\
This newly-defined method is very promising for next-generation large spectroscopic surveys from the ground and in space, such as Euclid and WFIRST.
}
\keywords{
        Methods: data analysis -
        Methods: statistics -
        Techniques: spectroscopic -
        Galaxies: distances and redshift -
        Surveys.
}  

  \authorrunning{Jamal, S. et al.}
  \titlerunning{Automated reliability assessment for spectroscopic redshift measurements}
  \maketitle


\section{Introduction}\label{sec:s1} 
{
        Next-generation experiments in Cosmology face the formidable challenge of understanding dark matter (DM) and dark energy (DE), two major components seemingly dominating the Universe content and evolution. \\
To improve our understanding of the Universe evolution history, the investigation of the distribution of galaxies over large volumes of the Universe at different cosmic times now constitutes  a key requirement for future observational programs such as Euclid (\citealp{laureijs_euclid_2011}), WFIRST (\citealp{green_wide-field_2012}), and LSST (\citealp{ivezic_lsst:_2008}) that will exploit cosmological probes such as 
        {Weak Lensing} (WL) and 
        {Galaxy Clustering} (GC: Baryon Acoustic Oscillations - BAO, Redshift Space Distortions - RSD) 
to define the role of the dark components (\citealp{albrecht_report_2006}).
 
In GC, the detection of the BAOs at the sound horizon scale ($r_s\approx$105 $\rm h^{-1}Mpc$) is used to investigate the role of  DE in the evolution of the expansion through measurements of the Hubble parameter $H(z)$ and the comoving angular distances $D_A(z)$ (\citealp{beutler_6df_2011}), 
while the detection of the distorsions in the redshift space is used to probe the structures' growth and DE models by measuring the parameter combination $g_{\theta}=f(z)\sigma_8(z)$, where $f(z)$ and $\sigma_8$ refer to the growth rate and the RMS amplitude (in a sphere of radius $8$ $\rm h^{-1}Mpc$) of the density fluctuations (\citealp{beutler_6df_2012}), respectively.
The WL is used to map the matter distribution (dark + visible) in the Universe and constrain the expansion history through precise measurements of shapes and distances of lensed galaxies (\citealp{huterer_weak_2002,linder_cosmic_2003}).

In Cosmology, the redshift $z$ is a fundamental quantity, which links distances and cosmic time through the use of a cosmological model.
Accurate redshift measurements are at the core of all modern experiments aiming at precision cosmology for a better understanding of the Universe content, focused on the dominant DM and DE components,
as the cosmological probes GC and WL that require precise redshift measurements 
to build robust statistical models to constrain the DE equation-of-state and investigate the content of the dark Universe (\citealp{abdalla_photometric_2008, wang_designing_2010}).
In particular, 3D galaxy distribution maps from GC measurements entail precise measurements of spectroscopic redshifts, 
while cosmic shear measurements in WL require, along with high-quality imaging and photometry, the selection of sources using redshift measurements for two reasons: First, the galaxies in front of the lens are not affected by the gravitational lensing but they dilute the signal of the galaxy source in the 
background, and second, the galaxies at the same redshift as the lens contribute to the intrinsic alignment that disrupts the WL measurements.

\begin{figure*}[htb!]
        \vspace{0.4cm}
        \centering 
        {\includegraphics[width=0.67\textwidth]{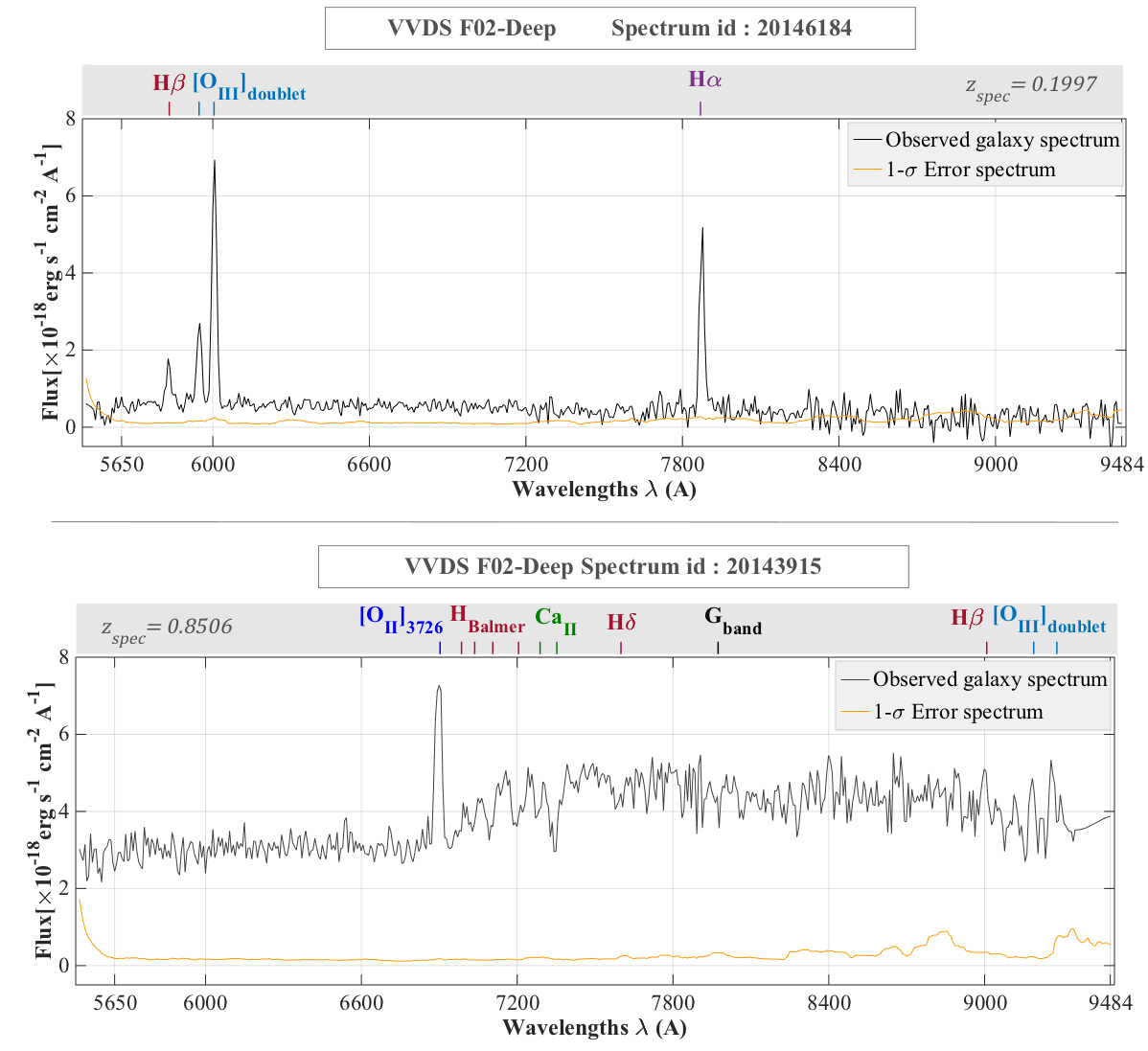}}
        \caption{VVDS-Deep galaxy spectra with identifiable spectral features at known redshifts: emission/absorption line, D4000A break.}
        \label{figure:onedSpec}
        \vspace{0.4cm}
\end{figure*}

As part of the future large-scale experiments in Cosmology designed to address the DE and DM origin, the Euclid mission is a M-Class ESA mission from the ESA Cosmic Vision program that aims to probe the expansion and the LSS growth histories in the Universe.
Through the combination of cosmological probes (Baryon Acoustic Oscillations - BAO, Redshift Space Distortions - RSD, WL, Clusters of galaxies, Supernovae - SNe), Euclid will achieve an unprecedented level of accuracy and control of systematic effects to derive precise measurements of the Hubble parameter $H(z)$, the linear growth rate of structures $\gamma$, the DE equation-of-state parameters ($\omega_p, \omega_a$), the non-Gaussianity amplitude $f_{NL}$ and the RMS fluctuation of the matter over-density $\sigma_8$, among other cosmological parameters (\citealp{laureijs_euclid_2011}).

By covering a large fraction of the sky (Wide: $\sim$15\,000 $\rm deg^2$, Deep: a total of 40 $\rm deg^2$), the mission will perform a photometric survey in the visible and three near-infrared bands to measure the weak gravitational lensing by imaging approximately 1.5 billion galaxies with a photometric redshift accuracy of $\sigma_z/(1+z) \leq 0.05$, in addition to a spectroscopic slitless survey of approximately $ 25 $ million galaxies with a redshift accuracy of $\sigma_z/(1+z) \leq 0.001$ in order to derive precise measurements of the galaxy power spectrum (\citealp{laureijs_euclid_2011}).
The wide-field Euclid survey will be particularly challenging because of the large-size sample of faint distant galaxies, for which the spectroscopic redshifts need to be automatically measured, and their corresponding reliability evaluated.

\noindent For large-scale surveys such as Euclid, the sheer amount of data requires the development of robust and fully automated data-processing pipelines to analyze the data, extract useful information (e.g., redshift) and ensure that all requirements are met.

Distinct approaches to estimate redshifts have been used in a broad range of galaxy surveys.
Photometric redshifts $z_{phot}$ are estimated using spectral energy distribution (SED) template fitting (e.g., Hyper-z \citealp{bolzonella_photometric_2000}, Le Phare \citealp{ilbert_accurate_2006}), classification with neural networks to produce a mapping between photometric observables and reference data (e.g., ANNz, \citealp{collister_annz:_2004}), or Bayesian inference to compute a posterior $z_{phot}$ PDF with prior information from integrated flux in filters, colour or magnitude: BPZ (\citealp{benitez_bayesian_1999}), ZEBRA  (\citealp{feldmann_zurich_2006}), EAZY (\citealp{brammer_eazy:_2008}).
On the other hand, spectroscopic redshifts $z_{spec}$ are estimated from the direct application of cross-correlation or chi-square-fitting methods between the observed data and a reference set of spectroscopic templates (\citealp{tonry_survey_1979}, \citealp{simkin_measurements_1974}, \citealp{schuecker_automated_1993}, \citealp{machado_darth_2013}), or using spectral feature detection (emission/absorption lines and continuum features including  spectral
discontinuities in the UV-visible domain such as the Lyman break or the Balmer and D4000A breaks) that can be very powerful (\citealp{schuecker_automated_1993}).
Some codes (EZ, \citealp{garilli_ez:_2010}) combine spectral lines detection with cross-correlation or chi-square fitting to inject prior knowledge about more plausible redshift solutions.

Despite their overall performances in redshift estimation, most algorithms in use today still suffer from numerous modeling and computational deficiencies, 
as the major recurrent issues with the $z_{spec}$ estimation algorithms remain the strong correlation between reliable spectral feature detection and the quality of the observed spectrum, the difficulty to define a representative set of reference templates, and the use of a pre-generated redshift grid $\Theta_z$ that might be beneficial for rapid and parallel processing but could induce a "bias" regarding the redshift space to probe.

In galaxy surveys, a key issue often overlooked is the necessary evaluation of the quality of a redshift measurement 
because 
spectroscopic redshift measurement methods may be affected by a number of known or unknown observational biases that may produce some errors in the output redshift, ranging all the way to a catastrophic measurement far from the real galaxy redshift.
Further, despite the general trend that consists in linking the reliability of a redshift measurement to the S/N of detected spectral features, the noise in the data usually presents a strongly non-linear dependency on the flux spectrum 
for various reasons (e.g., the wavelength-dependency of the background flux), which makes the definition of a precise redshift reliability criterion even more difficult.



A number of previous faint galaxy surveys have adopted redshift reliability assessments, either by using empirical thresholds applied to a single metric operator (\citealp{baldry_galaxy_2014}, \citealp{cool_prism_2013}), or by combining independent reliability assessments performed by more than two experienced astronomers in order to smooth-out the {observer bias} of each individual and produce a remarkably repetitive reliability assessment (\citealp{le_fevre_vimos_2013, le_fevre_vimos_2015}, \citealp{garilli_vimos_2014}, \citealp{guzzo_vimos_2014}).
%
All methods imply subjective information, either by selecting {"adequate}" thresholds from a constructed sample or by involving a human operator within the (visual) verification process that becomes largely unfeasible for samples over $10^5$ galaxies.
For massive spectroscopic surveys such as Euclid or WFIRST, there is a critical need for a fully automated reliability flag definition that will adapt to the observed data and display a greater use of all available information. 

In this paper, we propose to exploit a Bayesian framework for the spectroscopic redshift estimation to incorporate all sources of information and uncertainties of the estimation process (prior, data-model hypothesis), and produce a full $z_{spec}$ posterior PDF, that will be the starting point of our automated reliability flag definition.

\noindent To test the proposed methodology of assessing the redshift reliability, we use a new redshift estimation software called \textbf{AMAZED} {(\textbf{A}lgorithms for \textbf{M}assive \textbf{A}utomatic \textbf{Z} \textbf{E}valuation and \textbf{D}etermination)} developed as part of the Processing function (PF-SPE) in charge of the 1D spectroscopic data-processing pipeline of the Euclid space mission.

The paper is organized as follows.
After introducing the subject, we present the data used in this study in Section \ref{sec:s2}, and in Section \ref{sec:s3} we describe the Bayesian formalism of the spectroscopic redshift estimation.
Section \ref{sec:s4} is focused on the proposed automated reliability assessment method, where we first describe the principle, then present preliminary results of supervised and unsupervised classification techniques using the public database of the VIMOS VLT Deep Survey (VVDS, \citealp{le_fevre_vimos_2013}).
In Section \ref{sec:s5}, we present our results of redshift reliability predictions using preliminary simulations of Euclid spectra covering a wavelength range $[1.25 - 1.85]$\micro m, and we finally conclude in Section \ref{sec:s6}.
}

\section{Reference data}\label{sec:s2} 
{
\noindent To test the proposed method of assessing a redshift reliability, we use public data from the VIMOS VLT Deep Survey\footnote{\scriptsize\url{http://cesam.lam.fr/vvds/}} (VVDS) in this study.
The large VIMOS VLT Deep Survey  (\citealp{le_fevre_vimos_2013}) is a combination of three i-band magnitude limited surveys: Wide (17.5$\leq i_{AB}\leq$22.5; 8.6 $\rm deg^{2})$, Deep (17.5$\leq i_{AB}\leq$24; 0.6 $\rm deg^{2}$) and Ultra-Deep (23$\leq i_{AB}\leq$24.75; 512 $\rm arcmin^{2}$), that produced a total of 35526 spectroscopic galaxy redshifts between 0 and 6.7 (22434 in Wide, 12051 in Deep and 1041 in UDeep) with a spectral resolution ($R\simeq 230$, dispersion 7.14Å) approaching that of the upcoming Euclid mission ($R\geq380$ for a 0.5\arcsec object, dispersion 13.4Å) as illustrated in Figure~\ref{figure:onedSpec}.

The VIPGI ({VIMOS Interactive Pipeline and Graphical Interface}) data-processing software included background subtraction, decontamination, filtering and extraction of 1D spectra from 2D spectral images using sophisticated packages (\citealp{scodeggio_2005}).
The VIMOS 1D spectroscopic data was processed using the EZ software (\citealp{garilli_ez:_2010}) to compute spectroscopic redshift measurements and reliability flags by combining reliability assessments (visual checks) of at least two experienced astronomers (\citealp{le_fevre_vimos_2013}).
The VVDS project provides a reference sample with a range of redshifts and reliability flags well-suited for testing our methods in a broad parameter space. 

To evaluate our automated redshift reliability assessment method (\S \ref{sec:s4}), we use the VVDS data in two stages. 
First we exploit the existing redshift reliability flags of the VVDS data as a reference to assess the performances of supervised classification algorithms in predicting a similar redshift reliability label.
Then, after partitioning the VVDS data into distinct clusters of redshift reliability flags using unsupervised classification, we compare these results with the original VVDS redshift flags to evaluate the performances of the proposed methodology and unveil possible discrepancies.
}

\section{Spectroscopic redshift estimation}\label{sec:s3} 
{
\begin{figure*}[ht]  
        \vspace{0.4cm} \centering
        \includegraphics[width=0.675\textwidth]{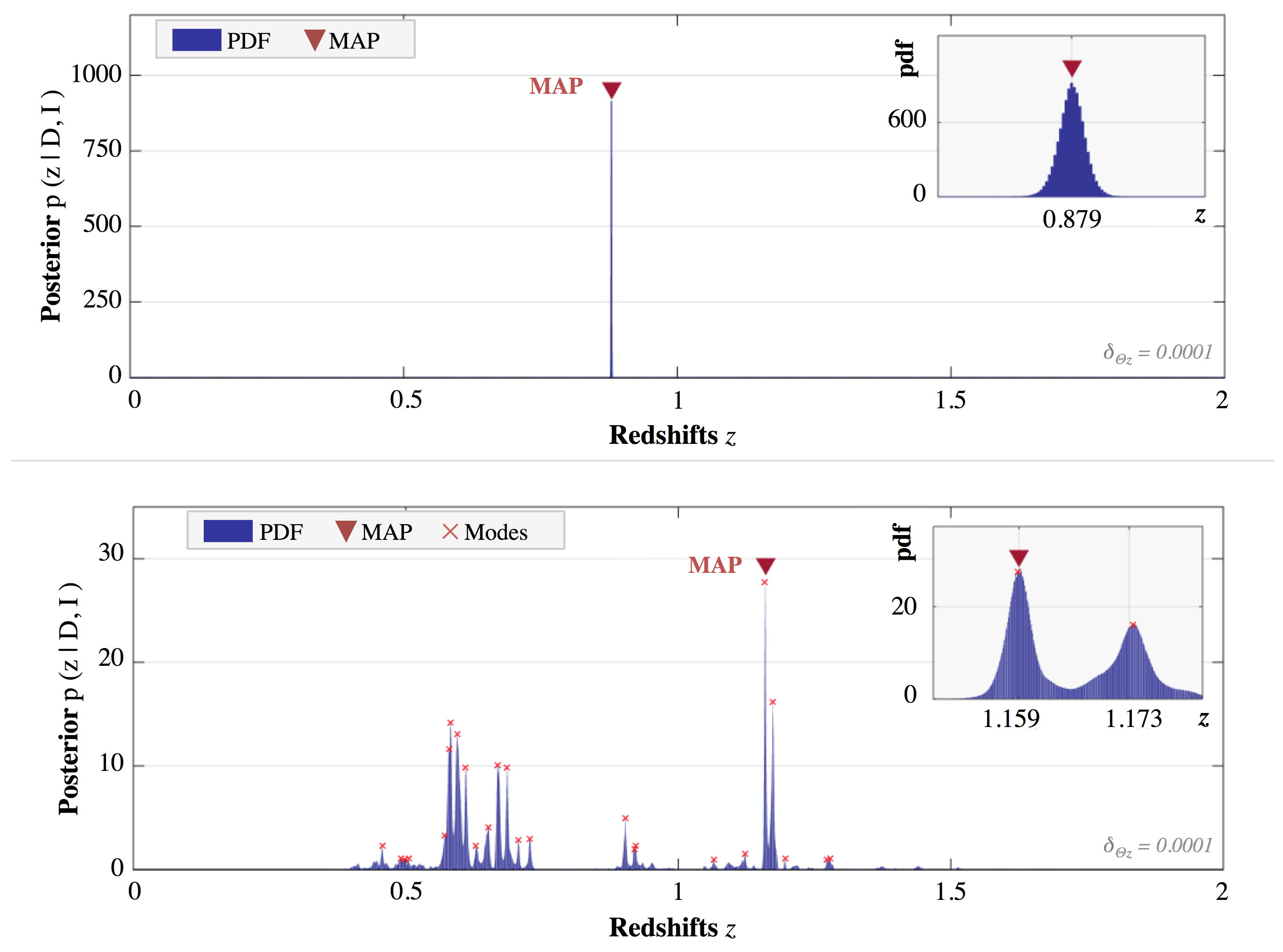}
        \caption{
                \color{dkgray2}Posterior redshift PDF of two VVDS-Deep spectra. \color{black}
                \textit{Top}: A unimodal zPDF characterizes a very reliable redshift measurement (a single peak at z=0.879).
                \textit{Bottom}: A multimodal zPDF refers to multiple redshift solutions (multiple peaks) possibly with similar probabilities 
                        associated to a diminished confidence level of the MAP estimate at z=1.159.
                The quantity $\delta_{\Theta_z}$ refers to the fixed step of the redshift grid used to compute the zPDFs.
        }
        \label{figure:pdf_modes}        
\end{figure*}

\subsection{Description}
{
To derive a redshift, the widely used template-based algorithms rely on the hypothesis that "there exists a reference template spectrum that is a true (and sufficient) representation of the observed data", implying that the observed spectrum can be described by at least one spectroscopic template of the reference library.
 
Using a set of rest-frame templates and a fixed grid of redshift candidates in $\Theta_z$, for each pair (redshift $z$, template $M_t$) we compute the Least-Square metric: 
\begin{equation}
    \label{fq}
        \chi^2{(z, t)}= {\sum_{i\in\Lambda} {\sigma_i}^{-2} (d_i - a\hspace{0.01cm}t_{i,z})^{2}} , z\in \Theta_{z}   
     \vspace{-0.2cm}
,\end{equation}   
 or the cross-correlation: 
 \begin{equation}
    \label{fq}
        xc{(z, t)}= \frac{1} { \sigma_d\hspace{0.05cm}\sigma_{t,z}} \sum_{i \in \Lambda} (d_i-\mu_{d})(t_{i,z}-\mu_{t,z} ){\sigma_i}^{-2}
                , z\in \Theta_{z}
,\end{equation}
where 
$d_i$ and $\sigma_i$ refer respectively to the observed flux and noise spectra at pixel $i$,
$t_{i,z}$ is the redshifted template interpolated at pixel $i$,
and ($\mu_{t,z}$;$\sigma_{t,z}$) and ($\mu_{d}$;$\sigma_{d}$) are the mean and standard-deviation of the redshifted template and the observed spectrum respectively.
The wavelength range in use $\Lambda$ contains $n$ datapoints, $\Theta_z$ refers to the redshift space to probe, 
and $a$ is a scale factor referring to the amplitude of the redshifted template that is usually computed at each trial from (weighted) least-square estimation. 

\noindent The estimated $z_{spec}$ results from a $\bf{joint}$-estimation of the pair $(z, M_t)$ and is performed by optimizing a chosen metric: maximization of the cross-correlation function or minimization of the chi-square operator.\\
In general, the accuracy of the template-based methods is tied to the representativeness and wavelength coverage of the spectroscopic templates $M_t$ in use.
}

\subsection{Bayesian inference} 
{
Assuming a linear and Gaussian data model with i.i.d. (independent and identically distributed)} residuals $\{N_i\}_{i\in\Lambda}$, the probability of observing the spectrum $\{D_i\}_{i\in\Lambda}$, at a redshift $z$ given a template model $M_t$ and any additional information $I$  is described by the likelihood function $\mathcal{L}(z,M_t )$ (cf. Appendix-\ref{subsec:app0}): 
\begin{equation}\label{fq}
        \begin{split}
                \mathcal{L}(z,M_t ) = {p}( D | z,M_t,I) = \prod_{i\in\Lambda}{p(N_i  | z,M_t,I)}  \hspace{0.60cm}\\
                = \prod_{1\leq i\leq n}{{(\sqrt{2\pi}\sigma_i)}^{-1}} \exp^{- \frac{1}{2} \chi^2{(z, t)} } \hspace{1.16cm}
        \end{split}
,\end{equation}
\begin{equation}\label{fq}
        \begin{split}
                \ell_{(z,M_t )} = \log{\big(\mathcal{L}(z,M_t ) \big)}  \hspace{3.2cm}\\
                = -\frac{1}{2}\chi^2{(z, t)}-\frac{N}{2}\log{(2\pi)} - \sum_{i\in\Lambda}{\log{(\sigma_i)}}
                \end{split}
.\end{equation}

\noindent Via the Bayes rule, the joint posterior distribution is:
\begin{equation}
    \label{fq}
        {p}( z ,M_t  | D ,I) = \frac{{p}( D | z ,M_t  ,I)  \times {\pi}{(z,t)}}{p(D | I ) }
,\end{equation}
\vspace{-0.2cm}
\begin{equation}
    \label{fq}
    \begin{split}
        \log{\big({p}(z ,M_t  | D ,I) \big)}= -\frac{1}{2}\chi^2{(z, t)} + \log{({\pi}{(z,t)})} \hspace{0.5cm} \\
        - \log{\left( \iint_{z, M_t}{{\pi}{(z,t)}\exp^{-\frac{1}{2}\chi^2{(z, t)}} \,dz\,dM_t}\right)} \hspace{-0.5cm}
        \end{split}
\end{equation}
where ${\pi}{(z,t)}$ is the joint-prior distribution of the pair $(z, M_t)$.

\noindent The 1D posterior distribution is obtained by marginalizing over $M_t$:
\begin{equation}
    \label{fq}
        {p}( z | D,I )   =\int_{M_t}{{p}( z ,M_t  | D ,I)  \,dM_t }
.\end{equation}
The "best" redshift $\widehat z_{spec}$ is the MAP\textit{(Maximum-A-Posteriori)} estimate:
\begin{equation}
        z_{\textrm{MAP}} = \textrm{argmax}_z \hspace{0.1cm} p( z | D,I )
.\end{equation} 

\indent This Bayesian formalism was not clearly stated for the spectroscopic redshift estimation. As for now, a posterior $z_{spec}$ PDF can be computed and prior information, if available, can easily be integrated.

\noindent Furthermore if the hypothesis of the datamodel is readjusted, the equations can be rapidly and accurately revised in the likelihood expression (cf. Appendix-\ref{subsec:app0}).

The template library used in this study includes a set of 9 continuum spectra of spiral, elliptical, starburst, and bulge galaxies, supplemented with 12 templates displaying different shapes and level for the continuum and the emission lines that were built by the VVDS team to take into account the diversity of galaxy spectra observed during the survey.\\
The spectroscopic templates that had only optical data were extended in the UV down to 912Å by exploiting the closest templates with UV data, and below 912Å by using nul flux spectra. In the infrared, a blackbody continuum was used to extrapolate the templates up to 20000Å.\\
This large wavelength coverage ensures that the intersection between the observed spectra and the templates is verified at each redshift trial.
}

\subsection{Numerical computation}
{
In the Bayesian inference, if our state of knowledge about a certain quantity $\theta$ is vague, a non-informative prior, such as the flat prior, is usually computed.
\begin{equation}
        \int_{\Delta \theta}  p{(\theta | data)} \,d\theta = 1 
.\end{equation}
Using a flat prior for redshift estimation implies that all redshifts and all templates are viewed as equiprobable solutions. The estimation algorithm will explore the full template library and the entire redshift grid and compute a (marginalized) posterior redshift PDF as displayed in Figure~\ref{figure:pdf_modes}.

\noindent If extra information about the pair $(z, M_t)$ is available, the joint prior will be more informative as it will display a refined structure in the $(z, M_t)$ space.
For example, to estimate photometric redshifts, integrated flux in filters, colour, or magnitude can be used as priors to efficiently probe the redshift space. In \citealp{benitez_bayesian_1999}, the joint-prior $p(z,T | m_0)$ provides additional information about the most eligible spectral objects, $T,$ with a magnitude, $m_0,$ that could be observed at certain redshifts, $z$.
\noindent However, for spectroscopic redshift estimation, there is no clear definition of a {(data-independent)} prior, a choice justified by the fact that spectroscopic data is more informative than photometry.
}  
}

\section{Reliability assessment}\label{sec:s4} 
{
As the size of massive surveys in astronomy continues to expand, assessing redshifts' reliability becomes increasingly challenging.
The need for fully automated reliability assessment methods is now part of the requirements for future surveys, and is justified by the fact that automation provides predictable and consistent performances while the behavior of a human operator remains unpredictable and often inconsistent 
and therefore can require several independent observers to smooth out personal biases. 

\noindent Moreover, the need for automation comes from the orders-of-magnitude increase in the total number of spectra that need to be processed.
Visual examination of all spectra in a survey (2dF, DEEP2, VVDS, VIPERS, zCOSMOS, VUDS, PRIMUS, etc.) is extremely difficult for samples containing $10^5$ objects or more, and will be completely impossible for next-generation spectroscopy surveys with more than $50 \times 10^6$ objects.

In general, existing approaches to automate the reliability assessment as well as the associated quality control in most engineering applications, such as the intrusion detection systems (IDS) that aim to evaluate the traffic quality by identifying any malicious activity or policy violation within a network, include: 
\begin{enumerate}
        \item \textbf{Anomaly detection systems} ({ADS}), where a component is labeled as an outlier if it deviates from an expected 
                behavior using a set of thresholds or reference data (\citealp{chandola_ads:_2009}, \citealp{patcha_ads:_2007}). 
                The ADS usually proceed by monitoring the system activity and detecting any sort of violation based on specific criteria or invariable 
                standards.
        \item \textbf{Supervised classification} that exploits prior knowledge of a referenced training set to predict a label (\citealp{shahid_imageQual_2014}). 
\end{enumerate}
\noindent Both methods deliver great performances in general, but still have some limitations: irrelevant thresholds to new data for the ADS, and poor representativity of the training set in classification, and so on.

\noindent To automate the redshift reliability assessment, reproducing the ADS reasoning scheme by setting empirical thresholds might not be the best option when dealing with massive surveys. However, the use of machine learning (ML) techniques can still be a viable option but first requires the search for a valid model and a coherent set of entries.

In this work, the method to automate the redshift reliability flag definition stems from an attempt to address questions about the meaning of a "{reliable}" redshift: 
\begin{enumerate}
        \item What guides an experienced astronomer to declare an estimated redshift as a plausible solution; apart from visual inspection of the data and its fitted template?
        \item Is there some disregarded information within the z-estimation process that      we can further exploit? 
        \item How can a system "{perceive}" the same information as a human does?
\end{enumerate}

Spectroscopic redshift measurements are obtained from $\chi^2$ minimization or maximization of the posterior probability $p(z|D,I)$ in Bayesian inference (cf. \S \ref{sec:s3}), and usually no further analysis of the computed functions is conducted afterwards. 
When computing the posterior redshift PDF, broadly two types of probability density function can be observed (cf. Figure~\ref{figure:pdf_modes}): a unimodal PDF versus a multimodal distribution.
In both cases, a pipeline will provide a redshift estimation $z_{\textrm{MAP}}$ but the estimated redshifts from these two different types of PDFs  definitely do not show the same level of reliability.
In fact, the multimodal PDF refers to numerous redshift candidates possibly with similar probabilities, while a strong unimodal PDF with a prominent peak and low dispersion depicts a more "{reliable}" redshift estimation of the data.

We exploit such characteristics of the posterior PDF to build a discretized descriptor space that will be the entry point for ML techniques to predict a reliability label. Our approach aims to build the "{experience}" of an automated system in order to assess the quality of a redshift measurement from the zPDF.
}

\subsection{Description} \label{subsec:s41}
{
In machine learning, the typical entries of the model are a response vector $\textbf{Y}$ and a feature matrix $\textbf{X}$:
\begin{equation}
        \bf{X} = 
                \begin{pmatrix}
                          \textit{\textbf{x}}_{1}\\ \vdots\\ \textit{\textbf{x}}_{M}\\
                \end{pmatrix}
                = 
        \hspace{-0.2cm}
                \begin{blockarray}{cccc}
                        &\textcolor{dkgray}{d_1} &  & \textcolor{dkgray}{d_P} \\
                        \begin{block}{c(ccc)}
                          	\textcolor{dkgray}{s_1}&x_{11}	&\cdots&x_{1P}\\
                                         				    & \vdots&\ddots&\vdots\\
                          	\textcolor{dkgray}{s_M}&x_{M1}&\cdots&x_{MP}\\
                        \end{block}
                \end{blockarray}
        \hspace{0.1cm};\hspace{0.1cm}
        \bf{Y} = 
                \begin{pmatrix}
                          y_{1}\\ \vdots\\y_{M}\\
                \end{pmatrix}
,\end{equation} 
where $\textit{\textbf{x}}_j= (x_{j,1}\dots x_{j,P})$ is the $P$-dimensional feature vector of the $j$-th observational data $(s_j)_{j \in\{1,\dots M\}}$, and $y_j$ is its response variable.

\noindent If the response vector $\bf{Y}$ of the model is unknown, the prediction of a label $y_j$ using only the distribution of $\bf{X}$ in the feature space refers to {clustering} (unsupervised classification). 
Otherwise, we talk about supervised classification whose goal is to define a mapping between the observable entries $\bf{X}$ and their associated response variables $\bf{Y}$ through a dual training/test scheme.
 
In ML, the design of the entry model is decisive. What could be the optimal selection of informative and independent features to accurately describe the zPDF? Can a single operator, such as the integral under the redshift solution $z_{\rm MAP}$ or the difference in probability between the first two peaks (modes), be a unique and sufficient descriptor?
No definite answers can be given, since this approach of "quantifying the  spectroscopic redshift reliability" from the zPDF is new. 
Each set of selected features will define a different descriptor space that a classifier could separate differently.

In this study, our selected ML entries are redshift reliability flags ($\textbf{Y}$) and descriptors of the zPDFs ($\textbf{X}$), where
the feature vector $\textit{\textbf{x}}_j$ associated to the observation $s_j=\{D\}$ consists of a list of eight tailored descriptors of the zPDF:
\begin{itemize}
        \item [-] The quantity $\textrm{P}(z_{\textrm{MAP}} | D, I) \approx p(z_{\textrm{MAP}} | D, I) \times \delta_{\Theta_z}$, 
                where $\delta_{\Theta_z}$ is the fixed step of the redshift grid.
        \item [-] The number of {significant} modes in the PDF. 
                The "{significance}" of a mode is determined by partitioning the set of detected peaks of the PDF into two categories (strong/weak) 
                based on their prominence and height in order to avoid including the extremely-low density peaks ($10^{-100}$ usually) that result 
                from the conversion of logPDFs into a linear scale.
        \item [-] The difference in probability of the first two {best} redshift solutions ($z_{MAP}, z_2$): 
                $\textrm{P}(z_{\textrm{MAP}} | D, I) - \textrm{P}(z_{2} | D, I) $
        \item [-] The dispersion $\sigma=[{\int (z-\bar{z})^2 p(z)dz}]^{1/2}$, 
                with $\bar{z}\hspace{-0.05cm}=\hspace{-0.05cm}{\int z p(z)d z}$.
        \item [-] The cumulative probability in the region $R_2^{*}$:\\ $\big[z_{\rm MAP} \pm \delta \big]$ where the parameter $\delta$ is chosen 
                equal to 0.001.
        \item [-] The characteristics of the $\rm CR^{*}$(restricted version of the {\textbf{C}redibility \textbf{R}egion} with 95$\%$ in probability): 
                number of z candidates, width $\Delta z$, cumulative probability.\\
                In Bayesian Inference, the CR is analogous to the frequentist CI ({\textbf{C}onfidence \textbf{I}nterval}).\\
                For a $100(1-\alpha)\%$ level of credibility, the CR is defined as: $\int_{CR} p(z|D,I)dz = 1-\alpha$.\\
                The restricted $\rm CR^{*}$ used sets (optional) maximal bounds to the search region around $z_{\rm MAP}$ to accelerate the operation.
\end{itemize} 

Displays of distinct zPDFs are presented in Figures \ref{figure:zpdf1} to \ref{figure:zpdf4}, where the descriptors listed above highlight interesting features of the zPDFs.
For example, it is possible to obtain a similar dispersion for two zPDFs but a different number of significant redshift modes (\ref{figure:zpdf1}),
or the other way around: multimodal zPDFs with a comparable number of redshift modes with different amplitudes, and a different dispersion  (cf. Figure \ref{figure:zpdf2}) or difference in probability between the first two best redshift solutions (cf. Figure \ref{figure:zpdf3}).
Also, unimodal zPDFs can vary as they can display wider or narrower restricted CR (cf. Figure \ref{figure:zpdf4}) or different values of the dispersion $\sigma$.

Using the eight listed key descriptors, we estimate that the main features of the zPDF can be inferred.
This design is not immutable. Supplementing the feature matrix with additional information about the observed spectra, $s$, or designing a different feature selection can also be explored.
}

\subsection{Classification}\label{subsec:s42}
{

\subsubsection{Model}\label{subsubsec:s421}
{
\indent The ML entries in this study are obtained from a collection of zPDFs computed from M spectra of the VVDS to which a reliability label $(y_i)_{i\in\{1,\dots M\}}$ is known to belong to one of the flags (\citealp{le_fevre_vimos_2005,le_fevre_vimos_2013}):
\begin{itemize}
        \item [-] Flag 1, "{Unreliable redshift}''. 
        \item [-] Flag 2, "{Reliable redshift}''.
        \item [-] Flag 9, "{Reliable redshift, detection of a single emission line}''.
        \item [-] Flag 3, "{Very reliable redshift with strong spectral features}''.
        \item [-] Flag 4, "{Very reliable redshift with obvious spectral features}''.
\end{itemize} \vspace{-0.1cm}

\noindent The redshift reliability flags in the VVDS are determined by confronting independent redshift measurements performed by several observers on the same spectra. 

\noindent By comparing the redshift measurements with internal duplicated observations or with published redshifts from different surveys, the VVDS spectroscopic redshift flags have been empirically paired with a probability for "a redshift to be correct":
the VVDS redshift reliability flags $\{1, 2, 9, 3, 4\}$ are associated with probabilities of [50-75]\%, [75-85]\%, $\sim$80\%, [95-100]\%, and 100\% ,  respectively, that the measured redshifts are correct.

Using supervised classification, the objective is to predict similar redshift reliability flags for new unlabeled data.
However, since the reproducibility of the VVDS redshift reliability flags is difficult because of their subjective definition and the confusion between "quality of a redshift" and "specific information about the data", we first decided to regroup the VVDS flags, as following: 
\begin{itemize}
        \item[-] "Class   0", consisting of the "VVDS flags 1” to depict the {uncertain} redshifts.
        \item[-] "Class +1", consisting of the "VVDS flags 2-9” to depict the {reliable} redshifts.
        \item[-] "Class +2", consisting of the "VVDS flags 3-4” to depict the {very reliable} redshifts.
\end{itemize}

\noindent A three-class classification problem is then set.
For multi-class problems, the ECOC ({Error-Correcting-Output-Codes}), as introduced in \citealp{dietterich_solving_1995}, are adapted for several learners, such as SVM ({Support Vector Machines}), Tree templates, and Ensemble classifiers.
A description of the ECOC is provided in Appendix-\ref{subsec:app1}.
}

\subsubsection{Preliminary tests} \label{subsubsec:s422}
{
Classification tests are conducted using a VVDS subset of 24519 spectra with a constraint on the redshift accuracy $|z_{\rm MAP} - z_{ref}| / (1+z_{ref}) \leq 10^{-3}$ for the VVDS flags $\{2, 9, 3, 4\}$. Our main objective is to build a descriptor space from a diverse set of zPDFs and evaluate the ability of the system to predict a redshift reliability label.\\
The dataset is decomposed into a "Training set" and a "Test set" (cf. Tables \ref{table:vvdsf02}, \ref{table:TrainSetTab_vvds} and \ref{table:TestSetTab_vvds}).
\begin{table}[h!]{
        \begin{center}{
        \caption{Description of the VVDS dataset used in this study}\vspace{-0.2cm}
        \mynewformat
        \begin{tabular}{ |c|c|c|c|c| } 
        \hline
         \scshape{Type} &  \scshape{z Reliability Flags}  &     \scshape{Counts}         &        \scshape{z Range} \\ 
        \hline
        \multirow{4}{4em}{\centering{Primary objects\\}} 
                &       {"Unreliable"}\hspace{0.45cm}1   &       6768            &       0.0070  -       5.2280  \\
                &       {"Reliable"}\hspace{0.7cm}9      &       632             &       0.0195  -       4.9285  \\
                &       {"Reliable"}\hspace{0.7cm}2      &       4743            &       0.0017  -       4.4345  \\
                &       {"Very reliable"}        3               &       6455            &       0.0266  -       4.5400  \\
                &      {"Very reliable"}         4               &       5921            &       0.0213  -       3.8352  \\
                \hline                          
        \end{tabular}
        \label{table:vvdsf02}
        }\end{center}
} {
\parbox{.5\linewidth}{
        \begin{center}
        \caption{Training set \\(total of 16346 VVDS spectra)}\vspace{-0.2cm}
        \mynewformat
        \hspace{-1cm}
         \begin{tabular}{ c|c|c|c| } 
        \cline{2-4}
        & \scshape{Label} &  \scshape{Counts}    &      \%      \\
        \cline{2-4}
        \multirow{4}{*} {\centering{\rotatebox[origin=c]{90}{\hspace{1cm}\scshape{\color{dkblue}Train set}\hspace{-0.1cm}}}}
                &       " 0 "   &       4512            &       27.60   \\
                &       "+1"    &       3583            &       21.92   \\
                &       "+2"    &       8251            &       50.48   \\
                \cline{2-4}
        \end{tabular}
        \label{table:TrainSetTab_vvds}
        \end{center}
} \hspace{-0.2cm}
\parbox{.5\linewidth}{
        \begin{center}
        \caption{Test set \\ (total of 8173 VVDS spectra)}\vspace{-0.2cm}
        \mynewformat
        \hspace{-1cm}
        \begin{tabular}{ c|c|c|c| } 
        \cline{2-4}
        & \scshape{Label} &  \scshape{Counts}    &      \%\\ 
        \cline{2-4}
        \multirow{4}{*} {\centering{\rotatebox[origin=c]{90}{\hspace{1cm}\scshape{\color{orng}Test set}\hspace{-00cm}}}}
                &       " 0 "   &       2256    &       27.60   \\
                &       "+1"    &       1792            &       21.93   \\
                &       "+2"    &       4125            &       50.47   \\
                \cline{2-4}
        \end{tabular}
        \label{table:TestSetTab_vvds}
        \end{center}
        } \vspace{-0.3cm}
}\end{table}

Different classifiers are tested in this study to carry out a careful analysis and avoid blindly trusting the results in cases of overfitting.
We assess that different techniques should provide a different but not very disparate level of performance.
Three classifiers are selected: 
        the SVMs ({Support-Machine Vectors}) with linear and Gaussian kernels, 
        an ensemble of bagging trees (referred to simply as TreeBagger) 
        and a GentleBoost ensemble of decision trees.
A general description of the classifiers and the multi-class measures is provided in Appendices \ref{subsec:app3} and \ref{subsec:app5}.

To evaluate the performance of a classifier, two tests are conducted: 
\begin{itemize}
        \item[-] \textbf{\color{dkgray2}Test 1} :  \color{black}{Resubstitution.} 
        \item[-] \textbf{\color{dkgray2}Test 2} :  \color{black}{Test prediction.}
\end{itemize} 

\noindent In the resubstitution, the "Training set" is reused as the "Test set" during the prediction phase.
\noindent Extremely low prediction errors are expected ($\lesssim1\%$ classification error rate): if a bijective relation exists between the observables $\bf{X}_{\rm train}$ and the response vector $\bf{Y}_{\rm train}$, the generated mapping from the training phase is supposedly accurate.
The predicted labels $\bf{Y}_{\rm pred}$ in resubstitution tests are therefore expected to resemble the true labels $\bf{Y}_{\rm train}$ with high accuracy, otherwise a clear mismatch between the features matrix $\textbf{X}$ and the response vector $\textbf{Y}$ of the ML model is reported.
In such a case, the predictions of the second test ("{Test prediction}") would be baseless, since the mapping produced from the training phase is truly unusable.
The overall performances reported in Tables \ref{table:cm_vvds_resub} and \ref{table:cm_vvds_pred} in addition to the confusion matrices (cf. Tables~\ref{table:cm_vvds1} to~\ref{table:cm_vvds8}) representing the fraction of the predicted labels versus the true classes in $\bf{Y}_{\rm test}$, support this conclusion.
Most classifiers seem unable to predict the true labels in resubstitution: non-zero off-diagonal elements in the matrices and a high error-rate,
implying that a correct mapping between the feature matrix and the existing VVDS redshift reliability flags cannot be produced.\\
We would like to point out the singular case of the TreeBagger that seems to generate a good mapping in resubstitution (error rate 0.08\% on average) in comparison with the SVMs that are commonly-known as robust classifiers (error rate $>$10\% in average).
It seems reasonable to consider that the observed dissimilarity between the different classifiers in resubstitution is due to the sensitivity of the bagging trees to several parameters as the number of learners or the trees depth that can coerce the training into focusing on irregular patterns and establish an erroneous mapping).
As anticipated from the resubstitution results (high error rate), we also find that the test predictions present a significant error rate ($\sim40$\% on average).%

To summarize, these first results of supervised classification show that trying to match the subjective VVDS flags with descriptors of the zPDF gives poor results.\\
The entries and hypotheses for ML have to be reexamined.
}
}       

\subsection{Clustering and fuzzy classification}\label{sec:s43}
{
From the previous results, doubts can be raised regarding the engineered zPDF feature space derived from a collection of 24519 VVDS spectra. 
However the selected set of descriptors seems to be a viable description that portrays an existing but hidden structure of the feature space.

Clustering, known as {unsupervised classification}, is used in this Section to unveil the intricate structure and bring into light some properties of the data in the descriptor space.

\subsubsection{Partitioning the descriptor space}\label{sec:s431}
        {
In unsupervised classification, prior knowledge about class membership is unavailable. 
Partitioning the descriptor space into K manifolds is realized by applying separation rules only to the feature matrix $\textbf{X}$. 

\noindent By representing the zPDFs feature matrix $\bf{X}$ in 3D (cf. Figure~\ref{figure:disp_X}), a simple bi-partitioning is introduced: 
\begin{itemize} 
        \item [-] \textbf{Group 1}: high dispersion and low $\textrm{P}(z_{\rm MAP}|D, I)$ referring to multimodal PDFs or platykurtic unimodal PDFs.
        \item [-] \textbf{Group 2}: medium dispersion and high $\textrm{P}(z_{\rm MAP}|D, I)$ depicting strongly peaked unimodal PDFs.
\end{itemize} 

\noindent In each category, we choose to reapply a bi-partitioning to decompose the data into a dichotomized pattern (cf. Figure~\ref{figure:dicho}).
This partitioning strategy, applied to the entire descriptor components and not only to the two descriptor components as in the displays, alongside with the number of clusters, the feature selection and the ML algorithms tested in this work as a novelty to automate the redshift reliability, are not immutable and can be readjusted according to the data in hand.
Further evaluations will be conducted on these aspects of ML to develop a robust and precise automated assessment of redshift reliability.

Using the classic clustering algorithm FCM ({Fuzzy C-Means}) to minimize the intraclass variance (cf. Appendix-\ref{subsec:app4}), the final groups identify distinct partitions in the feature space (cf. Figures \ref{figure:repres_cluFla} to \ref{figure:rep_pzmap}).
In this study, the selection of the number of clusters is an empirical process based on the analysis of the intermediate partitions and testing different configurations.
We assess that the final architecture is a viable solution amongst others.

\begin{itemize}
        \item[-] \color{dkgray}\textbf{"Cluster C1"}: \color{black}Highly dispersed PDFs 
                                with multiple equiprobable modes, $\textrm{P}(z_{MAP})\sim0.028\pm0.023$.
                                \vspace{0.1cm}
        \item[-] \color{dkgray}\textbf{"Cluster C2"}: \color{black}Less dispersed PDFs, 
                                with few modes and low probabilities $\textrm{P}(z_{MAP})\sim0.087\pm0.033$.
                                \vspace{0.1cm}
        \item[-] \color{dkgray}\textbf{"Cluster C3"}: \color{black}Low $\sigma$, 
                                {intermediate} probabilities\hspace{0.05cm}$\textrm{P}(z_{MAP})$ $\sim 0.166 \pm 0.035$.
                                \vspace{0.1cm}
        \item[-] \color{dkgray}\textbf{"Cluster C4"}: \color{black}Unimodal PDFs 
                                with low dispersion, higher probabilities $\textrm{P}(z_{MAP})\sim0.290\pm0.059$.
                                \vspace{0.1cm}
        \item[-] \color{dkgray}\textbf{"Cluster C5"}: \color{black}Strong unimodal PDFs 
                                with extremely low dispersion, better probabilities $\textrm{P}(z_{\rm MAP})\sim 0.618 \pm 0.204$.
\end{itemize} 

\noindent The coordinates of the clusters' centroids in the descriptor space are reported in Table~\ref{table:centroids}.
\begin{equation} 
        \left\{
        \begin{array}{ll}
                \text{Class } \{C_k\} \text{ centroid}
                        &       \hspace{-0.2cm}
                        \textit{\textbf{g}}_k=\frac{1}{M_k} \sum_{y_j\in C_k} \textit{\textbf{x}}_j      \\
                \text{Class } \{C_k\} \text{ variance}
                        &       \hspace{-0.2cm}
                        \textit{\textbf{V}}_k= \frac{1}{M_k} \sum_{y_j\in C_k} (\textit{\textbf{x}}_j - \textit{\textbf{g}}_k)^{\top} (\textit{\textbf{x}}_j - \textit{\textbf{g}}_k)
        \end{array}
        \right. 
,\end{equation}
where $M_k$ is the number of elements in cluster $C_k$.

\noindent Tables~\ref{table:dispersion} and \ref{table:dispersion_full} report the intraclass dispersion $\sqrt{W}$ and the interclass dispersion $\sqrt{B}$ that characterize the newly defined clusters:
\begin{equation}
        \left\{
        \begin{array}{lll}
                \text{Interclass variance}
                        & \hspace{-0.2cm}   
                        \textit{\textbf{B}} = \frac{1}{M}\sum_{k}M_k(\textit{\textbf{g}}_k - \textit{\textbf{g}})^{\top} (\textit{\textbf{g}}_k - \textit{\textbf{g}}) \\
                \text{Intraclass variance}
                &       \hspace{-0.2cm} 
                        \textit{\textbf{W}}= \frac{1}{M} \sum_{k} M_k \textit{\textbf{V}}_k  
        \end{array}
        \right.
,\end{equation}
with the total variance:
\begin{equation}
      		  \textit{\textbf{V}}= \frac{1}{N} \sum_{j=1}^{M} (\textit{\textbf{x}}_j - \textit{\textbf{g}})^{\top}(\textit{\textbf{x}}_j - \textit{\textbf{g}})  
                		= \textit{\textbf{B}} + \textit{\textbf{W}}     
,\end{equation}  
where $\textit{\textbf{g}}$ is the global centroid and $M$ is the full number of elements in the descriptor space.
\begin{table}[htp!]{
        \begin{center}
        \caption{Coordinates of the clusters' centroids in the descriptor space} \vspace{-0.2cm}
        \mynewformat
                \begin{tabular}{ |c|c|c|c|c|c| }
                \hline  
                \scshape Selected  &\multicolumn{5}{c|}{{\scshape{   $\text{Class } \{C_k\} \text{ centroid}$    }}}\\
                \cline{2-6}
                 \scshape descriptors&C1&C2&C3&C4&C5\\ %
                \hline
                Dispersion $\sigma$                                            &       0.524   &       0.049   &       0.005   &       0.002   &       5e-4            \\
                $\textrm{P}(z_{\rm MAP}|D, I)$                          	&       0.028   &       0.087   &       0.166   &       0.290   &       0.618   \\
                $Card$(z $\in$ $\rm CR^{*}$)                            	&       24.06   &       20.98   &       11.16   &       6.45            &       3.16            \\
                %
                Width $\Delta$z $\in$ $\rm CR^{*}$                    &       2.3e-3  &       2.0e-3  &       1.0e-3  &       5.5e-4  &       2.2e-4  \\
                $\sum_i$ \textrm{P}($z_i$ $\in$ $\rm CR^{*}$)   &       0.387   &       0.890   &       0.957   &       0.964   &       0.978   \\
                $\sum_i$ \textrm{P}($z_i$ $\in$ $R_2^*$)           &       0.364   &       0.884   &       0.998   &       1.0             &       1.0             \\
                %
                Significant peaks                                               	&       107.89  &       2.30            &       1.27            &       1.05            &       1.00            \\
                $\Delta$\textrm{P}(two "{best}" z)              		&       0.013   &       0.049   &       0.130   &       0.274   &       0.743   \\
                \hline \hline
                \cellcolor{lightGray}Nb elements $M_k$          	&       3156             &       6720            &       5677            &       4966            &       4030            \\
                \hline
                \end{tabular}
        \label{table:centroids}
        \end{center}
}
\vspace{0.15cm}
\centering \rule{8cm}{0.4pt}    
\vspace{-0.15cm}
{
        \begin{center}
        \caption{Intraclass dispersion in the descriptor space}\vspace{0.2cm}
        \mynewformat
                \begin{tabular}{ |c|c|c|c|c|c| }
                \hline  
                \scshape Selected  &\multicolumn{5}{c|}{{\scshape{  $\text{Class } \{C_k\} \text{ dispersion}$   }}}\\
                \cline{2-6}
                 \scshape descriptors&C1&C2&C3&C4&C5\\ %
                \hline
                Dispersion $\sigma$                                             	&       0.585   &       0.167   &       0.039   &       0.023   &       0.008   \\
                $\textrm{P}(z_{\rm MAP}|D, I)$                          		&       0.023   &       0.033   &       0.035   &       0.059   &       0.204   \\
                $Card$(z $\in$ $\rm CR^{*}$)                            		&       7.36     &       5.22     &       2.01     &       1.31     &       1.37            \\
                %
                Width $\Delta$z $\in$ $\rm CR^{*}$                      	&       7.4e-4  &       5.2e-4  &       2.0e-4  &       1.3e-4  &       1.4e-4  \\
                $\sum_i$ \textrm{P}($z_i$ $\in$ $\rm CR^{*}$)   		&       0.191   &       0.093   &       0.006   &       0.010   &       0.016   \\
                $\sum_i$ \textrm{P}($z_i$ $\in$ $R_2^*$)                	&       0.175   &       0.107   &       0.006   &       0.003   &       0.001   \\
                %
                Significant peaks                                               		&       718.76  &       1.87            &       0.50            &       0.23            &       0.03            \\
                $\Delta$\textrm{P}(two "{best}" z)               			&       0.030   &       0.045   &       0.063   &       0.071   &       0.245   \\
                \hline
                \end{tabular}
        \label{table:dispersion}
        \end{center}
        }
\vspace{0.15cm}
\centering \rule{8cm}{0.4pt}    
\vspace{-0.1cm}
	{
        \mynewformat
        \begin{center}
        \caption{Class dispersions in the descriptor space} \vspace{0.2cm}
                \begin{tabular}{ |c|c|c|c|c }
                \cline{1-4}     
                \scshape Selected  &\multicolumn{3}{c|}{\scshape{Variance $\textbf{V}$ =  $\textbf{B}$ + $\textbf{W}$ }}\\
                \cline{2-4}
                 \scshape descriptors   &       $\sqrt{\bf V}$  &       $\sqrt{\bf B}$     &       $\sqrt{\bf W}$ \\ %
                \cline{1-4} 
                Dispersion $\sigma$                                             	&       0.2851  &       0.1709  &       0.2282  \\
                $\textrm{P}(z_{\rm MAP}|D, I)$                          		&       0.2131  &       0.1929  &       0.0905  \\
                %
                $Card$(z $\in$ $\rm CR^{*}$)                            		&       8.64     &       7.65      &       4.01            \\
                Width $\Delta$z $\in$ $\rm CR^{*}$                      	&       8.6e-4  &       7.7e-4  &       4.0e-4  \\
                $\sum_i$ \textrm{P}($z_i$ $\in$ $\rm CR^{*}$)   		&       0.207   &       0.189  	 &       0.085  \\
                $\sum_i$ \textrm{P}($z_i$ $\in$ $R_2^*$)               	&       0.223   &       0.207   &       0.084   \\
                %
                Significant peaks                                               		&       260.28  &       35.63   &       257.83  \\
                $\Delta$\textrm{P}(two "{best}" z)               			&       0.271   &       0.247   &       0.112   \\              
                \cline{1-4}
                \end{tabular}
        \label{table:dispersion_full}
        \end{center}
        \vspace{-0.2cm}
}\end{table}

The variance tables show that the intraclass variance, $\bf W,$ is generally small in comparison to the interclass variance, $\bf B$, except for two descriptors (the dispersion and the number of modes). This results from the fact that the cluster C1 allows wider variations for these two components. Since the class C1 refers by definition to multimodal zPDFs associated to very unreliable redshift measurement, the results remain coherent.

\noindent Given the possibility that the clustering results might be unreliable due to inherent computational limitations or an incorrect modeling of the descriptor space, the full content of each partition $(C_k)_{ k\in\{1,2,3,4,5\}}$ is investigated. We find that, overall, the zPDFs within each class, $C_k$ , verify the properties listed above.
The newly defined partitions genuinely describe a homogeneous representation of the data in the feature space.
}

\subsubsection{Cluster analysis}\label{subsubsec:s434}
        {
In this section, we compare the initial VVDS redshift reliability flags and the new clusters in order to point out peculiar cases of misclassifications: unexplained discrepancies between the manually attributed flags in the VVDS database and those resulting from the unsupervised classification (cf. \S \ref{sec:s431}).

\vspace{0.1cm}
\noindent Two examples of misclassification are reported in Figures \ref{figure:flag1_inC5} and \ref{figure:flag9_inC1}:
\begin{enumerate}
        \item A misclassification of a "{VVDS Flag 1: unreliable redshift estimation}" as C5 (unimodal zPDF and very reliable $\widehat z_{spec}$) 
        is presented in  Figure \ref{figure:flag1_inC5}. The misclassification is due to the mismatch between the flux spectrum and its noise component, 
        where the latter seems very inadequate when considering the good quality of the data. A problem regarding the generation of the 1D data 
        (flux \& noise components) from the 2D$\rightarrow$1D extraction can be noted.\vspace{0.1cm}
        
        \item A different type of misclassification illustrated in Figure \ref{figure:flag9_inC1}, where a "{VVDS flag 9: secure redshift estimation 
        with an identifiable strong EL}" is identified as C1 (for very multimodal zPDFs and extremely unreliable $\widehat z_{spec}$). 
        This discrepancy between the VVDS flag and the new label from clustering could be ascribed to an imprecise computation of the zPDF 
        due to a lack of representative templates at the given redshift, or a biased evaluation of a human operator.
\end{enumerate}

To evaluate the misclassification rate for the entire VVDS dataset used in this study, Tables \ref{table:recap_project} and \ref{table:recap_project_perc} summarize the repartition of the {initial} VVDS flags $\{1;2;9;3;4\}$ within the predicted reliability clusters 
\footnote{The $\{C_k\}_{k\in\{1,2,3,4,5\}}$ redshift reliability flags obtained for the VVDS data are available at the CDS via anonymous ftp to cdsarc.u-strasbg.fr (130.79.128.5) or via http://cdsweb.u-strasbg.fr/cgi-bin/qcat?J/A+A/}.

\begin{table}[h!] {
        \begin{center} {  
        \caption{Repartition of the initial VVDS redshift reliability flags within the predicted labels ({in absolute values).}}\vspace{-0.2cm}
        \mynewformat
        \begin{tabular}{ l|c|c|c|c|c|c|c }                                                                                                                              
                \multicolumn{2}{c}{}     &      \multicolumn{5}{c}{{\scshape{\color{black}VVDS Initial Flags}}}&\\             
                \cline{3-7}                                                                                                                     
                \multicolumn{2}{c|}{}&                                                                                                                  
                        \color{black}F1&                                                                                                                
                        \color{black}F9&                                                                                                                
                        \color{black}F2&                                                                                                                
                        \color{black}F3&                                                                                                                
                        \color{black}F4&                                                                                                                
                         \textit{\color{black}Total}    \\                                                                                                      
                \hhline{~-|-----|-}                                                                                                                     
                \multirow{7}{*}                                                                                                                 
                {\centering{\rotatebox[origin=c]{90}{\hspace{0.7cm}\scshape{\color{black}Clusters}}}}           
                        &\color{black}C1        &\cellcolor{lightGreen}2776     &\cellcolor{lightRed2}39                        &\cellcolor{lightRed2}233               
                                                        &\cellcolor{lightRed}85         &\cellcolor{lightRed}23                 &\color{dkgray} 3156                    \\
                        \hhline{~-|-----|}                                                                                                              
                        &\color{black}C2        &\cellcolor{lightGreen}3023     &\cellcolor{lightRed1}252                       &\cellcolor{lightRed1}2055      
                                                        &\cellcolor{lightRed}1169               &\cellcolor{lightRed}221                        &\color{dkgray} 6720                    \\                      
                        \hhline{~-|-----|}                                                                                                              
                        &\color{black}C3        &\cellcolor{lightRed2}657               &\cellcolor{lightGreen2}212             &\cellcolor{lightGreen}1534
                                                        &\cellcolor{lightGreen}2345     &\cellcolor{lightGray}899                       &\color{dkgray} 5647                    \\
                        \hhline{~-|-----|}                                                                                                              
                        &\color{black}C4        &\cellcolor{lightRed}241                &\cellcolor{lightGray}104                       &\cellcolor{lightGray}750
                                                        &\cellcolor{lightGreen}2019     &\cellcolor{lightGreen}1852             &\color{dkgray} 4966                    \\
                        \hhline{~-|-----|}                                                                                                              
                        &\color{black}C5        &\cellcolor{lightRed}71         &\cellcolor{lightGray}25                        &\cellcolor{lightGray}171       
                                                        &\cellcolor{lightGreen2}837     &\cellcolor{lightGreen}2926             &\color{dkgray} 4030                    \\                 
                        \cline{2-8}                                                                                                             
                \multicolumn{1}{c}{} &                                                                                                                  
                \multicolumn{1}{c|}     {\textit{\color{dkgray}Total}} &                                                                                                                        
                \multicolumn{1}{c}      {\color{dkgray}         6768    } &                                                                                                       
                \multicolumn{1}{c}      {\color{dkgray} 632     } &                                                                                                     
                \multicolumn{1}{c}      {\color{dkgray} 4743    } &                                                                                                     
                \multicolumn{1}{c}      {\color{dkgray} 6455    } &                                                                                                     
                \multicolumn{1}{c}      {\color{dkgray} 5921    } &                                                                                                     
                \multicolumn{1}{|c}     {\color{dkgray}         24519 }\\                                                                                                       
        \end{tabular}
        \label{table:recap_project}                                                                                                                             
        }\end{center}                                                                                                                           
\centering \rule{8cm}{0.4pt}    
        \begin{center}{ 
        \caption{Repartition of the initial VVDS redshift reliability flags within the predicted labels ({in percent).}}\vspace{-0.cm}
        \mynewformat
        \begin{tabular}{ l|c|c|c|c|c|c| }                                                                                                                               
                \multicolumn{2}{c}{}     &      \multicolumn{5}{c}{{\scshape{\color{black}VVDS Initial Flags}}}\\                      
                \cline{3-7}                                                                                                                     
                \multicolumn{2}{c|}{}&                                                                                                                  
                        \color{black}F1&                                                                                                                
                        \color{black}F9&                                                                                                                
                        \color{black}F2&                                                                                                                
                        \color{black}F3&                                                                                                                
                        \color{black}F4 \\                                                                                                      
                \hhline{~-|-----|}                                                                                                                      
                \multirow{6}{*}                                                                                                                 
                {\centering{\rotatebox[origin=c]{90}{\hspace{0.5cm}\scshape{\color{black}Clusters}}}}   
                %
                        &\color{black}\text{  C1  }        &\cellcolor{lightGreen}41.0\%           &\cellcolor{lightRed2}6.2\%             &\cellcolor{lightRed2}4.9\%     
                                                        &\cellcolor{lightRed}   1.3\%           &\cellcolor{lightRed}0.4\%                      \\
                        \hhline{~-|-----|}                                                                                                                                              
                        &\color{black}\text{  C2  }         &\cellcolor{lightGreen}44.7\%           &\cellcolor{lightRed1}39.9\%            &\cellcolor{lightRed1}43.3\%    
                                                        &\cellcolor{lightRed}18.1\%             &\cellcolor{lightRed}3.7\%                      \\
                        \hhline{~-|-----|}                                                                                                                                              
                        &\color{black}\text{  C3  }        &\cellcolor{lightRed2}9.7\%             &\cellcolor{lightGreen2}33.5\%          &\cellcolor{lightGreen}32.3\%   
                                                        &\cellcolor{lightGreen}36.3\%           &\cellcolor{lightGray}  15.2\%          \\
                        \hhline{~-|-----|}                                                                                                                                              
                        &\color{black}\text{  C4  }        &\cellcolor{lightRed}3.6\%                      &\cellcolor{lightGray}16.5\%            &\cellcolor{lightGray}  15.8\%  
                                                        &\cellcolor{lightGreen}31.3\%           &\cellcolor{lightGreen}31.3\%           \\
                        \hhline{~-|-----|}                                                                                                                                              
                        &\color{black}\text{  C5  }        &\cellcolor{lightRed}1.0\%                      &\cellcolor{lightGray}  4.0\%           &\cellcolor{lightGray}  3.6\%   
                                                        &\cellcolor{lightGreen2}13.0\%          & \cellcolor{lightGreen}49.4\%            \\ 
                        \cline{2-7}                                                                                                
        \end{tabular}
        \label{table:recap_project_perc}        
        }\end{center}                                                                                                           
}\end{table}

\noindent We find that:
\begin{itemize}
        \item [-] The green cells represent the "{expected}" behavior: the cluster C1 is mainly composed of the unreliable redshift "VVDS flags 1" 
                ($\sim$86\%), while the majority of the "VVDS flags 4" are in C4/C5 ($\sim$81\%) and the "VVDS flags 3" are in C3/C4 ($\sim$68\%).
                \vspace{0.2cm}
        \item [-] The gray cells represent a "{gray} area": the clustering provides homogeneous partitioning in comparison with the VVDS flags, 
                as it properly incorporates the full information from the input data (cf. observed flux and its associated noise component).
                
                \noindent We find that the "VVDS flags 2-9" in C4/C5 ($\sim$20\% each) are associated with extremely bright objects with easily 
                identifiable spectral features that make the estimated redshifts very secure.
                
                \noindent On the other hand, the "VVDS flags 4" predicted in C3 ($\sim$15\%) are associated to noisier spectra with scarce spectral 
                features in comparison with the "VVDS flags 4" in C5. The redshift reliability level for these spectra is thereby diminished.
        
                \noindent Similarly, the prediction of "VVDS flag 2" in C2 ($\sim$43\%) is due to the degradation of data quality in comparison with the 
                "VVDS flags 2" located in C3.
                
                \noindent The main reason behind these discrepancies lies in having different observers conducting the redshift-quality checks, 
                as each person has their own understanding of a "redshift reliability" depending on their experience and knowledge of 
                \text{objectively} assessing whether a redshift is deemed a secure estimation or not.
                \vspace{0.2cm}
        \item [-] The red cells are associated with peculiar cases of "{abnormal}" zPDFs resulting from incorrect noise spectra and/or 
                human misclassification.
                In particular, the 71 cases listed of "VVDS flag 1" in C5 result from a mismatch between the flux and noise components;
                the noise component seems extremely low considering the reduced data quality.
                Having very low noise components contributes to reinforce that the flux information depicts a real observation even when it is not the case. 
                We obtain, finally, extremely peaked zPDFs that are predicted as C5.
                
                \noindent For the 23 spectra of "VVDS flag 4" in C1, 13 cases are related to highly dispersed multimodal zPDFs where a confusion 
                between the oxygen emission line $\rm[OII]_{3726A}$ and $\rm Ly\alpha$ is reported: both emission lines are strong candidates which 
                gives at least two {significant} modes detected in the zPDF. 
                Also, the fact that the associated peaks are very distant in the redshift space results in a high dispersion value, $\sigma,$ of the zPDF. 
                The prediction in C1 is highly driven by these characteristics.
                \noindent We also report four cases within these 23 spectra that are associated to low S/N spectra: an important noise component 
                annihilates the confidence in the flux vector and therefore produces highly multimodal zPDFs predicted in C1. 
                For the remaining six cases of "VVDS flag 4" in C1, they result from an excessively-high noise component that produces 
                very degenerate zPDFs, also predicted in C1.
\end{itemize}

The main result from the cluster analysis is that existing redshift reliability flags cannot be reproduced with a 100\% accuracy due to their subjective definition, however a general trend can be retrieved as the majority of the VVDS initial redshift flags can be described by one or two of the redshift reliability clusters $\{C_k\}_{k\in\{1,2,3,4,5\}}$.
}

\subsubsection{Re-using the clusters for redshift reliability label predictions}\label{subsubsec:s432}
{

\begin{sffamily}
        \noindent \color{dkgray2} { - \textit{Classification tests}} \vspace{0.2cm}
\end{sffamily}

\noindent The clustering results showed a great coherency between the automated definition of redshift reliability labels using the zPDF's features matrix and our understanding of "a redshift reliability".

\noindent The idea presented in this Section consists in re-using the new labels of the 24519 VVDS spectra as the response vector $\bf Y_{\rm train}$ in supervised classification, to train a classifier to predict redshift reliability labels for new unlabeled data.
For this purpose, classification tests are performed using the Training and Test sets in Tables~\ref{table:TrainSetTab_cluste} and \ref{table:TestSetTab_cluste}. The resubstitution and test predictions are also used to verify once again the accuracy of the partitioning and objectively assess whether the FCM dichotomized strategy produced "random results" or a "a true description of the zPDFs" in the descriptor space.

\begin{table}[h!]{
        \hspace{0.2cm}
        \parbox{.47\linewidth}{
                \begin{center}
                \caption{Training set \\(total of 16347 VVDS spectra)} \vspace{-0.2cm}
                \mynewformat
                 \hspace{-1cm}
                 \begin{tabular}{ c|c|c|c| } 
                \cline{2-4}
                & \scshape{Label} &  \scshape{Counts}    &      \%\\ 
                \cline{2-4}
                \multirow{4}{*} {\centering{\rotatebox[origin=c]{90}{\hspace{-0.2cm}\scshape{\color{dkblue}Train set}}}}
                        &       C1      &       2104    &       12.87   \\
                        &       C2      &       4480    &       27.41   \\
                        &       C3      &       3765    &       23.03   \\
                        &       C4      &       3311    &       20.25   \\
                        &       C5      &       2687    &       16.44   \\
                \cline{2-4}
                \end{tabular}
                \label{table:TrainSetTab_cluste}
                \end{center}
        }
        \hspace{-0.1cm}
        \parbox{.47\linewidth}{
                \begin{center}
                \caption{Test set \\(total of 8172 VVDS spectra)} \vspace{-0.2cm}
                \mynewformat
                 \hspace{-1cm}
                 \begin{tabular}{ c|c|c|c| } 
                \cline{2-4}
                & \scshape{Label} &  \scshape{Counts}    &      \%\\  
                \cline{2-4}
                \multirow{4}{*} {\centering{\rotatebox[origin=c]{90}{\hspace{0.cm}\scshape{\color{orng}Test set}\hspace{0.cm}}}}
                        &       C1      &       1052    &       12.87   \\
                        &       C2      &       2240    &       27.41   \\
                        &       C3      &       1882    &       23.03   \\
                        &       C4      &       1655    &       20.25   \\
                        &       C5      &       1343    &       16.43   \\
                \cline{2-4}
                \end{tabular}
                \label{table:TestSetTab_cluste}
                \end{center}
        }
}\end{table} 

Similar performances are observed for several classifiers in resubstitution, with extremely low off-diagonal elements in the confusion matrices and an average per-class error rate $\lesssim1\%$ (cf. Tables~\ref{table:cm_clust1} to~\ref{table:cm_clust4}, and Table \ref{table:cm_clust_resub}) for all four classifiers, which is a clear contrast with the results in \S~\ref{subsec:s42}. 
By having low resubstitution errors, the mapping is deemed a reliable reproduction of the input data, and the prediction of $\bf X_{\rm test}$ can be examined. 
We find in test predictions that the confusion matrices for several classifiers offer a good predictive power (average per-class error rate $<2\%$), with the Linear SVM scoring slightly lower results (cf. Tables~\ref{table:cm_clust5} to~\ref{table:cm_clust8}, and Table \ref{table:cm_clust_pred}).\\

}

{
\begin{sffamily}
        \noindent \color{dkgray2} { - \textit{Fuzzy approach}} \vspace{0.2cm}
\end{sffamily}
        
\noindent In ML, two main approaches exist: "hard" partitioning where an object is said to belong to a unique class ({binary membership}), and "soft/fuzzy" partitioning where the membership of an object to a class is expressed in terms of a probability between 0 and 1 (\citealp{wahba_soft_1998,wahba_soft_2002}).

\noindent In the classification tests, "soft" partitioning is used to compute the posterior class prediction probability in order to evaluate the classifier predictive power.
The class posterior probabilities $p( Label\hspace{0.1cm} |\left\{ C_1,C_2,C_3,C_4,C_5 \right\}; Descriptors)$ are obtained by minimizing the Kullback-Leibler divergence (\citealp{hastie_classification_1998}).

In the test predictions on the evaluated VVDS dataset, we find that most class prediction probabilities fall between 0.7 and 1.0 ({bright colors} in Figure~\ref{figure:perf_cluPostProb}). However, we estimate that it could be possible for new data to be assigned to a redshift reliability label with a lower probability, meaning that the classifier cannot project with certainty the unlabeled zPDF into the descriptor space as a result of an incorrect PDF (numerical limitations, degraded input spectra, etc.), or if it is located close to the margins of two or more clusters.
For such cases, a new class of "{Unidentified}" objects has to be set apart from the labeled clusters $\{C_k\}_{k\in\{1, 2, 3, 4, 5\}}$.
This particular point on the class prediction using soft partitioning is addressed further in the following section.
}

\begin{sidewaysfigure*}[h]
  \begin{minipage}[b]{0.5\linewidth}
        \centering
        {\includegraphics[width=0.95\textwidth]{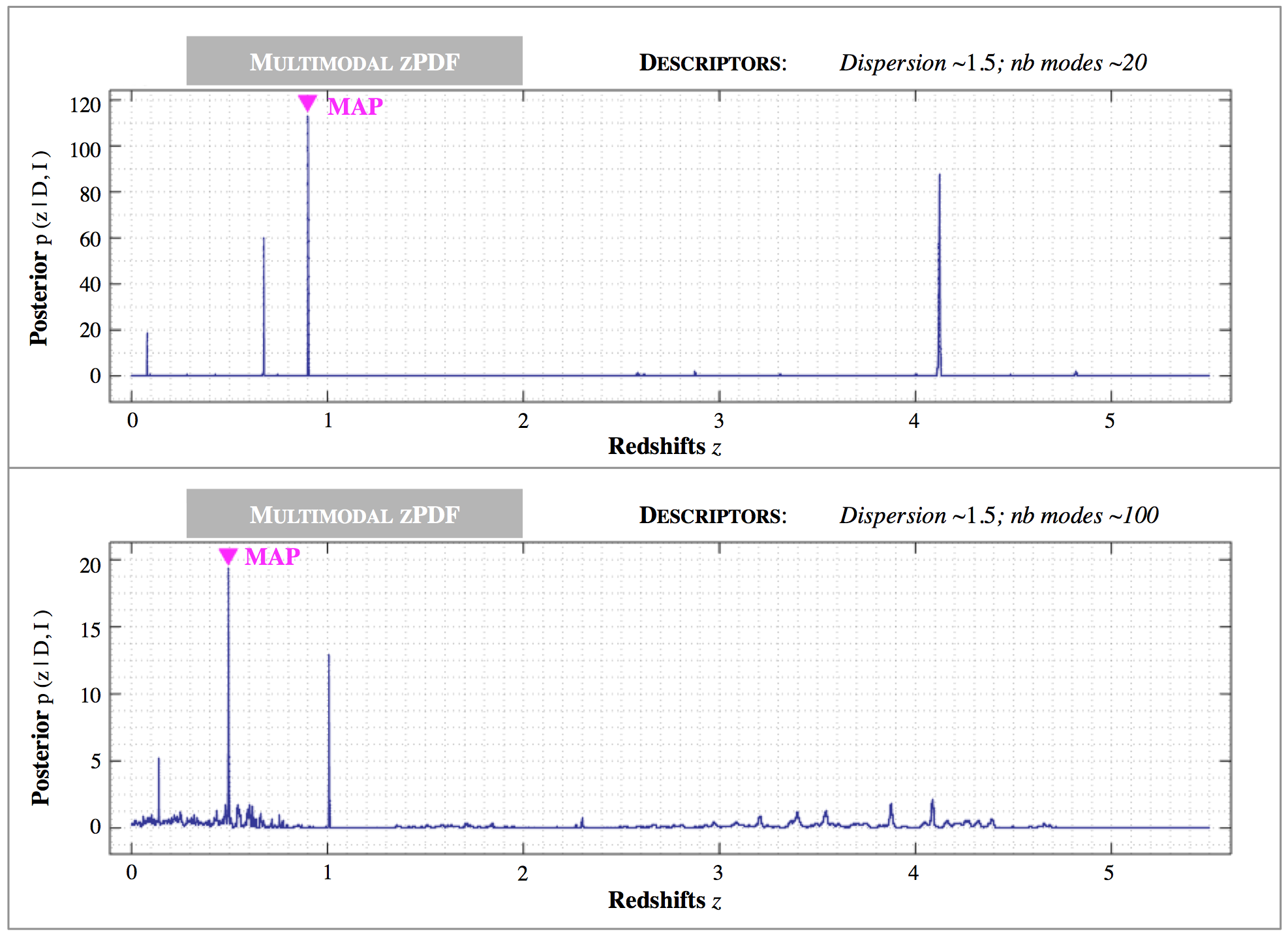}}
        \vspace{-0.2cm}
        \caption{ Display of two zPDFs with multiples modes and a similar dispersion. \vspace{0.2cm}}
        \label{figure:zpdf1}
         \end{minipage}
 \begin{minipage}[b]{0.5\linewidth}     
        \centering
        {\includegraphics[width=0.95\textwidth]{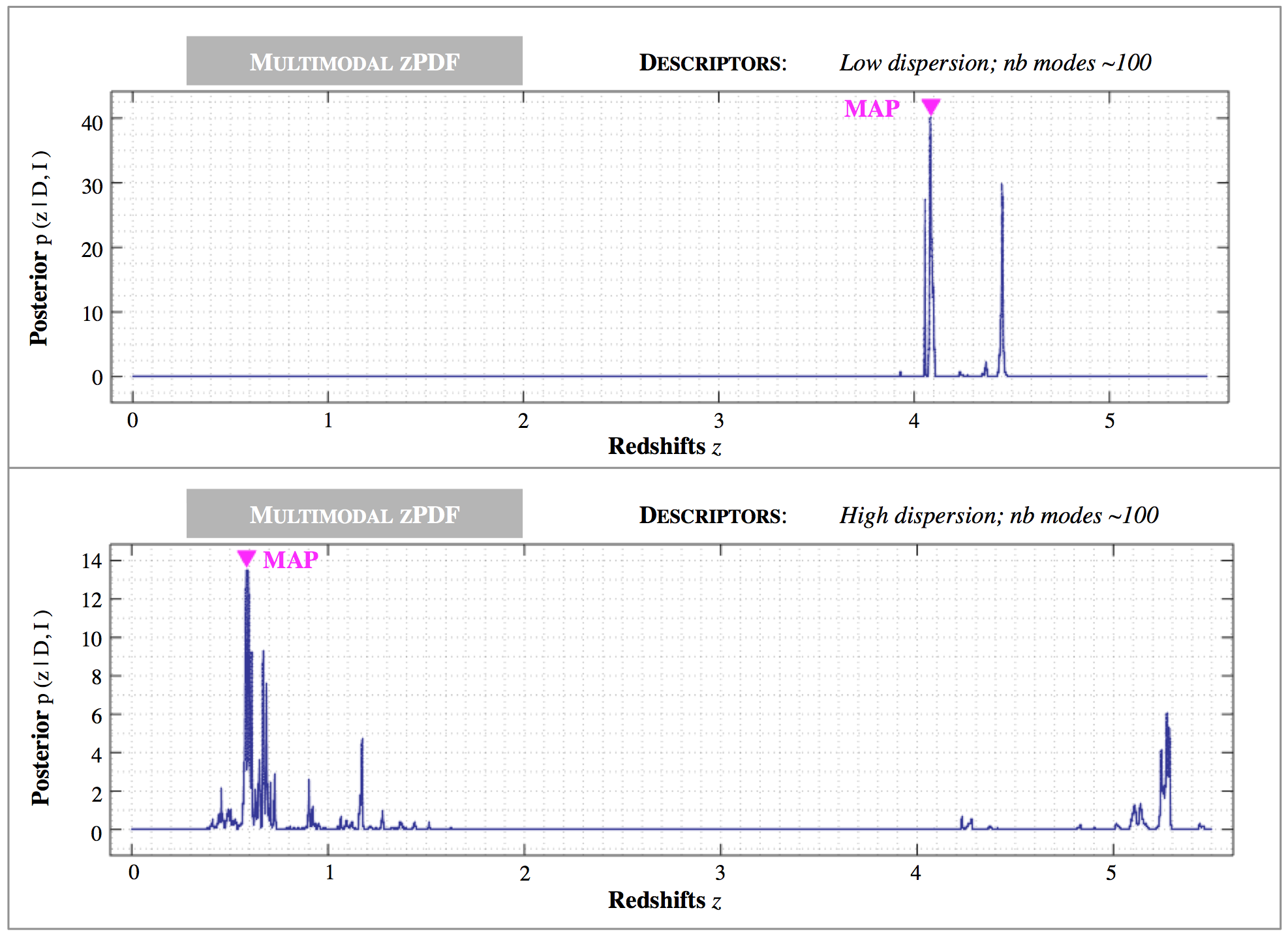}}
        \vspace{-0.2cm}
        \caption{Display of two zPDFs with multiples modes and a different dispersion.\vspace{0.2cm}}
        \label{figure:zpdf2}
    \end{minipage}
 \begin{minipage}[b]{0.5\linewidth}     
        \centering
        {\includegraphics[width=0.95\textwidth]{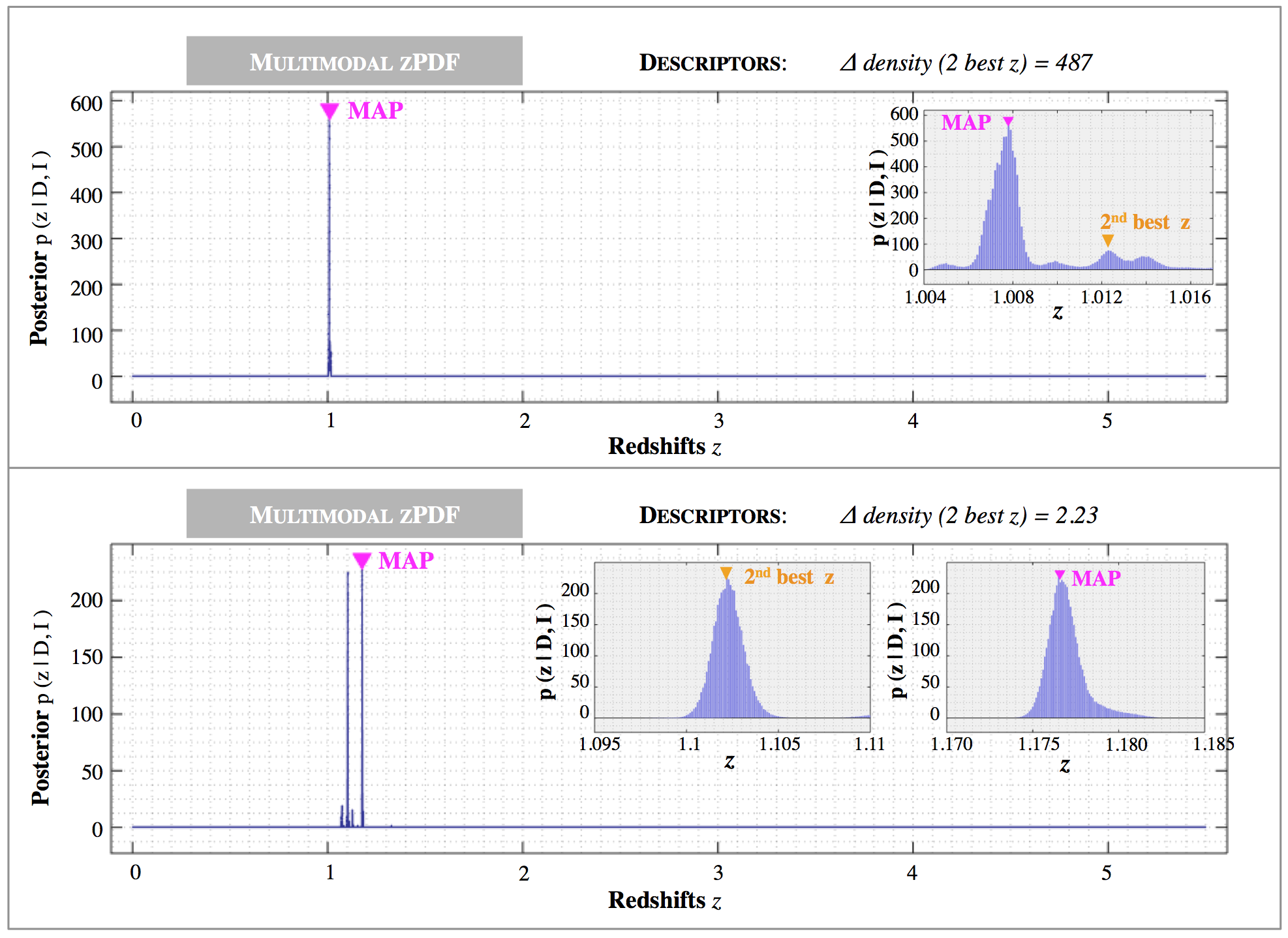}}
        \vspace{-0.2cm}
        \caption{Display of two zPDFs with multiples modes and different characteristics of the two best z solutions.}
        \label{figure:zpdf3}
    \end{minipage}
 \begin{minipage}[b]{0.5\linewidth}     
        \centering
        {\includegraphics[width=0.95\textwidth]{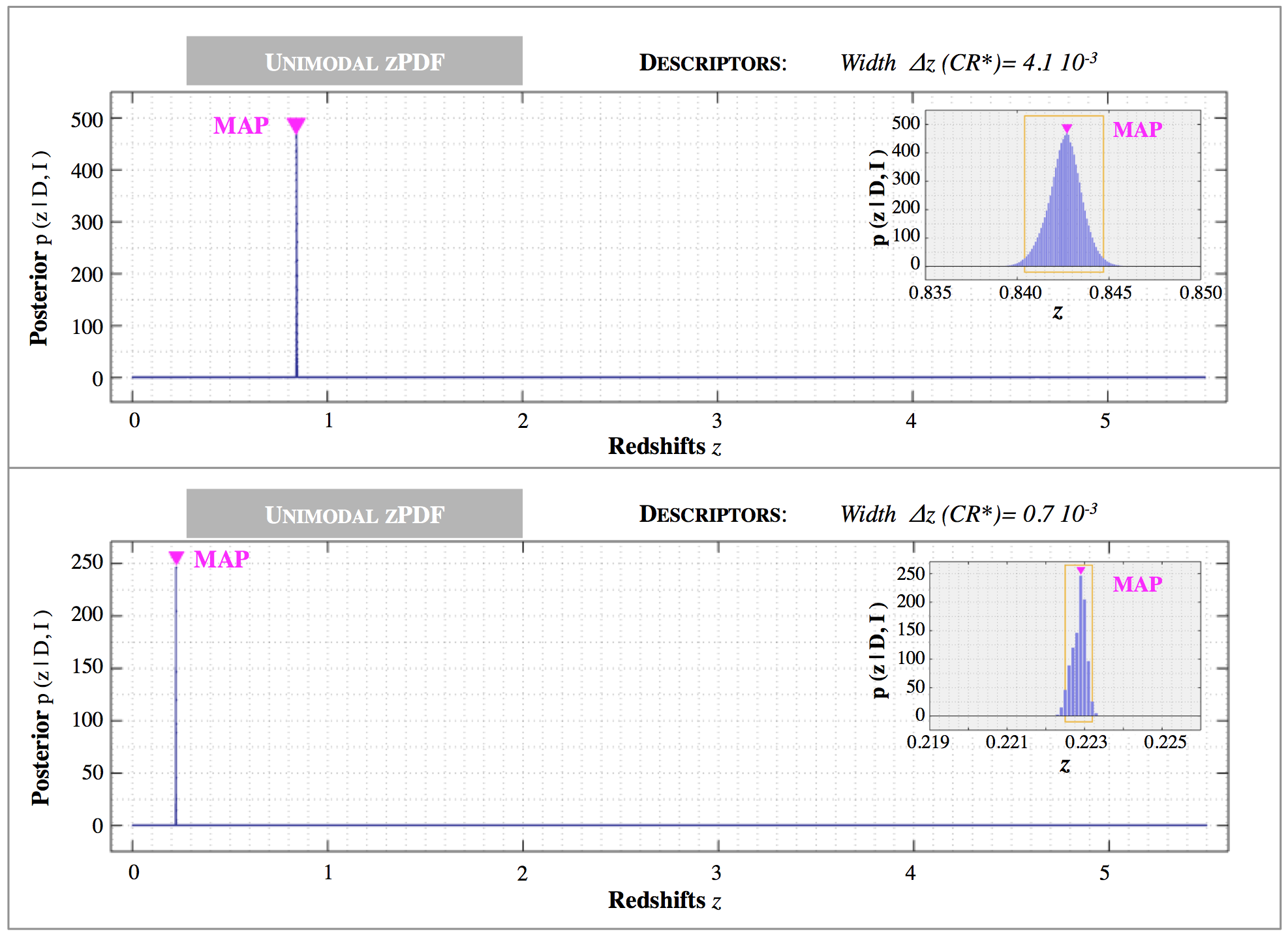}}
        \vspace{-0.2cm}
        \caption{Display of two zPDFs with a single mode and a different width of $\rm CR^{*}$. \vspace{0.2cm}}
        \label{figure:zpdf4}
    \end{minipage}
\end{sidewaysfigure*}

\begin{figure*}[h] 
         \vspace{1cm}
        \centering   
        {\includegraphics[width=0.97\textwidth]{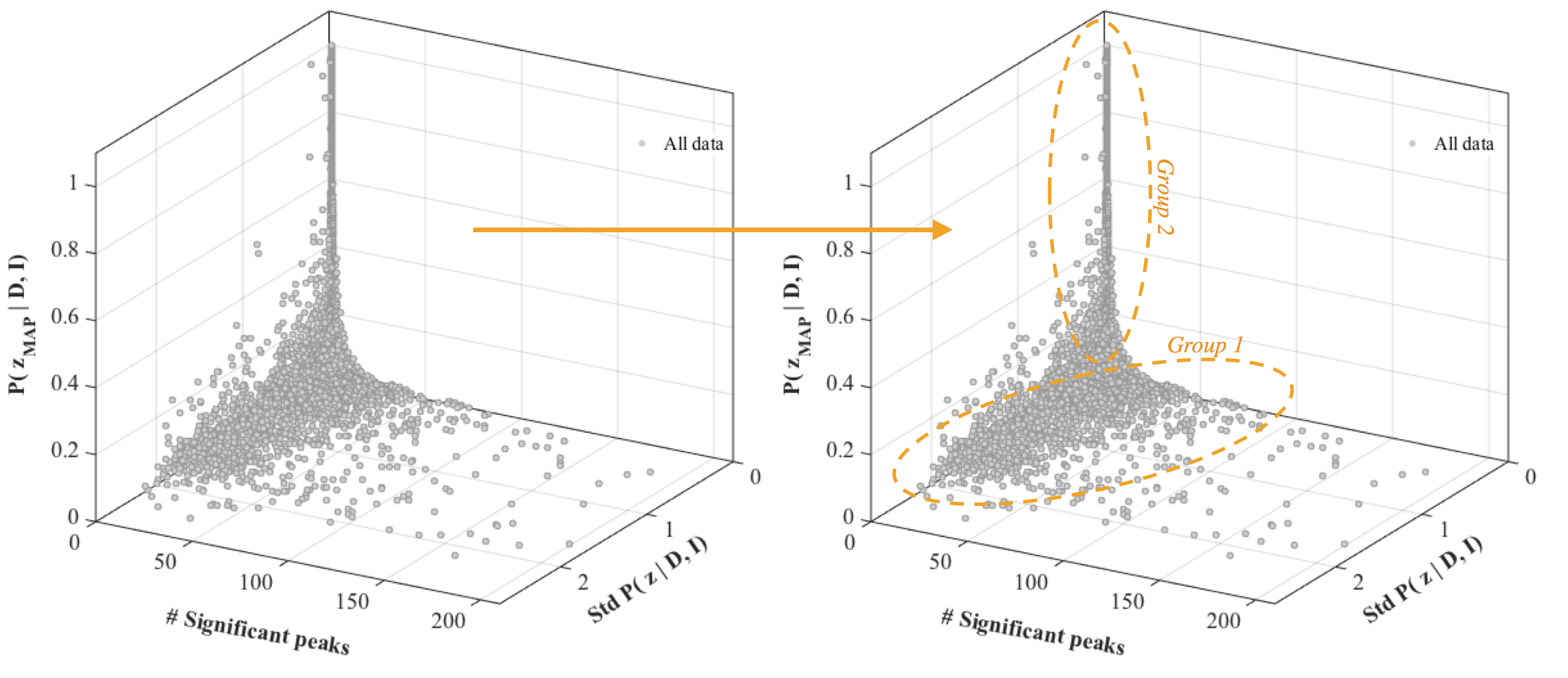}} 
        \vspace{-0.2cm}
        \caption{
                \color{dkgray2}3D representation of the feature matrix $\bf{X}$. \color{black}
                Broadly, two categories are noticeable. The first group refers to the zPDFs with high 
                dispersion, large $z_{MAP}$ peak and low probability $P(z_{\rm MAP}|D, I)$ that can be assimilated to multimodal PDFs or platykurtic 
                unimodal PDFs.
                The second group characterizes the zPDFs with medium-to-low dispersion, narrow $z_{MAP}$ peak and high $P(z_{\rm MAP}|D, I)$ 
                that depict strongly peaked unimodal PDFs.
        }
        \label{figure:disp_X}
\end{figure*}

\begin{figure*}[h] 
        \vspace{1cm} 
        \centering 
        \fcolorbox{gray}{white}         
        {\includegraphics[width=0.8\textwidth]{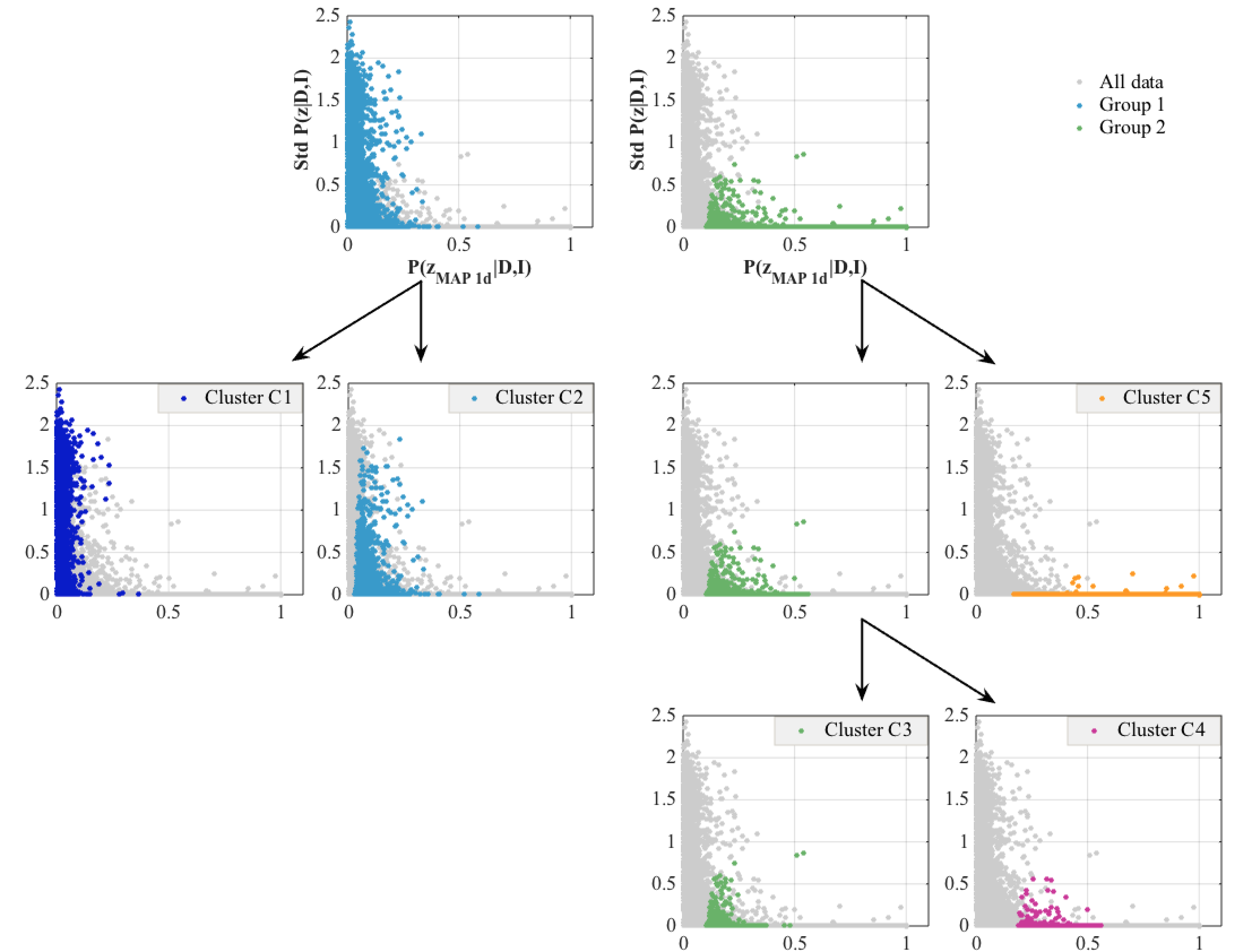}} 
        \caption{
                \color{dkgray2}Clustering the zPDFs features $\bf X$ in a dichotomized pattern. \color{black}
                The clustering strategy exploits the classic FCM algorithm at each step to decompose the input data into two sub-classes 
                using the entire set of descriptors. The final categories $\{C_k\}_{k\in\{1,2,3,4,5\}}$ are displayed in distinct colors.
        }
        \label{figure:dicho}
\end{figure*}

\begin{figure*}[h]  
        \centering 
        \vspace{5cm}
        {\includegraphics[width=1.0\textwidth]{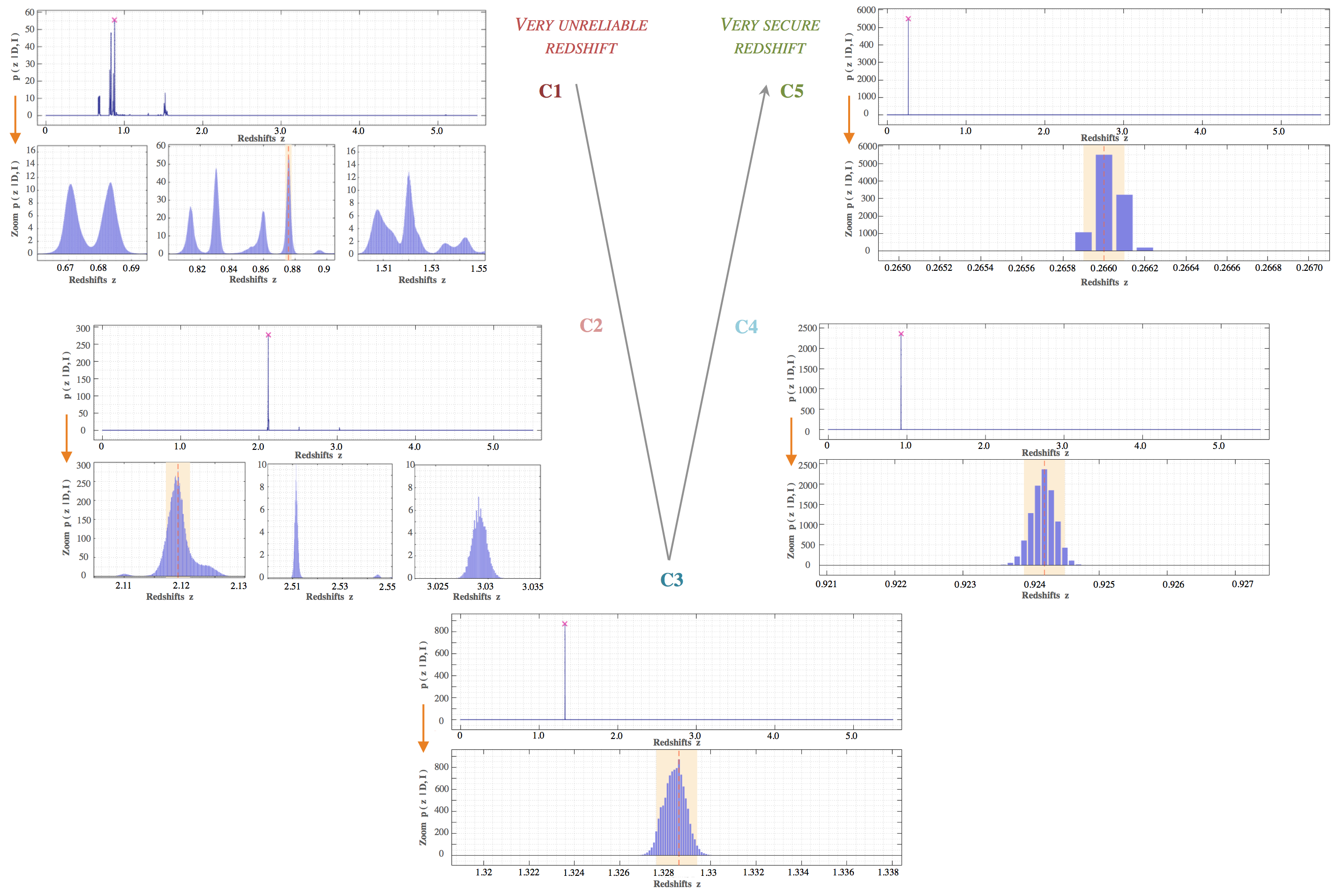}}
        \caption{
                \color{dkgray2}Representative zPDFs for each clusters. \color{black}
                Display of representative zPDFs in each cluster obtained from clustering. 
                The shift in the confidence level in the $\{C_k\}_{k\in\{1,2,3,4,5\}}$ clusters is apparent in the type of the zPDF: from multimodal zPDFs to unimodal zPDFs 
                with narrower $z_{\rm MAP}$ peaks, the confidence level ranges from "extremely unreliable redshift estimate" (C1) to "very 
                certain redshift estimate" (C5).
                }
        \label{figure:repres_cluFla}
\end{figure*}

\begin{figure*}[h]   
        \centering 
        \vspace{1cm}  
        {\includegraphics[width=0.57\textwidth]{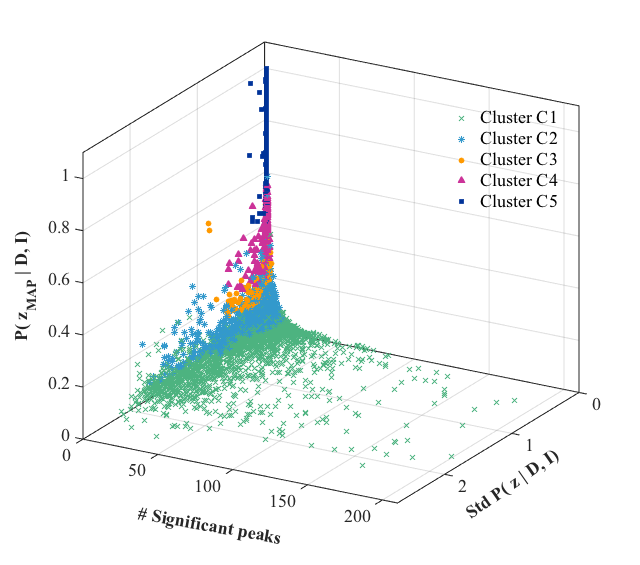}}
        \caption{
                \color{dkgray2}Clusters repartition in a selected 3D space. \color{black}
                The five zReliability clusters described in \S \ref{sec:s431} are associated to different types of 
                redshift PDFs, where the two extreme categories $C_1$ and $C_5,$ respectively, describe highly dispersed multimodal 
                zPDFs and peaked unimodal zPDFs.
                } 
        \label{figure:rep_clust3d} 
\end{figure*}


\begin{figure*}[h]
        \centering
        \vspace{2.0cm} 
        \hspace{-0.5cm}
        {\includegraphics[width=1.05\textwidth]{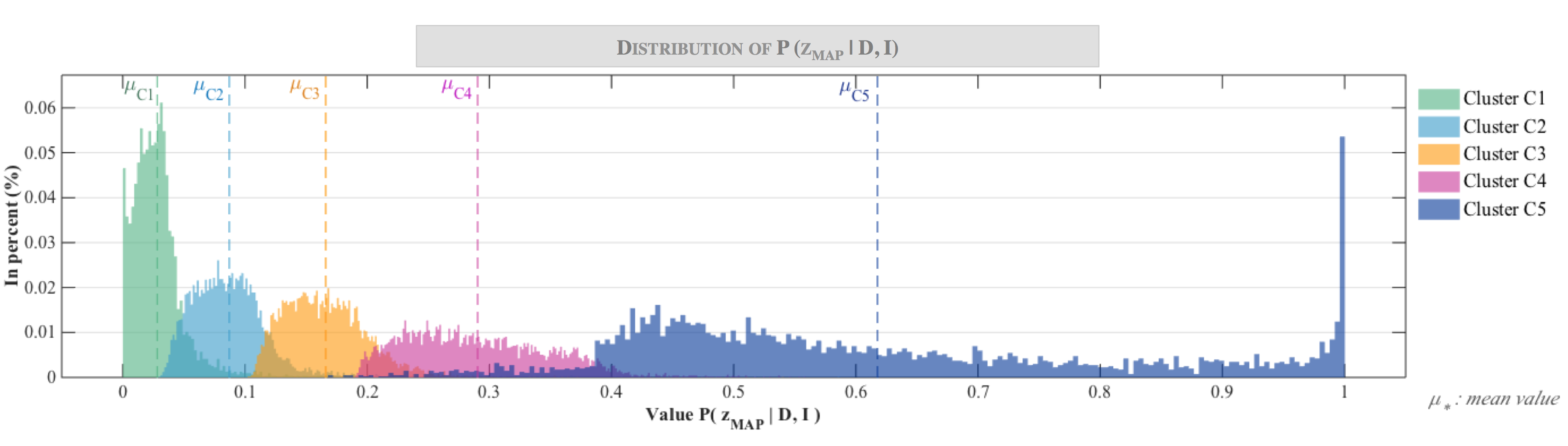}}
                \caption{
                \color{dkgray2}Distribution of the probability value $P(z_{MAP} | D, I)$ within the five partitions. \color{black}
                The distribution of the probability values (component of the descriptor space) $P(z_{MAP} | D, I)$ show distinct properties of the clusters.
                The five zReliability clusters (cf. \S \ref{sec:s431}) are associated with different categories of redshift PDFs.
        }
        \label{figure:rep_pzmap}
\end{figure*} 

\begin{figure*}[h]
        \vspace{1.5cm} \hspace{-1cm} 
        {\includegraphics[width=1.1\textwidth]{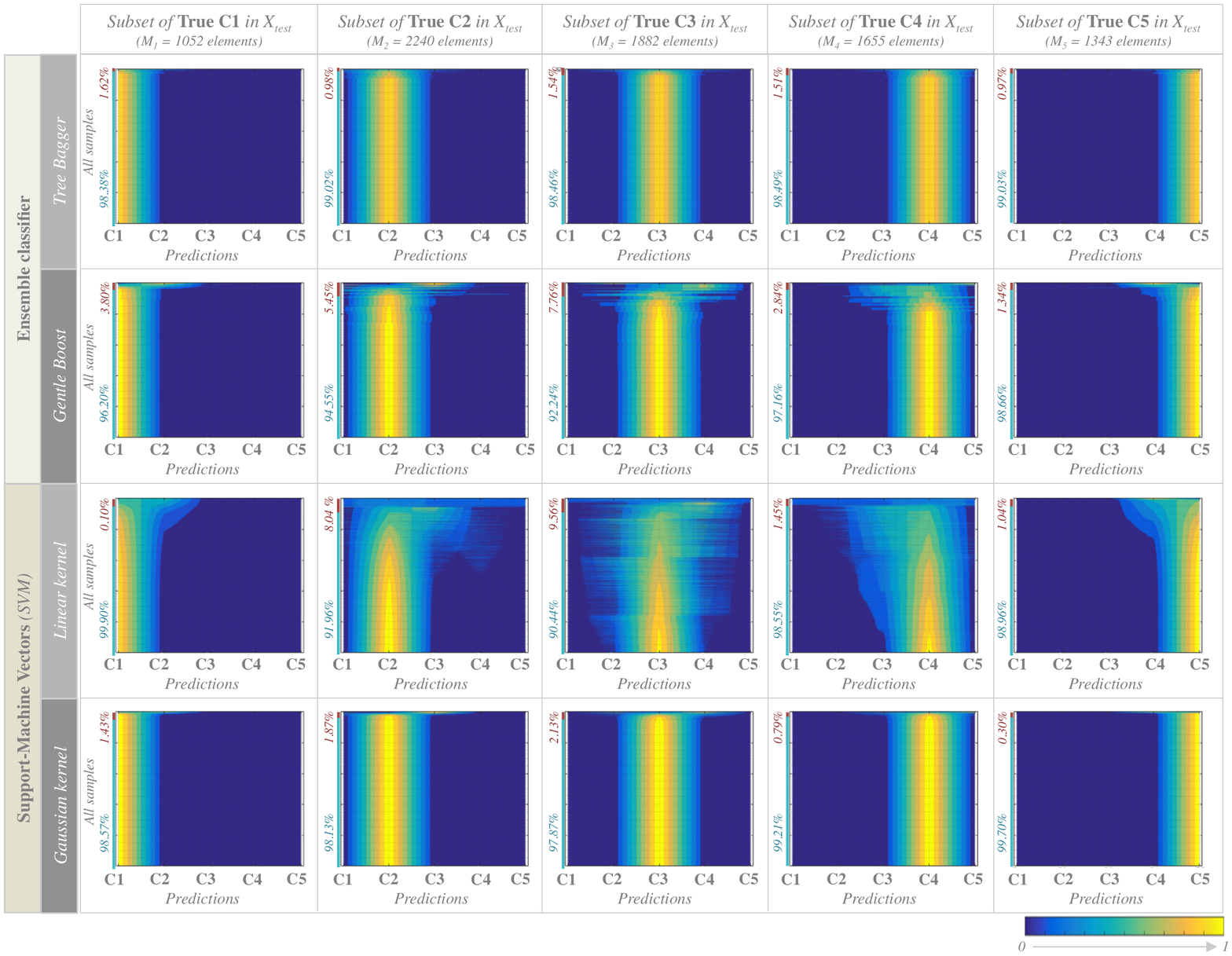}}
        \caption{
                \color{dkgray2}Class posterior probabilities. \color{black}
                The predictive power of several classifiers is displayed for each true class in $\bf Y_{\rm test}$. 
                Most prediction probabilities fall between $\sim$[70-100]\% ({bright colors}).
                For example: the Linear SVM correctly predicts $\sim$92\% (2060 elements) of the subset of "true C2" (around 2240 elements) 
                in $\bf Y_{\rm test}$ (cf. Table \ref{table:cm_clust7}) with class prediction probabilities between 0.7 and 1 ({bright colors}).
                }
        \centering \vspace{0.5cm} 
        {\includegraphics[width=0.45\textwidth]{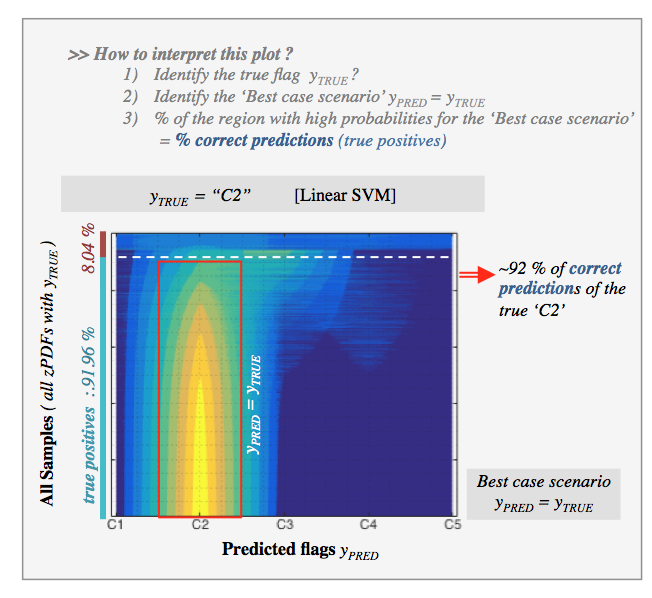}}
        \label{figure:perf_cluPostProb}
\end{figure*}

\begin{figure*}[ht] 
        \centering
        \vspace{-0.1cm} 
        {\includegraphics[width=0.50\textwidth]{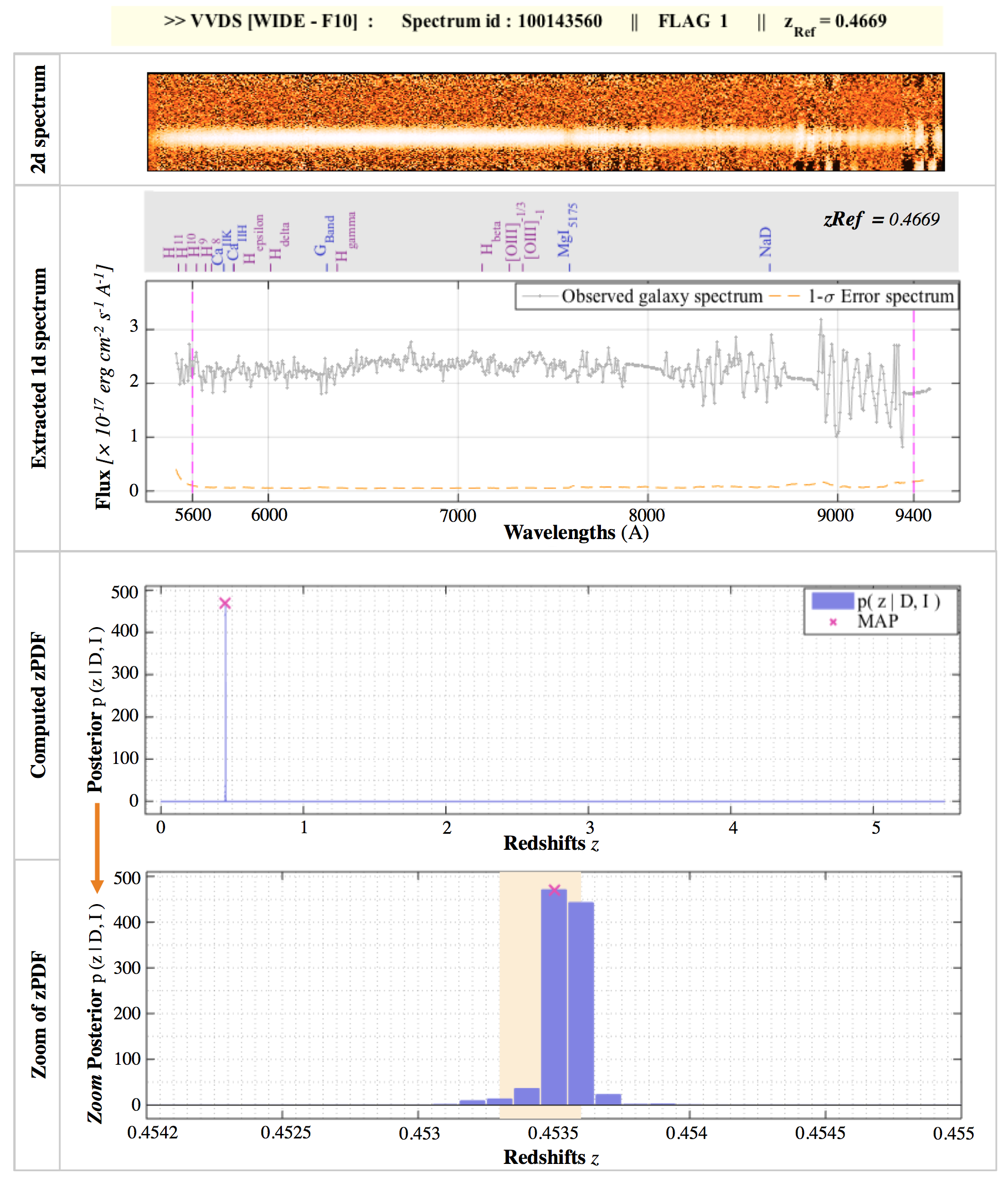}}
        \vspace{-0.1cm} 
        \caption{
                \color{dkgray2}Misclassification - case 1. \color{black}
                A "VVDS flag 1: {very unreliable redshift} " is predicted by the classifier in the new 
                category "C5" for very reliable redshifts. The spectrum displays very low noise components that reinforce the confidence in the 
                measured flux pixels. However the extracted 1D spectrum appears distorted considering the initial 2D spectrum. 
                The extraction 2D$\rightarrow$1D induced a bias in the estimation of the (falsely unimodal) zPDF.
                }
        \label{figure:flag1_inC5}
\end{figure*}

\begin{figure*}[ht]
        \centering
        \hspace{-0.2cm} 
        {\includegraphics[width=0.50\textwidth]{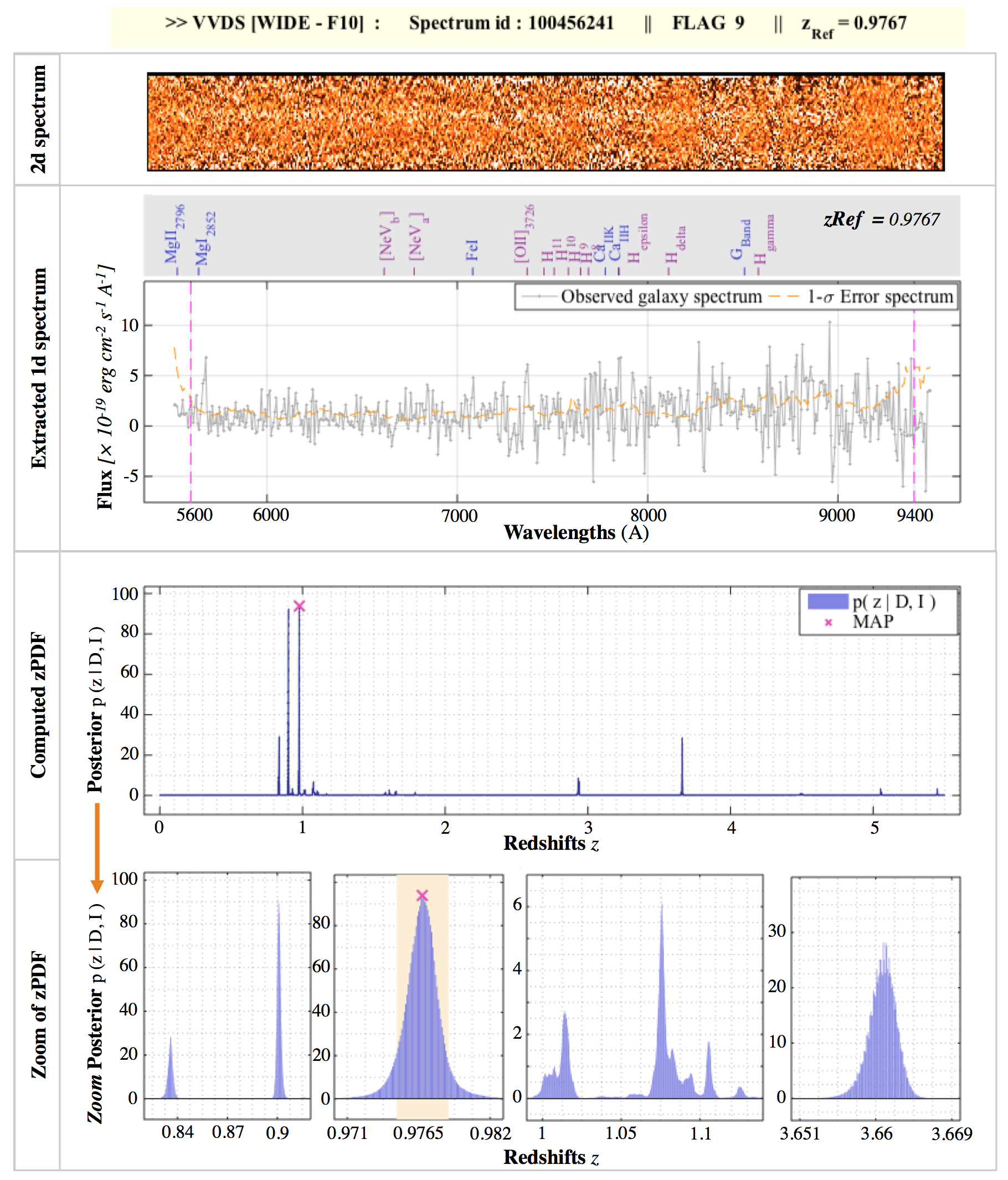}}
        \vspace{-0.1cm} 
        \caption{
                \color{dkgray2}Misclassification - case 2. \color{black}
                A "VVDS flag 9: {reliable redshift, detection of a single emission line}" 
                is predicted by the classifier in the new category "C1" for very unreliable redshifts. The spectrum displays a strong noise 
                component that annihilates the confidence in the measured flux pixels (especially the spectral emission line $\rm[OII]_{3726}$ 
                at 7365A). 
                Several redshift solutions are declared as plausible solutions (a multimodal zPDF).
        }
        \label{figure:flag9_inC1}
\end{figure*}

}

\section{Tests on mock simulations for the Euclid space mission}\label{sec:s5} 
{
An end-to-end simulation pipeline is currently under development for Euclid using catalogs of realistic input sources with spectro-photometric information and an instrumental model for the spectrophotometer NISP designed to perform slitless spectroscopy and imaging photometry in the near-infrared (NIR) wavelength domain.
For Euclid, observations of the same field will be obtained from the combination of three or more different roll angles (referring to different orientations of the grisms) in order to alleviate the superposition of overlapping spectra due to the slitless mode.

Using the pixel simulator software TIPS (\citealp{zoubian_2014}), 1D spectra are obtained from 2D dispersed images after subtracting the sky background from the raw data and combining co-added image stamps of different roll angles.
In these preliminary simulations for Euclid, a contamination model (zodiacal light, adjacent sources, etc.) is not included.

\noindent Table \ref{table:euc_sim_detail} reports the main characteristics of the simulated data of $\textrm{H}{\alpha}$ EL galaxies at redshifts in the range $0.95\leq z \leq1.40$.

\begin{table}[htp!]{
        \vspace{0.1cm} 
        \begin{center}
        \caption{Parameters of preliminary mock simulations for Euclid}\vspace{-0.2cm}
        \mynewformat
        \begin{tabular}{ |l|l|l|l| } 
        \hline
        \multirow{4}{*} {\centering{\rotatebox[origin=c]{90}{\scshape{\color{dkgray} Properties }}}}
                                &       Redshift range                                          &       [   0.95  ;       1.40    ]                               \\
                                &       Magnitude $J_{AB}$ range                        &       [   21.8  ;       24.5 ]                          \\
                                &       Extinction E(B-V)                                       &       [   0.00  ;       0.57    ]                               \\
                                &       log ($fH{\alpha})$ [erg $\rm s^{-1} cm^{-2}$]                 &       [-16.2  ;         -14.1\hspace{0.01cm}]   \\      
        \hline
        \multirow{4}{*} {\centering{\rotatebox[origin=c]{90}{\scshape{\color{dkgray}Simulator}}}} 
                                &\multirow{2}{*}{Source Size in arcsec, sigma}  
                                        &       0.10 ({Set S1})          \\
                                        &&      0.50 ({Set S2})          \\
                                \hhline{~|-|-|}                 
                                &\multirow{2}{*}{Sky Background in $\rm e^{-} s^{-1} pix^{-1}$}    
                                        &       0.8  ({Set S1})          \\
                                        &&      2.0  ({Set S2})          \\
        \hline          
        \end{tabular}
        \label{table:euc_sim_detail}
        \end{center}
        \vspace{-0.4cm}
} \end{table} 

The Euclid simulations are not associated with a redshift reliability flag, and thereby are qualified as "unlabeled data" in this work.
To test the performance of the redshift reliability assessment method, two sets of unlabeled spectra are used (S1,S2), with a total of 3169 spectra per set.\\
By varying the source size and the sky background level, the difference in data quality between the two datasets is noticeable: Figure \ref{figure:onedSpec_euc} displays sample spectra for each dataset.

\subsection{Reliability class predictions}
{
\noindent The redshift PDFs of the Euclid simulated datasets are computed using a constant prior in $\Theta_z$ (cf. Figures \ref{figure:s1_inC5_euc} and\ref{figure:s2_inC1_euc}), and projected into the mapping (cf. \S \ref{sec:s43}) using soft partitioning to predict redshift reliability labels.
Class prediction results are reported in Tables \ref{table:EucProjTab_cluste_absVal} and \ref{table:EucProjTab_cluste}.

\begin{table}[h!]
{
        \vspace{0.1cm}
        \begin{center}          
        \caption{zReliability predictions ({in absolute values}) of preliminary mock simulations for Euclid.}\vspace{-0.1cm}
        \mynewformat
        \begin{tabular}{ c|c|c|c|c|c|c| }       
                \cline{3-7}                                                                                                             
                \multicolumn{2}{c}{}     &      \multicolumn{5}{|c|}{  {\scshape{Predictions in absolute values}  }  } \\
                \cline{3-7}                                                                                             
                \multicolumn{1}{c}{}&
                        \scshape{Set}                   &                                                                       
                        \scshape{"C1"}                  &                                                                                       
                        \scshape{"C2"}                  &                                                                                       
                        \scshape{"C3"}                  &                                                                                       
                        \scshape{"C4"}                  &                                                                                       
                        \scshape{"C5"}                  \\                                                                              
                \hhline{~-|----|-}                                                                                              
                \multirow{6}{*} 
                        &       S1      &\color{dkmauve}        3       &\color{dkmauve}        61      &       313     &\color{trColo} 835     &\color{trColo} 1957    \\
                \hhline{~-|-----|}                                                                                                      
                        &       S2      &\color{trColo} 383     &\color{trColo} 1275    &       555     &       662     &\color{dkmauve}        294     \\
                \cline{2-7}                                                                                                                                             
                \end{tabular}                                                   
        \label{table:EucProjTab_cluste_absVal}
        \end{center}
}
{       
        \begin{center}          
        \caption{zReliability predictions ({in percent}) of preliminary mock simulations for Euclid.}\vspace{-0.cm}
        \mynewformat
        \begin{tabular}{ c|c|c|c|c|c|c| }       
                \cline{3-7}                                                                                                             
                \multicolumn{2}{c}{}     &      \multicolumn{5}{|c|}{  {\scshape{Predictions in \%}  }  } \\
                \cline{3-7}                                                                                             
                \multicolumn{1}{c}{}&
                        \scshape{Set}                   &                                                                       
                        \scshape{"C1"}                  &                                                                                       
                        \scshape{"C2"}                  &                                                                                       
                        \scshape{"C3"}                  &                                                                                       
                        \scshape{"C4"}                  &                                                                                       
                        \scshape{"C5"}                  \\                                                                              
                \hhline{~-|----|-}                                                                                              
                \multirow{6}{*} 
                        & S1    &       \color{dkmauve}0.09             &       \color{dkmauve}1.92             &       9.88    &       \color{trColo}26.35     &       \color{trColo}61.75     \\
                \hhline{~-|-----|}      
                        &S2     &       \color{trColo}12.09             &       \color{trColo}40.23     &       17.51   &       20.89   &       \color{dkmauve}9.28                     \\
                \cline{2-7}                                                                                                                                             
                \end{tabular}                                                   
        \label{table:EucProjTab_cluste}
        \end{center}
        \vspace{-0.1cm}
} \end{table}

The system computes predominantly multimodal zPDFs with high dispersion when the useful information cannot be retrieved from the data because of low S/N: the estimated redshifts are deemed unreliable, which explains the high percentage of S2 spectra in the clusters C1/C2 (\color{trColo}$\sim$52.3\%\color{black}).
In contrast, for high S/N data, the system identifies the majority of redshifts as very reliable: high percentage of S1 spectra in the clusters C4/C5 (\color{trColo}$\sim$88.1\%\color{black}).

\noindent Moreover, the highlighted cells (in \textcolor{dkmauve}{magenta}) within the result tables indicate two particular cases we anticipated to be null fractions when considering the data quality:
We denote on one hand few spectra in the dataset S1 (high S/N) that are associated with unreliable redshift measurements ($\sim2\%$ predicted in C1/C2), 
and on the other hand a small fraction of spectra in S2 (low S/N) that is linked to very reliable redshifts ($\sim9\%$ predicted in C5).
%
%
Such results can easily be understood by looking at the distribution in [$\log$($f$ $\rm{H\alpha}$), $J_{AB}$] of the input spectra (cf. Figure \ref{figure:s1s2_logha_mag_euc}).
We find that:
\begin{itemize}
        \item[-] Bright objects are mainly located in C5, while the majority of faint objects are predicted as C1/C2, in particular when the flux spectrum is 
                embedded in a strong noise (S2). This distribution can be assimilated to a shift $C1\rightarrow C5$ according to the intrinsic properties of the observed object.
                \vspace{0.05cm}
                
        \item[-] The difference in absolute values (cf. Table \ref{table:EucProjTab_cluste_absVal}) between the results in S1 and S2 is due to the increased 
                noise level from the sky background that injects a higher uncertainty in the observed flux spectrum. 
                The redshift reliability is decreased in S2 in comparison with less noisy data (S1).\\
                The repartition in absolute values seems to describe a shift $C5\rightarrow C1$ according to observational constraints (S/N). 
\end{itemize}
}
\vspace{-0.3cm}
\subsection{Redshift error distribution}
{
We further investigate the distribution of the redshift error $\varepsilon_z= |z_{\textrm{MAP}} - z_{ref}|/(1+z_{ref})$ within the predicted clusters (cf. Table \ref{table:EucErrzTab_cluste}).\\
We find that the majority of incorrect redshift estimations ($\varepsilon_z>10^{-3}$) are located in the clusters C1/C2 for "unreliable redshifts" since low S/N data are more likely to be associated with inaccurate redshift measurements.\\
For the two datasets, the fraction of spectra associated with low redshift error ($\varepsilon_z\leq10^{-3}$) is 
$\sim$100\%, $\sim$99\%, $\sim$95\%, and $<$70\% in C5/C4, C3, C2, and C1, respectively.

From this particular result, one approach would be to identify a possible correlation between the redshift reliability clusters and a specific range of redshift errors in order to define a probability for "{a redshift to be correct}" within the $\{C_k\}$ clusters, in a similar way to that used for VVDS.\\
In this direction, the next step will be to conduct similar tests on a wide basis of Euclid simulated datasets (with a contamination model) to statistically constrain the correlation between redshift errors and the redshift reliability clusters.

\begin{table}[h!] {
        \vspace{0.1cm}
        \begin{center}                                                                                                  
        \caption{Redshift error distribution within the predicted redshift reliability classes for preliminary mock simulations for Euclid. 
                The initial predictions are indicated in gray.}
        \vspace{-0.2cm}\hspace{0.2cm}
        \mynewformat
        \begin{tabular}{ c|c|c|c|c|c|c| }       
                \cline{3-7}                                                                                                             
                \multicolumn{2}{c}{}     &      \multicolumn{5}{|c|}{  {\scshape{Fraction of spectra with $\varepsilon_z \leq 10^{-3}$}  }  } \\
                \cline{3-7}                                                                                             
                \multicolumn{1}{c}{}&
                        \scshape{Set}                   &                                                                       
                        \scshape{"C1"}                  &                                                                                       
                        \scshape{"C2"}                  &                                                                                       
                        \scshape{"C3"}                  &                                                                                       
                        \scshape{"C4"}                  &                                                                                       
                        \scshape{"C5"}                  \\                                                                              
                \hhline{~-|----|-}                                                                                              
                \multirow{6}{*} 
                        &       S1      &       1\color{gray}/3         &       59\color{gray}/61               &       313\color{gray}/313     
                                        &       835\color{gray}/835     &       1957\color{gray}/1957   \\
                \hhline{~-|-----|}                                                                                                      
                        &       S2      &       260\color{gray}/383     &       1221\color{gray}/1275         &       550\color{gray}/555     
                                        &       662\color{gray}/662     &       294\color{gray}/294             \\
                        \cline{2-7}                                                                                                                                             
                \end{tabular}                           
                \label{table:EucErrzTab_cluste}
        \end{center}
        \vspace{-0.3cm}
} \end{table}

}

\subsection{Fuzzy approach for label prediction}
{
As previously stated in \S \ref{subsubsec:s432}, soft partitioning in ML provides extra information about the classifier predictive power that can be affected by several factors as possible outliers in the training set or numerical limitations associated to the zPDF computation.

In this study, we find that the majority of S1 and S2 spectra are associated with class probability predictions higher than 99\%, as in the example of Table~\ref{table:example_posterProba2}.
However, peculiar cases, related to class predictions falling within the margins of two or more reliability clusters are detected, as in the example reported in Table \ref{table:example_posterProba} where the class posterior probabilities of the cluster C4 are quite close to the predicted class C3.

\begin{table}[h] {
        \vspace{0.1cm}
        \begin{center}  
        \caption{ Class posterior probabilities of two simulated Euclid spectra. \\
                         In green, the probability associated with the predicted class.}  \vspace{-0.2cm} 
        \mynewformat
        \begin{tabular}{ c|c|c|c|c|c|c|c| }     
                \cline{4-8}                                                                                                             
                \multicolumn{3}{c}{}     &      \multicolumn{5}{|c|}{  {\scshape{Class probability(in \%)}  }  } \\
                \cline{3-8}                                                                                             
                \multicolumn{1}{c}{}&
                        \scshape{Set}                   &                                                                                       
                        \scshape{ Spectrum id}  &                                                                       
                        \scshape{C1}                    &                                                                                       
                        \scshape{C2}                    &                                                                                       
                        \scshape{C3}                    &                                                                                       
                        \scshape{C4}                    &                                                                                       
                        \scshape{C5}                    \\                                                                              
                \hhline{~-|-----|-}                                                                                             
                \multirow{7}{*}                                                                                         
                        &       S1      &       53678850                &       0.00            &       0.17            &       0.02            &       0.01            &       \color{trColo}99.81             \\
                \hhline{~-|------|}     
                        &       S2      &       56932048                &       \color{trColo}99.88             &       0.05            &       0.02            &       0.03            &       0.03            \\
                \cline{2-8}                                                                                                                                             
                  \end{tabular}                                                         
        \label{table:example_posterProba2}
        \end{center}
        \vspace{-0.6cm}
} \end{table}
\begin{table}[h] {
        \vspace{0.1cm}
        \begin{center}                                                                                                  
        \caption{Class posterior probabilities for a simulated spectrum in S2.\\ 
                        In green, the probability associated with the predicted class}\vspace{-0.2cm}
        \mynewformat
        \begin{tabular}{ c|c|c|c|c|c|c|c| }     
                \cline{4-8}                                                                                                             
                \multicolumn{3}{c}{}     &      \multicolumn{5}{|c|}{  {\scshape{Class probability(in \%)}  }  } \\
                \cline{3-8}                                                                                             
                \multicolumn{1}{c}{}&
                        \scshape{Set}                   &                                                                                       
                        \scshape{Spectrum id}   &                                                                       
                        \scshape{C1}                    &                                                                                       
                        \scshape{C2}                    &                                                                                       
                        \scshape{C3}                    &                                                                                       
                        \scshape{C4}                    &                                                                                       
                        \scshape{C5}                    \\                                                                              
                \hhline{~-|-----|-}                                                                                             
                \multirow{7}{*}                                                                                         
                        &       S2      &       114440656       &       17.36   &       8.09            &       \color{trColo2}29.43            &       \color{trColo}32.10     &       13.03\\
                \cline{2-8}                                                                                                                                             
                  \end{tabular}                         
        \label{table:example_posterProba}
        \end{center}
} \end{table}

In this study, the predictions associated to lower-class probabilities are extremely few, with a confusion clearly stated between adjacent clusters (C1 with C2 or C4 with C5 for example). \\
A confusion entailing predicting a label C4/C5 as C1 (and vice-versa) could have been more problematic and can result from an erroneous computation of the zPDF or an incorrect spectroscopic data (flux and noise components).
We estimate that soft partitioning can be used to unveil such peculiar cases and improve the clustering by identifying possible outliers in the descriptor space that can be assigned to the "{Unidentified}" class independently from the $\{C_k\}_{k\in\{1,\dots5\}}$ clusters.
}

\subsection{Discussion}
{
The results obtained using preliminary mock simulations for Euclid show that the new automated reliability redshift definition can be used to quantify the reliability level of spectroscopic redshift measurements. This method could be useful for cosmological studies that require accurate redshift measurements.
By using 1D spectra of newly released Euclid simulations, upcoming studies will focus on the correlation between the distribution of redshift errors and the redshift reliability clusters to define the probability for "a redshift to be correct" in the $\{C_k\}_{k\in\{1,\dots5\}}$ clusters in a similar approach as in VVDS.
}

 \begin{figure*}[h] 
        \vspace{0.1cm}
        \centering 
        {\includegraphics[width=1.0\textwidth]{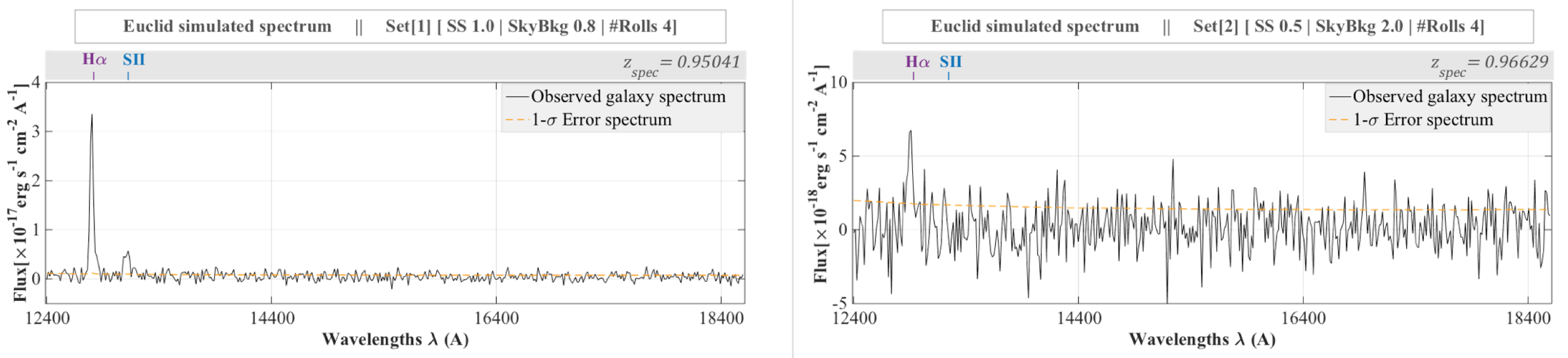}}
        \vspace{-0.2cm} 
        \caption{
                \color{dkgray2}Simulated Euclid spectra. \color{black}
                \textit{Left}: simulated galaxy spectrum (id: 53678850) in the dataset S1 with an identifiable $\rm H\alpha$ line at 12803A.
                \textit{Right}: simulated galaxy spectrum (id: 56932048) in the dataset S2 with a $\rm H\alpha$ emission line at 12908A.
        } 
        \label{figure:onedSpec_euc}
        \vspace{0.5cm}

  \begin{minipage}[b]{0.49\linewidth}
         \centering
        \fcolorbox{gray}{white}         
        {\includegraphics[width=0.95\textwidth]{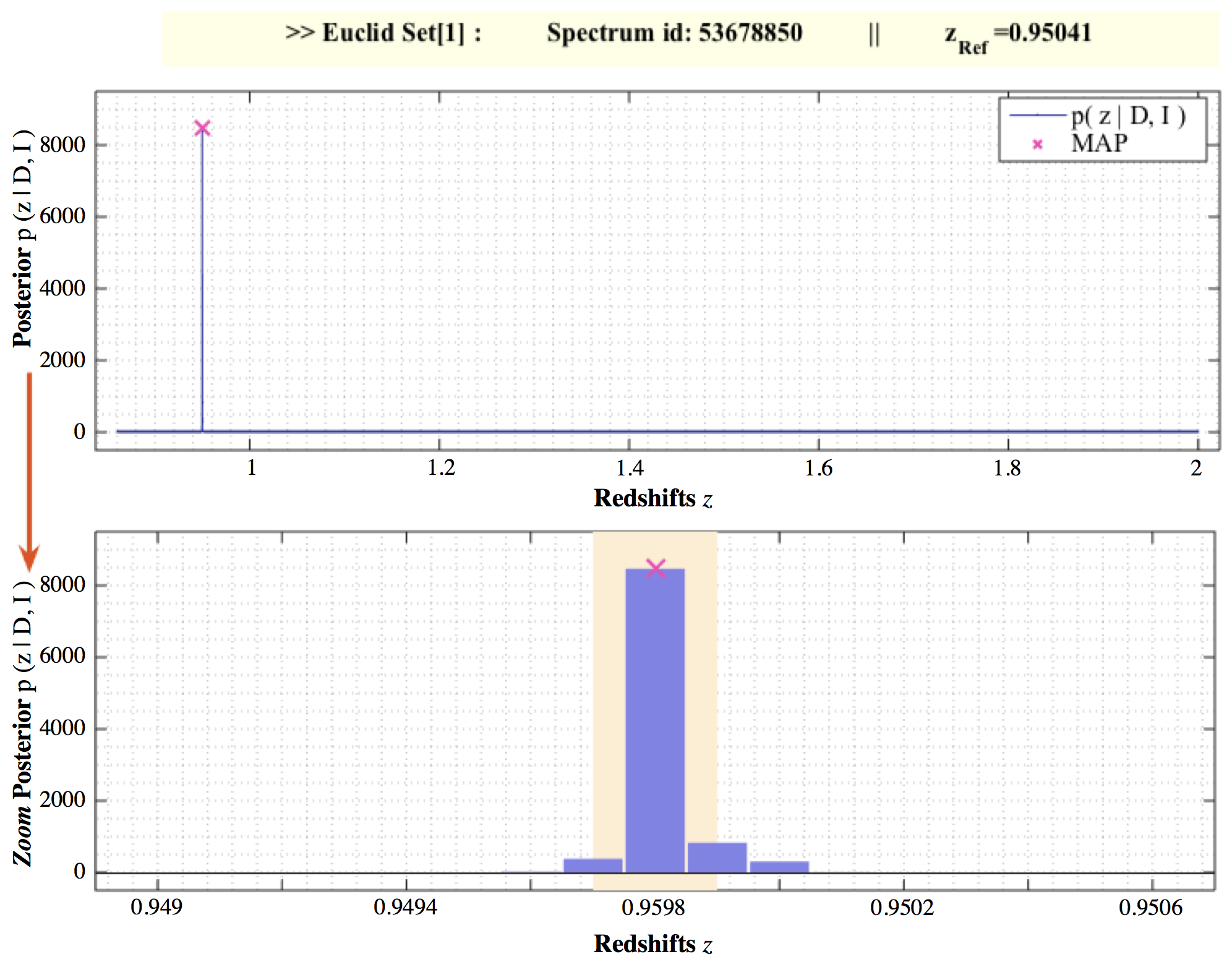}}
        \caption{
                \color{dkgray2}Computed zPDF for a galaxy spectrum in the dataset S1. \color{black} 
                The redshift probability density function is computed using a constant prior over $\Theta_z$.
                }
        \label{figure:s1_inC5_euc}
         \end{minipage}
\hspace{0.2cm}
 \begin{minipage}[b]{0.49\linewidth}    
         \centering
        \fcolorbox{gray}{white}         
        {\includegraphics[width=0.95\textwidth]{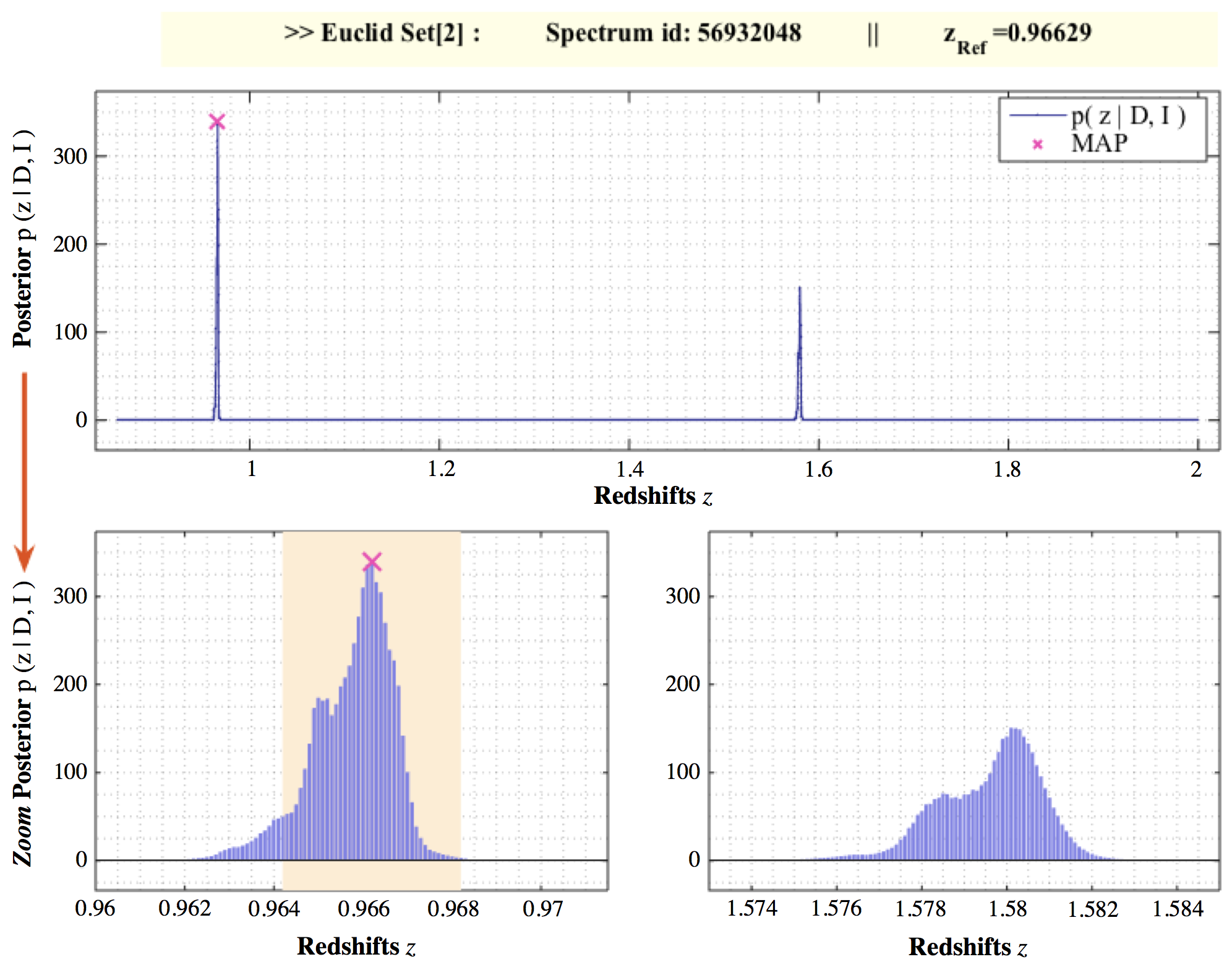}}
        \caption{
                \color{dkgray2}Computed zPDF for a galaxy spectrum in the dataset S2. \color{black} 
                The redshift probability density function is computed using a constant prior over $\Theta_z$.
        }
        \label{figure:s2_inC1_euc}
    \end{minipage}
\end{figure*}

\begin{figure*}[ht]
         \centering
        {\includegraphics[width=0.99\textwidth]{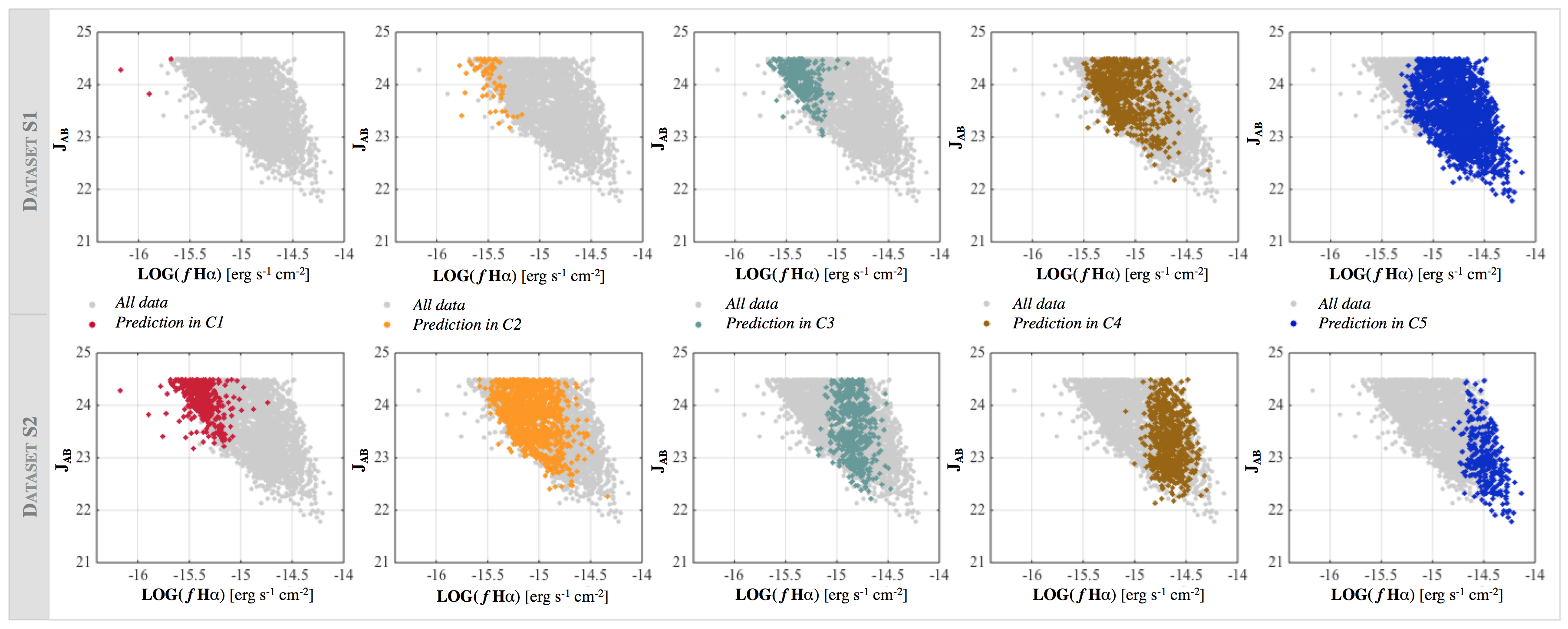}}
        \caption{
                \color{dkgray2} $\log$($f$ $\rm{H\alpha}$), $J_{AB}$ distribution of the reliability class predictions for unlabeled simulated galaxy 
                spectra for Euclid. \color{black}
                The number of faint objects predicted in C1/C2 increases when the noise component in the data is important ($S1\rightarrow S2$). 
                The increased sky background injects a strong noise component of the spectra that annihilates the confidence in a measured redshift, 
                resulting in multimodal zPDFs with high dispersion (C1/C2).
                In contrast, extremely bright objects with an identifiable $\rm H\alpha$ line are located in C4/C5, because the redshift 
                estimation is deemed very reliable when distinct spectral features are found.
        }
        \label{figure:s1s2_logha_mag_euc}
\end{figure*}

}

\section{Summary and Conclusions}\label{sec:s6}
{
By mapping the posterior PDF $p(z|D,I)$ into a discretized feature space and exploiting ML algorithms, we are able to design a new automated method that correlates relevant characteristics of the posterior zPDF, such as the dispersion of the probability distribution and the number of significant modes, with a reliability assessment of the estimated redshift.

\noindent The proposed methodology consists of three steps: 
\begin{enumerate}
        \item Using a set of representative spectra, compute the redshift posterior PDFs $p(z|D, I)$ and extract a set 
                        of features to build the descriptor matrix $\bf X$.
        \item Generate a reliable partitioning $\bf Y$ of the feature space using clustering techniques and prior knowledge, if available.
        \item Use the partitioning to train a classifier that will predict a quality label for new unlabeled observations.
\end{enumerate} 

Using the zPDFs descriptors, we first tried to bypass the first two steps by exploiting existing reliability flags to train a classifier ({supervised classification}), but the results obtained (\S \ref{subsec:s42}) justify the need for new homogeneous partitions [steps 1 and 2] of the feature space because the reproducibility of the existing quality flags cannot be achieved due to their subjective definition:
the combination of several visual checks performed by different observers cannot derive homogeneous and objective criteria of redshift reliability for an automated system to learn from.

\noindent The results of unsupervised classification in \S \ref{sec:s43} displayed great coherency in describing distinct categories of zPDFs: the multimodal zPDFs with equiprobable redshift solutions and high dispersion, versus the unimodal zPDFs with a narrower peak around the $z_{\textrm{MAP}}$ solution, each depicting a different level of reliability for the measured redshift.

\noindent To predict a redshift reliability flag for unlabeled data (\S \ref{sec:s5}), our methodology consists in projecting the unlabeled zPDF [step 3] into the mapping generated from known zPDF descriptors $\textbf{X}$ and their associated z reliability labels \textbf{Y} to predict the class membership.

\noindent A fuzzy approach can also be used to predict the class prediction probability and provide relevant information about the classifier performance and possible discrepancies in the input data.

To conclude, the proposed method to automate the redshift reliability assessment is simple and flexible; the only requirement being robust redshift estimation algorithms with representative templates and a good computational efficiency to produce accurate redshift PDFs.
For the spectroscopic redshift estimation, the use of the Bayesian framework allows to incorporate multiple sources of information as a prior and any readjustment of the data/model hypotheses into the estimation process and produce a posterior zPDF.


\noindent In this work, we have demonstrated that by using a simple entry model and a few ML-algorithms that exploit descriptors of the redshift PDF, it is possible to capture an accurate description of the spectroscopic redshift reliability. 
This approach paves the way for fully automated processing pipelines of large spectroscopic samples as for next-generation large-scale galaxy surveys.
We expect to further develop and test our method for the needs of the Euclid space mission when large simulations of realistic spectra become available. 
Advanced techniques in ML, such as neural networks and deep learning, will be explored to build a complex learning scheme.
}

\section*{Acknowledgments} 
{ 
\small \it
We thank the referee for helpful comments and multiple suggestions that significantly improved this paper.

\noindent This work is part of a thesis co-funded by the University Aix-Marseille and the Centre National d'Etudes Spatiale (CNES) in the context of the upcoming Euclid mission of the European Space Agency, as developed by the Euclid Consortium.

\noindent The authors would like to thank the members of the AMAZED team who provided helpful feedback and insightful discussions on this topic. 
This work benefitted from access to databases managed by the CESAM (Astrophysical Data Center of Marseille) in LAM (Laboratoire d'Astrophysique de Marseille).

\noindent We would also like to express our thanks to all the contributors whose fruitful comments and revision of this manuscript helped us expand our understanding and improve the paper.
}

{
        \bibliographystyle{aa}
        \setstretch{0.8}
        \bibliography{bib2.bib}

\begin{thebibliography}{41}
\expandafter\ifx\csname natexlab\endcsname\relax\def\natexlab#1{#1}\fi

\bibitem[{Abdalla {et~al.}(2008)Abdalla, Amara, Capak, Cypriano, Lahav, \&
  Rhodes}]{abdalla_photometric_2008}
Abdalla, F.~B., Amara, A., Capak, P., {et~al.} 2008, MNRAS, 387, 969,
  \url{http://adsabs.harvard.edu/abs/2008MNRAS.387..969A}

\bibitem[{Albrecht {et~al.}(2006)Albrecht, Bernstein, Cahn, Freedman, Hewitt,
  Hu, Huth, Kamionkowski, Kolb, Knox, Mather, Staggs, \&
  Suntzeff}]{albrecht_report_2006}
Albrecht, A., Bernstein, G., Cahn, R., {et~al.} 2006
  [\eprint{astro-ph/0609591}], \url{https://arxiv.org/abs/astro-ph/0609591}

\bibitem[{Baldry {et~al.}(2014)Baldry, Alpaslan, Bauer, Bland-Hawthorn, Brough,
  Cluver, Croom, Davies, Driver, Gunawardhana, Holwerda, Hopkins, Kelvin,
  Liske, López-Sánchez, Loveday, Norberg, Peacock, Robotham, \&
  Taylor}]{baldry_galaxy_2014}
Baldry, I.~K., Alpaslan, M., Bauer, A.~E., {et~al.} 2014, MNRAS, 441, 2440,
  \url{http://adsabs.harvard.edu/abs/2014MNRAS.441.2440B}

\bibitem[{Benitez(1999)}]{benitez_bayesian_1999}
Benitez, N. 1999, in ASP conference series, Vol. 536, 571--583,
  \url{http://adsabs.harvard.edu/abs/1999ASPC..191...31B}

\bibitem[{Beutler {et~al.}(2011)Beutler, Blake, Colless, Jones, Staveley-Smith,
  Campbell, Parker, Saunders, \& Watson}]{beutler_6df_2011}
Beutler, F., Blake, C., Colless, M., {et~al.} 2011, MNRAS, 416, 3017,
  \url{http://adsabs.harvard.edu/abs/2011MNRAS.416.3017B}

\bibitem[{Beutler {et~al.}(2012)Beutler, Blake, Colless, Jones, Staveley-Smith,
  Poole, Campbell, Parker, Saunders, \& Watson}]{beutler_6df_2012}
Beutler, F., Blake, C., Colless, M., {et~al.} 2012, MNRAS, 423, 3430,
  \url{http://adsabs.harvard.edu/abs/2012MNRAS.423.3430B}

\bibitem[{Bolzonella {et~al.}(2000)Bolzonella, Miralles, \&
  Pelló}]{bolzonella_photometric_2000}
Bolzonella, M., Miralles, J.-M., \& Pelló, R. 2000, A\&A, 363, 476,
  \url{http://adsabs.harvard.edu/abs/2000A%26A...363..476B}

\bibitem[{Brammer {et~al.}(2008)Brammer, van Dokkum, \&
  Coppi}]{brammer_eazy:_2008}
Brammer, G.~B., van Dokkum, P.~G., \& Coppi, P. 2008, ApJS, 686, 1503,
  \url{http://adsabs.harvard.edu/abs/2008ApJ...686.1503B}

\bibitem[{Chandola {et~al.}(2009)Chandola, Banerjee, \&
  Kumar}]{chandola_ads:_2009}
Chandola, V., Banerjee, A., \& Kumar, V. 2009, {ACM} Comput. Surv., 41, 15:1,
  \url{http://doi.acm.org/10.1145/1541880.1541882}

\bibitem[{Collister \& Lahav(2004)}]{collister_annz:_2004}
Collister, A.~A. \& Lahav, O. 2004, PASP, 116, 345,
  \url{http://adsabs.harvard.edu/abs/2004PASP..116..345C}

\bibitem[{Cool {et~al.}(2013)Cool, Moustakas, Blanton, Burles, Coil,
  Eisenstein, Wong, Zhu, Aird, Bernstein, Bolton, Hogg, \&
  Mendez}]{cool_prism_2013}
Cool, R.~J., Moustakas, J., Blanton, M.~R., {et~al.} 2013, ApJ, 767, 118,
  \url{http://adsabs.harvard.edu/abs/2013ApJ...767..118C}

\bibitem[{Cristianini \& Shawe-Taylor(2000)}]{cristianini_svm_2000}
Cristianini, N. \& Shawe-Taylor, J. 2000, "An Introduction to Support Vector
  Machines: And Other Kernel-based Learning Methods" (Cambridge University
  Press)

\bibitem[{Dietterich(2000)}]{dietterich_ensemble_2000}
Dietterich, T.~G. 2000, in Multiple Classifier Systems (Springer), 1--15,
  \url{https://link.springer.com/chapter/10.1007/3-540-45014-9_1}

\bibitem[{Dietterich \& Bakiri(1995)}]{dietterich_solving_1995}
Dietterich, T.~G. \& Bakiri, G. 1995, JAIR, 2, 263,
  \url{http://www.jair.org/papers/paper105.html}

\bibitem[{Fawcett(2006)}]{fawcett_introduction_2006}
Fawcett, T. 2006, Pattern Recognition Letters, 27, 861,
  \url{http://linkinghub.elsevier.com/retrieve/pii/S016786550500303X}

\bibitem[{Feldmann {et~al.}(2006)Feldmann, Carollo, Porciani, Lilly, Capak,
  Taniguchi, Le~F\`{e}vre, Renzini, Scoville, Ajiki, Aussel, Contini,
  {McCracken}, Mobasher, Murayama, Sanders, Sasaki, Scarlata, Scodeggio,
  Shioya, Silverman, Takahashi, Thompson, \& Zamorani}]{feldmann_zurich_2006}
Feldmann, R., Carollo, C.~M., Porciani, C., {et~al.} 2006, MNRAS, 372, 565,
  \url{http://adsabs.harvard.edu/abs/2006MNRAS.372..565F}

\bibitem[{Garilli {et~al.}(2010)Garilli, Fumana, Franzetti, Paioro, Scodeggio,
  Le~F\`{e}vre, Paltani, \& Scaramella}]{garilli_ez:_2010}
Garilli, B., Fumana, M., Franzetti, P., {et~al.} 2010, PASP, 122, 827,
  \url{http://adsabs.harvard.edu/abs/2010PASP..122..827G}

\bibitem[{Garilli {et~al.}(2014)Garilli, Guzzo, Scodeggio, Bolzonella, Abbas,
  Adami, Arnouts, Bel, Bottini, Branchini, Cappi, Coupon, Cucciati, Davidzon,
  Lucia, Torre, Franzetti, Fritz, Fumana, Granett, Ilbert, Iovino, Krywult,
  Brun, Fèvre, Maccagni, Małek, Marulli, {McCracken}, Paioro, Polletta,
  Pollo, Schlagenhaufer, Tasca, Tojeiro, Vergani, Zamorani, Zanichelli, Burden,
  Porto, Marchetti, Marinoni, Mellier, Moscardini, Nichol, Peacock, Percival,
  Phleps, \& Wolk}]{garilli_vimos_2014}
Garilli, B., Guzzo, L., Scodeggio, M., {et~al.} 2014, 562, A23,
  \url{http://dx.doi.org/10.1051/0004-6361/201322790}

\bibitem[{Green {et~al.}(2012)Green, Schechter, Baltay, Bean, Bennett, Brown,
  Conselice, Donahue, Fan, Gaudi, Hirata, Kalirai, Lauer, Nichol, Padmanabhan,
  Perlmutter, Rauscher, Rhodes, Roellig, Stern, Sumi, Tanner, Wang, Weinberg,
  Wright, Gehrels, Sambruna, Traub, Anderson, Cook, Garnavich, Hillenbrand,
  Ivezic, Kerins, Lunine, {McDonald}, Penny, Phillips, Rieke, Riess, van~der
  Marel, Barry, Cheng, Content, Cutri, Goullioud, Grady, Helou, Jackson, Kruk,
  Melton, Peddie, Rioux, \& Seiffert}]{green_wide-field_2012}
Green, J., Schechter, P., Baltay, C., {et~al.} 2012 [\eprint{1208.4012}],
  \url{http://arxiv.org/abs/1208.4012}

\bibitem[{Guzzo {et~al.}(2014)Guzzo, Scodeggio, Garilli, Granett, Fritz, Abbas,
  Adami, Arnouts, Bel, Bolzonella, Bottini, Branchini, Cappi, Coupon, Cucciati,
  Davidzon, Lucia, Torre, Franzetti, Fumana, Hudelot, Ilbert, Iovino, Krywult,
  Brun, Fèvre, Maccagni, Małek, Marulli, {McCracken}, Paioro, Peacock,
  Polletta, Pollo, Schlagenhaufer, Tasca, Tojeiro, Vergani, Zamorani,
  Zanichelli, Burden, Porto, Marchetti, Marinoni, Mellier, Moscardini, Nichol,
  Percival, Phleps, \& Wolk}]{guzzo_vimos_2014}
Guzzo, L., Scodeggio, M., Garilli, B., {et~al.} 2014, 566, A108,
  \url{http://dx.doi.org/10.1051/0004-6361/201321489}

\bibitem[{Hastie \& Tibshirani(1998)}]{hastie_classification_1998}
Hastie, T. \& Tibshirani, R. 1998, AOS, 26, 451,
  \url{http://projecteuclid.org/euclid.aos/1028144844}

\bibitem[{Huterer(2002)}]{huterer_weak_2002}
Huterer, D. 2002, Physical Review D, 65,
  \url{http://adsabs.harvard.edu/abs/2002PhRvD..65f3001H}

\bibitem[{Ilbert {et~al.}(2006)Ilbert, Arnouts, {McCracken}, Bolzonella,
  Bertin, Le~F\`{e}vre, Mellier, Zamorani, Pellò, Iovino, Tresse, Le~Brun,
  Bottini, Garilli, Maccagni, Picat, Scaramella, Scodeggio, Vettolani,
  Zanichelli, Adami, Bardelli, Cappi, Charlot, Ciliegi, Contini, Cucciati,
  Foucaud, Franzetti, Gavignaud, Guzzo, Marano, Marinoni, Mazure, Meneux,
  Merighi, Paltani, Pollo, Pozzetti, Radovich, Zucca, Bondi, Bongiorno,
  Busarello, de~La~Torre, Gregorini, Lamareille, Mathez, Merluzzi, Ripepi,
  Rizzo, \& Vergani}]{ilbert_accurate_2006}
Ilbert, O., Arnouts, S., {McCracken}, H.~J., {et~al.} 2006, A\&A, 457, 841,
  \url{http://adsabs.harvard.edu/abs/2006A%26A...457..841I}

\bibitem[{Ivezic {et~al.}(2008)Ivezic, Tyson, Abel, Acosta, Allsman,
  {AlSayyad}, Anderson, Andrew, Angel, Angeli, Ansari, Antilogus, Arndt,
  Astier, Aubourg, Axelrod, Bard, Barr, Barrau, Bartlett, Bauman, Beaumont,
  Becker, Becla, Beldica, Bellavia, Blanc, Blandford, Bloom, Bogart, Borne,
  Bosch, Boutigny, Brandt, Brown, Bullock, Burchat, Burke, Cagnoli, Calabrese,
  Chandrasekharan, Chesley, Cheu, Chiang, Claver, Connolly, Cook, Cooray,
  Covey, Cribbs, Cui, Cutri, Daubard, Daues, Delgado, Digel, Doherty, Dubois,
  Dubois-Felsmann, Durech, Eracleous, Ferguson, Frank, Freemon, Gangler,
  Gawiser, Geary, Gee, Geha, Gibson, Gilmore, Glanzman, Goodenow, Gressler,
  Gris, Guyonnet, Hascall, Haupt, Hernandez, Hogan, Huang, Huffer, Innes,
  Jacoby, Jain, Jee, Jernigan, Jevremovic, Johns, Jones, Juramy-Gilles, Juric,
  Kahn, Kalirai, Kallivayalil, Kalmbach, Kantor, Kasliwal, Kessler, Kirkby,
  Knox, Kotov, Krabbendam, Krughoff, Kubanek, Kuczewski, Kulkarni, Lambert,
  Guillou, Levine, Liang, Lim, Lintott, Lupton, Mahabal, Marshall, Marshall,
  May, {McKercher}, Migliore, Miller, Mills, Monet, Moniez, Neill, Nief,
  Nomerotski, Nordby, O'Connor, Oliver, Olivier, Olsen, Ortiz, Owen, Pain,
  Peterson, Petry, Pierfederici, Pietrowicz, Pike, Pinto, Plante, Plate, Price,
  Prouza, Radeka, Rajagopal, Rasmussen, Regnault, Ridgway, Ritz, Rosing,
  Roucelle, Rumore, Russo, Saha, Sassolas, Schalk, Schindler, Schneider,
  Schumacher, Sebag, Sembroski, Seppala, Shipsey, Silvestri, Smith, Smith,
  Strauss, Stubbs, Sweeney, Szalay, Takacs, Thaler, Van~Berg, Berk, Vetter,
  Virieux, Xin, Walkowicz, Walter, Wang, Warner, Willman, Wittman, Wolff,
  Wood-Vasey, Yoachim, Zhan, \& Collaboration}]{ivezic_lsst:_2008}
Ivezic, Z., Tyson, J.~A., Abel, B., {et~al.} 2008 [\eprint{0805.2366}],
  \url{http://arxiv.org/abs/0805.2366}

\bibitem[{Laureijs {et~al.}(2011)Laureijs, Amiaux, Arduini, Auguères,
  Brinchmann, Cole, Cropper, Dabin, Duvet, Ealet, Garilli, Gondoin, Guzzo,
  Hoar, Hoekstra, Holmes, Kitching, Maciaszek, Mellier, Pasian, Percival,
  Rhodes, Criado, Sauvage, Scaramella, Valenziano, Warren, Bender, Castander,
  Cimatti, Le~F\`{e}vre, Kurki-Suonio, Levi, Lilje, Meylan, Nichol, Pedersen,
  Popa, Lopez, Rix, Rottgering, Zeilinger, Grupp, Hudelot, Massey, Meneghetti,
  Miller, Paltani, Paulin-Henriksson, Pires, Saxton, Schrabback, Seidel, Walsh,
  Aghanim, Amendola, Bartlett, Baccigalupi, Beaulieu, Benabed, Cuby, Elbaz,
  Fosalba, Gavazzi, Helmi, Hook, Irwin, Kneib, Kunz, Mannucci, Moscardini, Tao,
  Teyssier, Weller, Zamorani, Osorio, Boulade, Foumond, Di~Giorgio, Guttridge,
  James, Kemp, Martignac, Spencer, Walton, Blümchen, Bonoli, Bortoletto,
  Cerna, Corcione, Fabron, Jahnke, Ligori, Madrid, Martin, Morgante, Pamplona,
  Prieto, Riva, Toledo, Trifoglio, Zerbi, Abdalla, Douspis, Grenet, Borgani,
  Bouwens, Courbin, Delouis, Dubath, Fontana, Frailis, Grazian, Koppenhöfer,
  Mansutti, Melchior, Mignoli, Mohr, Neissner, Noddle, Poncet, Scodeggio,
  Serrano, Shane, Starck, Surace, Taylor, Verdoes-Kleijn, Vuerli, Williams,
  Zacchei, Altieri, Sanz, Kohley, Oosterbroek, Astier, Bacon, Bardelli, Baugh,
  Bellagamba, Benoist, Bianchi, Biviano, Branchini, Carbone, Cardone, Clements,
  Colombi, Conselice, Cresci, Deacon, Dunlop, Fedeli, Fontanot, Franzetti,
  Giocoli, Garcia-Bellido, Gow, Heavens, Hewett, Heymans, Holland, Huang,
  Ilbert, Joachimi, Jennins, Kerins, Kiessling, Kirk, Kotak, Krause, Lahav, van
  Leeuwen, Lesgourgues, Lombardi, Magliocchetti, Maguire, Majerotto, Maoli,
  Marulli, Maurogordato, {McCracken}, {McLure}, Melchiorri, Merson, Moresco,
  Nonino, Norberg, Peacock, Pello, Penny, Pettorino, Di~Porto, Pozzetti,
  Quercellini, Radovich, Rassat, Roche, Ronayette, Rossetti, Sartoris,
  Schneider, Semboloni, Serjeant, Simpson, Skordis, Smadja, Smartt, Spano,
  Spiro, Sullivan, Tilquin, Trotta, Verde, Wang, Williger, Zhao, Zoubian, \&
  Zucca}]{laureijs_euclid_2011}
Laureijs, R., Amiaux, J., Arduini, S., {et~al.} 2011 [\eprint{1110.3193}],
  \url{http://arxiv.org/abs/1110.3193}

\bibitem[{Le~F\`{e}vre {et~al.}(2013)Le~F\`{e}vre, Cassata, Cucciati, Garilli,
  Ilbert, Brun, Maccagni, Moreau, Scodeggio, Tresse, Zamorani, Adami, Arnouts,
  Bardelli, Bolzonella, Bondi, Bongiorno, Bottini, Cappi, Charlot, Ciliegi,
  Contini, Torre, Foucaud, Franzetti, Gavignaud, Guzzo, Iovino, Lemaux,
  López-Sanjuan, {McCracken}, Marano, Marinoni, Mazure†, Mellier, Merighi,
  Merluzzi, Paltani, Pellò, Pollo, Pozzetti, Scaramella, Tasca, Vergani,
  Vettolani, Zanichelli, \& Zucca}]{le_fevre_vimos_2013}
Le~F\`{e}vre, O., Cassata, P., Cucciati, O., {et~al.} 2013, A\&A, 559, A14,
  \url{http://dx.doi.org/10.1051/0004-6361/201322179}

\bibitem[{Le~F\`{e}vre {et~al.}(2015)Le~F\`{e}vre, Tasca, Cassata, Garilli,
  Le~Brun, Maccagni, Pentericci, Thomas, Vanzella, Zamorani, Zucca, Amorin,
  Bardelli, Capak, Cassarà, Castellano, Cimatti, Cuby, Cucciati, de~la Torre,
  Durkalec, Fontana, Giavalisco, Grazian, Hathi, Ilbert, Lemaux, Moreau,
  Paltani, Ribeiro, Salvato, Schaerer, Scodeggio, Sommariva, Talia, Taniguchi,
  Tresse, Vergani, Wang, Charlot, Contini, Fotopoulou, López-Sanjuan, Mellier,
  \& Scoville}]{le_fevre_vimos_2015}
Le~F\`{e}vre, O., Tasca, L., Cassata, P., {et~al.} 2015, A\&A, 576, A79,
  \url{http://adsabs.harvard.edu/abs/2015A%26A...576A..79L}

\bibitem[{Le~F\`{e}vre {et~al.}(2005)Le~F\`{e}vre, Vettolani, Garilli, Tresse,
  Bottini, Brun, Maccagni, Picat, Scaramella, Scodeggio, Zanichelli, Adami,
  Arnaboldi, Arnouts, Bardelli, Bolzonella, Cappi, Charlot, Ciliegi, Contini,
  Foucaud, Franzetti, Gavignaud, Guzzo, Ilbert, Iovino, {McCracken}, Marano,
  Marinoni, Mathez, Mazure, Meneux, Merighi, Paltani, Pellò, Pollo, Pozzetti,
  Radovich, Zamorani, Zucca, Bondi, Bongiorno, Busarello, Lamareille, Mellier,
  Merluzzi, Ripepi, \& Rizzo}]{le_fevre_vimos_2005}
Le~F\`{e}vre, O., Vettolani, G., Garilli, B., {et~al.} 2005, A\&A, 439, 845,
  \url{http://dx.doi.org/10.1051/0004-6361:20041960}

\bibitem[{Linder \& Jenkins(2003)}]{linder_cosmic_2003}
Linder, E.~V. \& Jenkins, A. 2003, MNRAS, 346, 573,
  \url{http://adsabs.harvard.edu/abs/2003MNRAS.346..573L}

\bibitem[{Machado {et~al.}(2013)Machado, Leonard, Starck, Abdalla, \&
  Jouvel}]{machado_darth_2013}
Machado, D.~P., Leonard, A., Starck, J.-L., Abdalla, F.~B., \& Jouvel, S. 2013,
  A\&A, 560, A83, \url{http://adsabs.harvard.edu/abs/2013A%26A...560A..83M}

\bibitem[{Patcha \& Park(2007)}]{patcha_ads:_2007}
Patcha, A. \& Park, J.-M. 2007, Computer Networks, 51, 3448,
  \url{http://linkinghub.elsevier.com/retrieve/pii/S138912860700062X}

\bibitem[{Schuecker(1993)}]{schuecker_automated_1993}
Schuecker, P. 1993, ApJS, 84, 39,
  \url{http://adsabs.harvard.edu/abs/1993ApJS...84...39S}

\bibitem[{Scodeggio {et~al.}(2005)Scodeggio, Franzetti, Garilli, Zanichelli,
  Paltani, Maccagni, Bottini, Le~Brun, Contini, Scaramella, Adami, Bardelli,
  Zucca, Tresse, Ilbert, Foucaud, Iovino, Merighi, Zamorani, Gavignaud, Rizzo,
  {McCracken}, Le~F\`{e}vre, Picat, Vettolani, Arnaboldi, Arnouts, Bolzonella,
  Cappi, Charlot, Ciliegi, Guzzo, Marano, Marinoni, Mathez, Mazure, Meneux,
  Pellò, Pollo, Pozzetti, \& Radovich}]{scodeggio_2005}
Scodeggio, M., Franzetti, P., Garilli, B., {et~al.} 2005, PASP, 117, 1284,
  \url{http://adsabs.harvard.edu/abs/2005PASP..117.1284S}

\bibitem[{Shahid {et~al.}(2014)Shahid, Rossholm, Lövström, \&
  Zepernick}]{shahid_imageQual_2014}
Shahid, M., Rossholm, A., Lövström, B., \& Zepernick, H.-J. 2014, EURASIP
  Journal on Image Video Processing, 2014, 40,
  \url{https://link.springer.com/article/10.1186/1687-5281-2014-40}

\bibitem[{Simkin(1974)}]{simkin_measurements_1974}
Simkin, S.~M. 1974, A\&A, 31, 129,
  \url{http://adsabs.harvard.edu/abs/1974A%26A....31..129S}

\bibitem[{Tonry \& Davis(1979)}]{tonry_survey_1979}
Tonry, J. \& Davis, M. 1979, ApJ, 84, 1511,
  \url{http://adsabs.harvard.edu/abs/1979AJ.....84.1511T}

\bibitem[{Vapnik(2000)}]{vapnik_learn_2000}
Vapnik, V.~N. 2000, "The Nature of Statistical Learning Theory" (Springer),
  \url{http://link.springer.com/10.1007/978-1-4757-3264-1}

\bibitem[{Wahba(1998)}]{wahba_soft_1998}
Wahba, G. 1998, "Support Vector Machines, Reproducing Kernel Hilbert Spaces and
  the Randomized {GACV}" (MIT Press),
  \url{http://pages.stat.wisc.edu/~wahba/talks1/nips.97/}

\bibitem[{Wahba(2002)}]{wahba_soft_2002}
Wahba, G. 2002, {PNAS}, 99, 16524,
  \url{http://www.pnas.org/content/99/26/16524}

\bibitem[{Wang {et~al.}(2010)Wang, Percival, Cimatti, Mukherjee, Guzzo, Baugh,
  Carbone, Franzetti, Garilli, Geach, Lacey, Majerotto, Orsi, Rosati, Samushia,
  \& Zamorani}]{wang_designing_2010}
Wang, Y., Percival, W., Cimatti, A., {et~al.} 2010, Monthly Notices of the
  Royal Astronomical Society, 409, 737,
  \url{http://adsabs.harvard.edu/abs/2010MNRAS.409..737W}

\bibitem[{Zoubian {et~al.}(2014)Zoubian, K{\"u}mmel, Kermiche, Apostolakos,
  Chapon, Ealet, Franzetti, Garilli, Jullo, \& Paioro}]{zoubian_2014}
Zoubian, J., K{\"u}mmel, M., Kermiche, S., {et~al.} 2014, in ASP conference
  series, Vol. 485, 509,
  \url{http://adsabs.harvard.edu/abs/2014ASPC..485..509Z}

\end{thebibliography}
}

\newpage 
{
\newgeometry{top=1in,bottom=1in,right=1.0in,left=0.5in} {
\setcounter{table}{0} \renewcommand{\thetable}{A.\arabic{table}} 
\begin{table*}[h]                                                                                                               
\vspace{3cm}                                                                                                            
\centering\footnotesize \textbf{\colorbox{gray!20}{ \color{dkgray} Confusion matrices using modified VVDS flags }} \\                                                                                                                                                                                                        
\parbox{.45\linewidth}{                                                                                                         
        \begin{center}                                                                                                  
        \scriptsize                                                                                                     
\begin{tabular}{ l|c|c|c|c|c }                                                                                                  
                \multicolumn{2}{c}{}     &      \multicolumn{3}{c}{{\scshape{\color{trColo}True}}}&\\                                                                           
                \cline{3-5}                                                                                             
                \multicolumn{2}{c|}{}&                                                                                          
                        \color{trColo}"  0 "&                                                                                   
                        \color{trColo}"+1"&                                                                                     
                        \color{trColo}"+2"&                                                                                     
                         \textit{\color{dkgray}Total}   \\                                                                              
                \hhline{~-|---|-}                                                                                               
                \multirow{4}{*}                                                                                         
                {\centering{\rotatebox[origin=c]{90}{\hspace{0.5cm}\scshape{\color{prColo}Predicted}}}}                                                                                         
                &       \color{prColo}" 0 "     &\cellcolor{lightGreen} 4503    &       2       &       1       &       \color{dkgray}  4506    \\      
                \hhline{~-|---|}                                                                                                
                &       \color{prColo}"+1"       &      9       &\cellcolor{lightGreen} 3581    &       203     &       \color{dkgray}  3793    \\
\end{tabular}                                                                                         
        \caption{\color{dkblue}[Resubstitution prediction]\color{black}- Bagging Trees}                                                                                                  
        \label{table:cm_vvds1}                                                                                                  
        \end{center}            
}\hspace{0.7cm}         
\parbox{.45\linewidth}{                                                                                                         
        \begin{center}                                                                                                  
        \scriptsize                                                                                                     
        \begin{tabular}{ l|c|c|c|c|c }                                                                                                  
                \multicolumn{2}{c}{}     &      \multicolumn{3}{c}{{\scshape{\color{trColo}True}}}&\\                                                                           
                \cline{3-5}                                                                                             
                \multicolumn{2}{c|}{}&                                                                                          
                        \color{trColo}"  0 "&                                                                                   
                        \color{trColo}"+1"&                                                                                     
                        \color{trColo}"+2"&                                                                                     
                         \textit{\color{dkgray}Total}   \\                                                                              
                \hhline{~-|---|-}                                                                                               
                \multirow{4}{*}                                                                                         
                {\centering{\rotatebox[origin=c]{90}{\hspace{0.5cm}\scshape{\color{prColo}Predicted}}}}                                                                                         
                &       \color{prColo}" 0 "     &\cellcolor{lightGreen} 3520    &       271     &       0       &       \color{dkgray}  3791    \\      
                \hhline{~-|---|}                                                                                                
                &       \color{prColo}"+1"       &      988     &\cellcolor{lightGreen} 3070    &       1752    &       \color{dkgray}  5810    \\
\end{tabular}                                                                                         
        \caption{\color{dkblue}[Resubstitution prediction]\color{black} - Gentle Boost}                                                                                                   
        \label{table:cm_vvds2}                                                                                                  
        \end{center}            
}\hspace{0.7cm} 
\parbox{.45\linewidth}{                                                                                                         
        \begin{center}                                                                                                  
        \scriptsize                                                                                                     
        \begin{tabular}{ l|c|c|c|c|c }                                                                                                  
                \multicolumn{2}{c}{}     &      \multicolumn{3}{c}{{\scshape{\color{trColo}True}}}&\\                                                                           
                \cline{3-5}                                                                                             
                \multicolumn{2}{c|}{}&                                                                                          
                        \color{trColo}"  0 "&                                                                                   
                        \color{trColo}"+1"&                                                                                     
                        \color{trColo}"+2"&                                                                                     
                         \textit{\color{dkgray}Total}   \\                                                                              
                \hhline{~-|---|-}                                                                                               
                \multirow{4}{*}                                                                                         
                {\centering{\rotatebox[origin=c]{90}{\hspace{0.5cm}\scshape{\color{prColo}Predicted}}}}                                                                                         
                &       \color{prColo}" 0 "     &\cellcolor{lightGreen} 3654    &       457     &       1       &       \color{dkgray}  4112    \\      
                \hhline{~-|---|}                                                                                                
                &       \color{prColo}"+1"       &      851     &\cellcolor{lightGreen} 2281    &       625     &       \color{dkgray}  3757    \\
\end{tabular}                                                                                         
        \caption{\color{dkblue}[Resubstitution prediction]\color{black} - SVM (Linear kernel)}                                                                                                    
        \label{table:cm_vvds3}                                                                                                  
        \end{center}                    
}\hspace{0.7cm} 
\parbox{.45\linewidth}{                                                                                                         
        \begin{center}                                                                                                  
        \scriptsize                                                                                                     
        \begin{tabular}{ l|c|c|c|c|c }                                                                                                  
                \multicolumn{2}{c}{}     &      \multicolumn{3}{c}{{\scshape{\color{trColo}True}}}&\\                                                                           
                \cline{3-5}                                                                                             
                \multicolumn{2}{c|}{}&                                                                                          
                        \color{trColo}"  0 "&                                                                                   
                        \color{trColo}"+1"&                                                                                     
                        \color{trColo}"+2"&                                                                                     
                         \textit{\color{dkgray}Total}   \\                                                                              
                \hhline{~-|---|-}                                                                                               
                \multirow{4}{*}                                                                                         
                {\centering{\rotatebox[origin=c]{90}{\hspace{0.5cm}\scshape{\color{prColo}Predicted}}}}                                                                                         
                &       \color{prColo}" 0 "     &\cellcolor{lightGreen} 3646    &       372     &       2       &       \color{dkgray}  4020    \\      
                \hhline{~-|---|}                                                                                                
                &       \color{prColo}"+1"       &      866     &\cellcolor{lightGreen} 2823    &       1431    &       \color{dkgray}  5120    \\
\end{tabular}                                                                                         
        \caption{\color{dkblue}[Resubstitution prediction]\color{black}- SVM (Gaussian kernel)}                                                                                                  
        \label{table:cm_vvds4}                                                                                                  
        \end{center}                                                                                                    
                                                                                                                
}\hfill 
                                                                                                        
\centering \rule{14cm}{0.4pt}                                                                                                           
\parbox{.45\linewidth}{                                                                                                         
        \begin{center}                                                                                                  
        \scriptsize                                                                                                     
        \begin{tabular}{ l|c|c|c|c|c }                                                                                                  
                \multicolumn{2}{c}{}     &      \multicolumn{3}{c}{{\scshape{\color{trColo}True}}}&\\                                                                           
                \cline{3-5}                                                                                             
                \multicolumn{2}{c|}{}&                                                                                          
                        \color{trColo}"  0 "&                                                                                   
                        \color{trColo}"+1"&                                                                                     
                        \color{trColo}"+2"&                                                                                     
                         \textit{\color{dkgray}Total}   \\                                                                              
                \hhline{~-|---|-}                                                                                               
                \multirow{4}{*}                                                                                         
                {\centering{\rotatebox[origin=c]{90}{\hspace{0.5cm}\scshape{\color{prColo}Predicted}}}}                                                                                         
                &       \color{prColo}" 0 "     &\cellcolor{lightGreen} 722     &       271     &       442     &       \color{dkgray}  1435    \\      
                \hhline{~-|---|}                                                                                                
                &       \color{prColo}"+1"       &      974     &\cellcolor{lightGreen} 256     &       1424    &       \color{dkgray}  2654    \\
\end{tabular}                                                                                         
        \caption{\color{dkgreen2}[Test prediction]\color{black} - Bagging Trees}                                                                                                  
        \label{table:cm_vvds5}                                                                                                  
        \end{center}            
}\hspace{0.7cm} 
\parbox{.45\linewidth}{                                                                                                         
        \begin{center}                                                                                                  
        \scriptsize                                                                                                     
        \begin{tabular}{ l|c|c|c|c|c }                                                                                                  
                \multicolumn{2}{c}{}     &      \multicolumn{3}{c}{{\scshape{\color{trColo}True}}}&\\                                                                           
                \cline{3-5}                                                                                             
                \multicolumn{2}{c|}{}&                                                                                          
                        \color{trColo}"  0 "&                                                                                   
                        \color{trColo}"+1"&                                                                                     
                        \color{trColo}"+2"&                                                                                     
                         \textit{\color{dkgray}Total}   \\                                                                              
                \hhline{~-|---|-}                                                                                               
                \multirow{4}{*}                                                                                         
                {\centering{\rotatebox[origin=c]{90}{\hspace{0.5cm}\scshape{\color{prColo}Predicted}}}}                                                                                         
                &       \color{prColo}" 0 "     &\cellcolor{lightGreen} 447     &       272     &       253     &       \color{dkgray}  972     \\      
                \hhline{~-|---|}                                                                                                
                &       \color{prColo}"+1"       &      1398    &\cellcolor{lightGreen} 381     &       1799    &       \color{dkgray}  3578    \\
\end{tabular}                                                                                         
        \caption{\color{dkgreen2}[Test prediction]\color{black} - Gentle Boost}                                                                                                  
        \label{table:cm_vvds6}                                                                                                  
        \end{center}            
}\hspace{0.7cm} 
\parbox{.45\linewidth}{                                                                                                         
        \begin{center}                                                                                                  
        \scriptsize                                                                                                     
        \begin{tabular}{ l|c|c|c|c|c }                                                                                                  
                \multicolumn{2}{c}{}     &      \multicolumn{3}{c}{{\scshape{\color{trColo}True}}}&\\                                                                           
                \cline{3-5}                                                                                             
                \multicolumn{2}{c|}{}&                                                                                          
                        \color{trColo}"  0 "&                                                                                   
                        \color{trColo}"+1"&                                                                                     
                        \color{trColo}"+2"&                                                                                     
                         \textit{\color{dkgray}Total}   \\                                                                              
                \hhline{~-|---|-}                                                                                               
                \multirow{4}{*}                                                                                         
                {\centering{\rotatebox[origin=c]{90}{\hspace{0.5cm}\scshape{\color{prColo}Predicted}}}}                                                                                         
                &       \color{prColo}" 0 "     &\cellcolor{lightGreen} 559     &       272     &       330     &       \color{dkgray}  1161    \\      
                \hhline{~-|---|}                                                                                                
                &       \color{prColo}"+1"       &      956     &\cellcolor{lightGreen} 141     &       1317    &       \color{dkgray}  2414    \\
\end{tabular}                                                                                         
        \caption{\color{dkgreen2}[Test prediction]\color{black} - SVM (Linear kernel)}                                                                                                        
        \label{table:cm_vvds7}                                                                                                  
        \end{center}            
}\hspace{0.7cm} 
\parbox{.45\linewidth}{                                                                                                         
        \begin{center}                                                                                                  
        \scriptsize                                                                                                     
        \begin{tabular}{ l|c|c|c|c|c }                                                                                                  
                \multicolumn{2}{c}{}     &      \multicolumn{3}{c}{{\scshape{\color{trColo}True}}}&\\                                                                           
                \cline{3-5}                                                                                             
                \multicolumn{2}{c|}{}&                                                                                          
                        \color{trColo}"  0 "&                                                                                   
                        \color{trColo}"+1"&                                                                                     
                        \color{trColo}"+2"&                                                                                     
                         \textit{\color{dkgray}Total}   \\                                                                              
                \hhline{~-|---|-}                                                                                               
                \multirow{4}{*}                                                                                         
                {\centering{\rotatebox[origin=c]{90}{\hspace{0.5cm}\scshape{\color{prColo}Predicted}}}}         
                &       \color{prColo}" 0 "     &\cellcolor{lightGreen} 494     &       270     &       289     &       \color{dkgray}  1053    \\      
                \hhline{~-|---|}                                                                                                
                &       \color{prColo}"+1"       &      1300    &\cellcolor{lightGreen} 336     &       1638    &       \color{dkgray}  3274    \\
\end{tabular}                                                                                         
        \caption{\color{dkgreen2}[Test prediction]\color{black} - SVM (Gaussian kernel)}                                                                                                        
        \label{table:cm_vvds8}                                                                                                  
        \end{center}                                                                                                    
}                                                                                                                                               
\end{table*}    
}\restoregeometry
}

{
\setcounter{table}{8} \renewcommand{\thetable}{A.\arabic{table}} 
\begin{table*}[h]                                                                                                               
\vspace{-0.1cm}                                                                                                            
\centering\footnotesize \textbf{\colorbox{gray!20}{ \color{dkgray} General performances using modified VVDS flags }} \\     
\vspace{-0.5cm}           
\parbox{1.0\linewidth}{ 
\begin{center}  
\parbox{.5\linewidth}{                                                                                                          
        \begin{center}                                                                                                  
        \scriptsize             
        \begin{tabular}{|*{5}{c|}} 
                \multicolumn{2}{c}{}     &      \multicolumn{3}{c}{{\textit{\color{dkgray}Measures per class}}}\\                                                                          
                \cline{3-5}                                                                                             
                \multicolumn{2}{c|}{}           &                                                                                               
                        \color{trColo}"  0 "    &                                                                                       
                        \color{trColo}"+1"      &                                                                                       
                        \color{trColo}"+2"      \\                                                                              
                        \hline                                                                                                          
                                \multirow{5}{*}{\centering{\rotatebox[origin=c]{90}{ \scshape{Tree Bagger}}}}                                                                                        
                                        &       \multirow{1}{*}{Accuracy}               &       99.93\% &       98.69\% &       98.75\% \\      \cline{2-5}
                                        &       \multirow{1}{*}{Precision}              &       99.93\% &       94.41\% &       100\%   \\      \cline{2-5}
                                        &       \multirow{1}{*}{Sensitivity}            &       99.80\% &       99.94\% &       97.53\% \\      \cline{2-5}
                                        &       \multirow{1}{*}{Specificity}            &       99.97\% &       98.34\% &       100\%   \\      \cline{2-5}
                                        &       \multirow{1}{*}{F-score}                &       99.87\% &       97.10\% &       98.75\% \\      
                        \hhline{=====}                                                                                                  
                                \multirow{5}{*}{\centering{\rotatebox[origin=c]{90}{ \scshape{Gentle Boost}}}}                                                                                       
                                        &       \multirow{1}{*}{Accuracy}               &       92.27\% &       80.10\% &       87.78\% \\      \cline{2-5}
                                        &       \multirow{1}{*}{Precision}              &       92.85\% &       52.84\% &       96.35\% \\      \cline{2-5}
                                        &       \multirow{1}{*}{Sensitivity}            &       78.01\% &       85.68\% &       78.77\% \\      \cline{2-5}
                                        &       \multirow{1}{*}{Specificity}            &       97.71\% &       78.53\% &       96.96\% \\      \cline{2-5}
                                        &       \multirow{1}{*}{F-score}                &       84.79\% &       65.37\% &       86.68\% \\      
                        \hhline{=====}                                                                                                  
                                \multirow{5}{*}{\centering{\rotatebox[origin=c]{90}{ \scshape{SVM (linear)}}}}                                                                                       
                                        &       \multirow{1}{*}{Accuracy}               &       91.95\% &       83.01\% &       90.96\% \\      \cline{2-5}
                                        &       \multirow{1}{*}{Precision}              &       88.86\% &       60.71\% &       89.95\% \\      \cline{2-5}
                                        &       \multirow{1}{*}{Sensitivity}            &       80.98\% &       63.66\% &       92.41\% \\      \cline{2-5}
                                        &       \multirow{1}{*}{Specificity}            &       96.13\% &       88.44\% &       89.47\% \\      \cline{2-5}
                                        &       \multirow{1}{*}{F-score}                &       84.74\% &       62.15\% &       91.16\% \\      
                        \hhline{=====}                                                                                                  
                                \multirow{5}{*}{\centering{\rotatebox[origin=c]{90}{ \scshape{SVM (rbf)}}}}                                                                                  
                                        &       \multirow{1}{*}{Accuracy}               &       92.41\% &       81.30\% &       88.86\% \\      \cline{2-5}
                                        &       \multirow{1}{*}{Precision}              &       90.70\% &       55.14\% &       94.62\% \\      \cline{2-5}
                                        &       \multirow{1}{*}{Sensitivity}            &       80.81\% &       78.79\% &       82.63\% \\      \cline{2-5}
                                        &       \multirow{1}{*}{Specificity}            &       96.84\% &       82\%            &       95.21\% \\      \cline{2-5}
                                        &       \multirow{1}{*}{F-score}                &       85.47\% &       64.87\% &       88.22\% \\      
                                                                                                                        
                \cline{1-5}
        \end{tabular}                                                                                                           
        \end{center}                                                                                                    
}
\hspace{0.7cm} 
\parbox{.35\linewidth}{ 
        \begin{center}                                                                                                  
        \scriptsize             
        \begin{tabular}{|*{3}{c|}} 
                \multicolumn{1}{c}{}{}\\
                \multicolumn{1}{c}{}     &      \multicolumn{2}{c}{{\textit{\color{dkgray}Average per-class}}}\\
                        \hline                                                                          
                                \multirow{5}{*}{\centering{\rotatebox[origin=c]{90}{ \scshape{Tree Bagger}}}}                                                        
                                        &       \multirow{1}{*}{Accuracy}               &       99.12\% \\      \cline{2-3}
                                        &       \multirow{1}{*}{Error rate}             &       \color{prColo}{0.88\%}  \\      \cline{2-3}
                                        &       \multirow{1}{*}{Precision}              &       98.11\% \\      \cline{2-3}
                                        &       \multirow{1}{*}{Sensitivity}            &       99.09\% \\      \cline{2-3}
                                        &       \multirow{1}{*}{F-score}                &       98.60\% \\      
                        \hhline{===}                                                            
                                \multirow{5}{*}{\centering{\rotatebox[origin=c]{90}{ \scshape{Gentle Boost}}}}                                                       
                                        &       \multirow{1}{*}{Accuracy}               &       86.72\% \\      \cline{2-3}
                                        &       \multirow{1}{*}{Error rate}             &       \color{prColo}{13.28\%} \\      \cline{2-3}
                                        &       \multirow{1}{*}{Precision}              &       80.68\% \\      \cline{2-3}
                                        &       \multirow{1}{*}{Sensitivity}            &       80.82\% \\      \cline{2-3}
                                        &       \multirow{1}{*}{F-score}                &       80.75\% \\      
                        \hhline{===}                                                            
                                \multirow{5}{*}{\centering{\rotatebox[origin=c]{90}{ \scshape{SVM (linear)}}}}                                                       
                                        &       \multirow{1}{*}{Accuracy}               &       88.64\% \\      \cline{2-3}
                                        &       \multirow{1}{*}{Error rate}             &       \color{prColo}{11.36\%} \\      \cline{2-3}
                                        &       \multirow{1}{*}{Precision}              &       79.84\% \\      \cline{2-3}
                                        &       \multirow{1}{*}{Sensitivity}            &       79.02\% \\      \cline{2-3}
                                        &       \multirow{1}{*}{F-score}                &       79.43\% \\      
                                \hhline{===}                                                    
                                \multirow{5}{*}{\centering{\rotatebox[origin=c]{90}{ \scshape{SVM (rbf)}}}}                                                  
                                        &       \multirow{1}{*}{Accuracy}               &       87.52\% \\      \cline{2-3}
                                        &       \multirow{1}{*}{Error rate}             &       \color{prColo}{12.48\%} \\      \cline{2-3}
                                        &       \multirow{1}{*}{Precision}              &       80.15\% \\      \cline{2-3}
                                        &       \multirow{1}{*}{Sensitivity}            &       80.74\% \\      \cline{2-3}
                                        &       \multirow{1}{*}{F-score}                &       80.45\% \\      
                \cline{1-3}
        \end{tabular}                                                                                   
        \end{center}                                                                                                    
}
\end{center}
}
\vspace{-0.7cm}
\caption{\color{dkblue}[Resubstitution prediction]\color{black}- Measures from confusion matrices \vspace{0.1cm}}
\label{table:cm_vvds_resub}     
        
\centering \rule{14cm}{0.4pt}                                                                                                         
\parbox{1.0\linewidth}{  
\vspace{-0.5cm}  
\begin{center}          
\parbox{.5\linewidth}{                                                                                                          
        \begin{center}                                                                                                  
        \scriptsize             
        \begin{tabular}{|*{5}{c|}} 
                \multicolumn{2}{c}{}     &      \multicolumn{3}{c}{{\textit{\color{dkgray}Measures per class}}}\\                                                                          
                \cline{3-5}                                                                                             
                \multicolumn{2}{c|}{}           &                                                                                               
                        \color{trColo}"  0 "    &                                                                                       
                        \color{trColo}"+1"      &                                                                                       
                        \color{trColo}"+2"      \\                                                                              
                        \hline                                                                                                          
                                \multirow{5}{*}{\centering{\rotatebox[origin=c]{90}{ \scshape{Tree Bagger}}}}                                                                                        
                                        &       \multirow{1}{*}{Accuracy}               &       72.51\% &       51.87\% &       54.84\% \\      \cline{2-5}
                                        &       \multirow{1}{*}{Precision}              &       50.31\% &       9.65\%  &       55.31\% \\      \cline{2-5}
                                        &       \multirow{1}{*}{Sensitivity}            &       32\%            &       14.29\% &       54.76\% \\      \cline{2-5}
                                        &       \multirow{1}{*}{Specificity}            &       87.95\% &       62.42\% &       54.92\% \\      \cline{2-5}
                                        &       \multirow{1}{*}{F-score}                &       39.12\% &       11.52\% &       55.04\% \\      
                        \hhline{=====}                                                                                                  
                                \multirow{5}{*}{\centering{\rotatebox[origin=c]{90}{ \scshape{Gentle Boost}}}}                                                                                       
                                        &       \multirow{1}{*}{Accuracy}               &       71.44\% &       43.62\% &       55.93\% \\      \cline{2-5}
                                        &       \multirow{1}{*}{Precision}              &       45.99\% &       10.65\% &       57.22\% \\      \cline{2-5}
                                        &       \multirow{1}{*}{Sensitivity}            &       19.81\% &       21.26\% &       50.25\% \\      \cline{2-5}
                                        &       \multirow{1}{*}{Specificity}            &       91.13\% &       49.90\% &       61.71\% \\      \cline{2-5}
                                        &       \multirow{1}{*}{F-score}                &       27.70\% &       14.19\% &       53.51\% \\      
                        \hhline{=====}                                                                                                  
                                \multirow{5}{*}{\centering{\rotatebox[origin=c]{90}{ \scshape{SVM (linear)}}}}                                                                                       
                                        &       \multirow{1}{*}{Accuracy}               &       71.87\% &       51.99\% &       53.91\% \\      \cline{2-5}
                                        &       \multirow{1}{*}{Precision}              &       48.15\% &       5.84\%  &       53.89\% \\      \cline{2-5}
                                        &       \multirow{1}{*}{Sensitivity}            &       24.78\% &       7.87\%  &       60.07\% \\      \cline{2-5}
                                        &       \multirow{1}{*}{Specificity}            &       89.83\% &       64.38\% &       47.63\% \\      \cline{2-5}
                                        &       \multirow{1}{*}{F-score}                &       32.72\% &       6.70\%  &       56.82\% \\      
                        \hhline{=====}                                                                                                  
                                \multirow{5}{*}{\centering{\rotatebox[origin=c]{90}{ \scshape{SVM (rbf)}}}}                                                                                  
                                        &       \multirow{1}{*}{Accuracy}               &       71.60\% &       46.24\% &       56.26\% \\      \cline{2-5}
                                        &       \multirow{1}{*}{Precision}              &       46.91\% &       10.26\% &       57.15\% \\      \cline{2-5}
                                        &       \multirow{1}{*}{Sensitivity}            &       21.90\% &       18.75\% &       53.28\% \\      \cline{2-5}
                                        &       \multirow{1}{*}{Specificity}            &       90.55\% &       53.96\% &       59.29\% \\      \cline{2-5}
                                        &       \multirow{1}{*}{F-score}                &       29.86\% &       13.26\% &       55.15\% \\      
                \cline{1-5}
        \end{tabular}                                                                                                           
        \end{center}                                                                                                    
}
\hspace{0.7cm} 
\parbox{.35\linewidth}{ 
        \begin{center}                                                                                                  
        \scriptsize             
        \begin{tabular}{|*{3}{c|}} 
                \multicolumn{1}{c}{}{}\\
                \multicolumn{1}{c}{}     &      \multicolumn{2}{c}{{\textit{\color{dkgray}Average per-class}}}\\
                        \hline                                                                          
                                \multirow{5}{*}{\centering{\rotatebox[origin=c]{90}{ \scshape{Tree Bagger}}}}                                                        
                                        &       \multirow{1}{*}{Accuracy}               &       59.74\% \\      \cline{2-3}
                                        &       \multirow{1}{*}{Error rate}             &       \color{prColo}{40.26\%} \\      \cline{2-3}
                                        &       \multirow{1}{*}{Precision}              &       38.42\% \\      \cline{2-3}
                                        &       \multirow{1}{*}{Sensitivity}            &       33.68\% \\      \cline{2-3}
                                        &       \multirow{1}{*}{F-score}                &       35.90\% \\      
                        \hhline{===}                                                            
                                \multirow{5}{*}{\centering{\rotatebox[origin=c]{90}{ \scshape{Gentle Boost}}}}                                                       
                                        &       \multirow{1}{*}{Accuracy}               &       57\%            \\      \cline{2-3}
                                        &       \multirow{1}{*}{Error rate}             &       \color{prColo}{43\%}            \\      \cline{2-3}
                                        &       \multirow{1}{*}{Precision}              &       37.95\% \\      \cline{2-3}
                                        &       \multirow{1}{*}{Sensitivity}            &       30.44\% \\      \cline{2-3}
                                        &       \multirow{1}{*}{F-score}                &       33.79\% \\      
                        \hhline{===}                                                            
                                \multirow{5}{*}{\centering{\rotatebox[origin=c]{90}{ \scshape{SVM (linear)}}}}                                                       
                                        &       \multirow{1}{*}{Accuracy}               &       59.26\% \\      \cline{2-3}
                                        &       \multirow{1}{*}{Error rate}             &       \color{prColo}{40.74\%} \\      \cline{2-3}
                                        &       \multirow{1}{*}{Precision}              &       35.96\% \\      \cline{2-3}
                                        &       \multirow{1}{*}{Sensitivity}            &       30.91\% \\      \cline{2-3}
                                        &       \multirow{1}{*}{F-score}                &       33.24\% \\      
                                \hhline{===}                                                    
                                \multirow{5}{*}{\centering{\rotatebox[origin=c]{90}{ \scshape{SVM (rbf)}}}}                                                  
                                        &       \multirow{1}{*}{Accuracy}               &       58.03\% \\      \cline{2-3}
                                        &       \multirow{1}{*}{Error rate}             &       \color{prColo}{41.97\%} \\      \cline{2-3}
                                        &       \multirow{1}{*}{Precision}              &       38.11\% \\      \cline{2-3}
                                        &       \multirow{1}{*}{Sensitivity}            &       31.31\% \\      \cline{2-3}
                                        &       \multirow{1}{*}{F-score}                &       34.38\% \\      
                \cline{1-3}
        \end{tabular}                                                                                                           
        \end{center}                                                                                                    
}
\end{center}
}\vspace{-0.7cm}
\caption{\color{dkgreen2}[Test prediction]\color{black}- Measures from confusion matrices \vspace{0.1cm}}
\label{table:cm_vvds_pred}                                                                                                                              
\end{table*}    
}

{
\newgeometry{top=1in,bottom=1in,right=1.0in,left=0.5in}
{
\setcounter{table}{0} \renewcommand{\thetable}{B.\arabic{table}} 
\begin{table*}[h]                                                                                                                                       
\vspace{1.25cm}                                                                                                                                 
\centering\footnotesize \textbf{\colorbox{gray!20}{ \color{dkgray} Confusion matrices using partition labels }} \\                                                                                                                           
\scriptsize                                                                                                                                     
\parbox{.5\linewidth}{  
        \begin{center}{                                                                                                                         
\begin{tabular}{ l|c|c|c|c|c|c|c }                                                                                                                              
                \multicolumn{2}{c}{}     &      \multicolumn{5}{c}{{\scshape{\color{trColo}True}}}&\\                                                                                                   
                \cline{3-7}                                                                                                                     
                \multicolumn{2}{c|}{}&                                                                                                                  
                        \color{trColo}C1&                                                                                                               
                        \color{trColo}C2&                                                                                                               
                        \color{trColo}C3&                                                                                                               
                        \color{trColo}C4&                                                                                                               
                        \color{trColo}C5&                                                                                                               
                         \textit{\color{dkgray}Total}   \\
\end{tabular}
                }
        \caption{\color{dkblue}[Resubstitution prediction]\color{black} - Bagging Trees}                  
        \label{table:cm_clust1}                                                                                                                         
        \end{center}                                                                                                                            
}\hfill 
\parbox{.45\linewidth}{                                                                                                                                 
        \begin{center}                                                                                                                          
        \begin{tabular}{ l|c|c|c|c|c|c|c }                                                                                                                              
                \multicolumn{2}{c}{}     &      \multicolumn{5}{c}{{\scshape{\color{trColo}True}}}&\\                                                                                                   
                \cline{3-7}                                                                                                                     
                \multicolumn{2}{c|}{}&                                                                                                                  
                        \color{trColo}C1&                                                                                                               
                        \color{trColo}C2&                                                                                                               
                        \color{trColo}C3&                                                                                                               
                        \color{trColo}C4&                                                                                                               
                        \color{trColo}C5&                                                                                                               
                         \textit{\color{dkgray}Total}   \\
\end{tabular}                                                                                                                           
        \caption{\color{dkblue}[Resubstitution prediction]\color{black} - Gentle Boost}                                                                                                                           
        \label{table:cm_clust2}                                                                                                                         
        \end{center}                                                                                                                            
}\hfill
\parbox{.45\linewidth}{                                                                                                                                 
        \begin{center}                                                                                                                          
        \begin{tabular}{ l|c|c|c|c|c|c|c }                                                                                                                              
                \multicolumn{2}{c}{}     &      \multicolumn{5}{c}{{\scshape{\color{trColo}True}}}&\\                                                                                                   
                \cline{3-7}                                                                                                                     
                \multicolumn{2}{c|}{}&                                                                                                                  
                        \color{trColo}C1&                                                                                                               
                        \color{trColo}C2&                                                                                                               
                        \color{trColo}C3&                                                                                                               
                        \color{trColo}C4&                                                                                                               
                        \color{trColo}C5&                                                                                                               
                         \textit{\color{dkgray}Total}   \\
\end{tabular}                                                                                                                           
        \caption{\color{dkblue}[Resubstitution prediction]\color{black} - SVM (Linear kernel)}                                                                                                                            
        \label{table:cm_clust3}                                                                                                                         
        \end{center}                                                                                                                            
}\hfill
\parbox{.45\linewidth}{                                                                                                                                 
        \begin{center}                                                                                                                          
        \begin{tabular}{ l|c|c|c|c|c|c|c }                                                                                                                              
                \multicolumn{2}{c}{}     &      \multicolumn{5}{c}{{\scshape{\color{trColo}True}}}&\\                                                                                                   
                \cline{3-7}                                                                                                                     
                \multicolumn{2}{c|}{}&                                                                                                                  
                        \color{trColo}C1&                                                                                                               
                        \color{trColo}C2&                                                                                                               
                        \color{trColo}C3&                                                                                                               
                        \color{trColo}C4&                                                                                                               
                        \color{trColo}C5&                                                                                                               
                         \textit{\color{dkgray}Total}   \\
\end{tabular}                                                                                                                           
        \caption{\color{dkblue}[Resubstitution prediction]\color{black} - SVM (Gaussian kernel)}                                                                                                                          
        \label{table:cm_clust4}                                                                                                                         
        \end{center}                                                                                                                            
                                                                                                                                        
}\hfill                                                                                                                                 

\centering \rule{14cm}{0.4pt}                                                                                                                                   
\parbox{.45\linewidth}{                                                                                                                                 
        \begin{center}                                                                                                                          
        \begin{tabular}{ l|c|c|c|c|c|c|c }                                                                                                                              
                \multicolumn{2}{c}{}     &      \multicolumn{5}{c}{{\scshape{\color{trColo}True}}}&\\                                                                                                   
                \cline{3-7}                                                                                                                     
                \multicolumn{2}{c|}{}&                                                                                                                  
                        \color{trColo}C1&                                                                                                               
                        \color{trColo}C2&                                                                                                               
                        \color{trColo}C3&                                                                                                               
                        \color{trColo}C4&                                                                                                               
                        \color{trColo}C5&                                                                                                               
                         \textit{\color{dkgray}Total}   \\
\end{tabular}                                                                                                                   
        \caption{\color{dkgreen2}[Test prediction]\color{black} - Bagging Trees}                                                                                                                          
        \label{table:cm_clust5}                                                                                                                         
        \end{center}                                                                                                                            
}\hfill
\parbox{.45\linewidth}{                                                                                                                                 
        \begin{center}                                                                                                                          
        \begin{tabular}{ l|c|c|c|c|c|c|c }                                                                                                                              
                \multicolumn{2}{c}{}     &      \multicolumn{5}{c}{{\scshape{\color{trColo}True}}}&\\                                                                                                   
                \cline{3-7}                                                                                                                     
                \multicolumn{2}{c|}{}&                                                                                                                  
                        \color{trColo}C1&                                                                                                               
                        \color{trColo}C2&                                                                                                               
                        \color{trColo}C3&                                                                                                               
                        \color{trColo}C4&                                                                                                               
                        \color{trColo}C5&                                                                                                               
                         \textit{\color{dkgray}Total}   \\
\end{tabular}                                                                                                                           
        \caption{\color{dkgreen2}[Test prediction]\color{black} - Gentle Boost}                                                                                                                          
        \label{table:cm_clust6}                                                                                                                         
        \end{center}                                                                                                                            
}\hfill 
\parbox{.45\linewidth}{                                                                                                                                 
        \begin{center}                                                                                                                          
        \begin{tabular}{ l|c|c|c|c|c|c|c }                                                                                                                              
                \multicolumn{2}{c}{}     &      \multicolumn{5}{c}{{\scshape{\color{trColo}True}}}&\\                                                                                                   
                \cline{3-7}                                                                                                                     
                \multicolumn{2}{c|}{}&                                                                                                                  
                        \color{trColo}C1&                                                                                                               
                        \color{trColo}C2&                                                                                                               
                        \color{trColo}C3&                                                                                                               
                        \color{trColo}C4&                                                                                                               
                        \color{trColo}C5&                                                                                                               
                         \textit{\color{dkgray}Total}   \\
\end{tabular}                                                                                                                           
        \caption{\color{dkgreen2}[Test prediction]\color{black} - SVM (Linear kernel)}
        \label{table:cm_clust7}                                                                                                                         
        \end{center}                                                                                                                            
}\hfill
\parbox{.45\linewidth}{                                                                                                                                 
        \begin{center}                                                                                                                          
        \begin{tabular}{ l|c|c|c|c|c|c|c }                                                                                                                              
                \multicolumn{2}{c}{}     &      \multicolumn{5}{c}{{\scshape{\color{trColo}True}}}&\\                                                                                                   
                \cline{3-7}                                                                                                                     
                \multicolumn{2}{c|}{}&                                                                                                                  
                        \color{trColo}C1&                                                                                                               
                        \color{trColo}C2&                                                                                                               
                        \color{trColo}C3&                                                                                                               
                        \color{trColo}C4&                                                                                                               
                        \color{trColo}C5&                                                                                                               
                         \textit{\color{dkgray}Total}   \\
\end{tabular}                                                                                                                           
        \caption{\color{dkgreen2}[Test prediction]\color{black} - SVM (Gaussian kernel)}                                                                                                                                
        \label{table:cm_clust8}                                                                                                                         
        \end{center}                                                                                                                            
}                                                                                                                                       
\end{table*}                                                                                                                                                                                                            
}\restoregeometry
}

{
\setcounter{table}{8} \renewcommand{\thetable}{B.\arabic{table}} 
\begin{table*}[h]                                                                                                               
\vspace{-0.1cm}                                                                                                             
\centering\footnotesize \textbf{\colorbox{gray!20}{ \color{dkgray} General performances using partition labels }} \\                                                                                                       
\vspace{-0.5cm}   
\parbox{1.0\linewidth}{
\begin{center}          
\parbox{.55\linewidth}{                                                                                                         
        \begin{center}                                                                                                  
        \scriptsize             
        \begin{tabular}{|*{7}{c|}} 
                \multicolumn{2}{c}{}     &      \multicolumn{5}{c}{{\textit{\color{dkgray}Measures per class}}}\\                                                                          
                \cline{3-7}                                                                                             
                \multicolumn{2}{c|}{}   &                                                                                                                       
                        \color{trColo}C1        &                                                                                                               
                        \color{trColo}C2        &                                                                                                               
                        \color{trColo}C3        &                                                                                                               
                        \color{trColo}C4        &                                                                                                               
                        \color{trColo}C5        \\                                                                              
                \hline                                                                                                                                                          
                        \multirow{5}{*}{\centering{\rotatebox[origin=c]{90}{ \scshape{Tree Bagger}}}}                        
                                        &       \multirow{1}{*}{Accuracy}       &       99.98\% &       99.95\% &       99.94\% &       99.97\% &       100\%   \\      \cline{2-7}             
                                        &       \multirow{1}{*}{Precision}      &       99.86\% &       99.91\% &       99.92\% &       99.91\% &       100\%   \\      \cline{2-7}             
                                        &       \multirow{1}{*}{Sensitivity}    &       100\%   &       99.91\% &       99.81\% &       99.94\% &       100\%   \\      \cline{2-7}             
                                        &       \multirow{1}{*}{Specificity}    &       99.98\% &       99.97\% &       99.98\% &       99.98\% &       100\%   \\      \cline{2-7}             
                                        &       \multirow{1}{*}{F-score}        &       99.93\% &       99.91\% &       99.87\% &       99.92\% &       100\%   \\                      
                        \hhline{=======}                                                                                                                                                
                                \multirow{5}{*}{\centering{\rotatebox[origin=c]{90}{ \scshape{Gentle Boost}}}}                                                       
                                        &       \multirow{1}{*}{Accuracy}       &       99.55\% &       98.81\% &       98.20\% &       98.62\% &       99.66\% \\      \cline{2-7}             
                                        &       \multirow{1}{*}{Precision}      &       98.24\% &       98.62\% &       96.99\% &       94.83\% &       98.70\% \\      \cline{2-7}             
                                        &       \multirow{1}{*}{Sensitivity}    &       98.24\% &       97.01\% &       95.14\% &       98.55\% &       99.26\% \\      \cline{2-7}             
                                        &       \multirow{1}{*}{Specificity}    &       99.74\% &       99.49\% &       99.12\% &       98.63\% &       99.74\% \\      \cline{2-7}             
                                        &       \multirow{1}{*}{F-score}        &       98.24\% &       97.81\% &       96.06\% &       96.65\% &       98.98\% \\                      
                        \hhline{=======}                                                                                                                                                
                                \multirow{5}{*}{\centering{\rotatebox[origin=c]{90}{ \scshape{SVM (linear)}}}}                                               
                                        &       \multirow{1}{*}{Accuracy}       &       99.93\% &       98.80\% &       97.60\% &       98.53\% &       99.81\% \\      \cline{2-7}             
                                        &       \multirow{1}{*}{Precision}      &       99.62\% &       98.84\% &       95.35\% &       94.47\% &       99.52\% \\      \cline{2-7}             
                                        &       \multirow{1}{*}{Sensitivity}    &       99.86\% &       96.76\% &       94.16\% &       98.49\% &       99.33\% \\      \cline{2-7}             
                                        &       \multirow{1}{*}{Specificity}    &       99.94\% &       99.57\% &       98.63\% &       98.53\% &       99.90\% \\      \cline{2-7}             
                                        &       \multirow{1}{*}{F-score}        &       99.74\% &       97.79\% &       94.75\% &       96.44\% &       99.42\% \\                      
                        \hhline{=======}                                                                                                                                                
                                \multirow{5}{*}{\centering{\rotatebox[origin=c]{90}{ \scshape{SVM (rbf)}}}}                                  
                                        &       \multirow{1}{*}{Accuracy}       &       99.89\% &       99.56\% &       99.44\% &       99.65\% &       99.87\% \\      \cline{2-7}             
                                        &       \multirow{1}{*}{Precision}      &       99.43\% &       99.64\% &       98.68\% &       98.77\% &       99.59\% \\      \cline{2-7}             
                                        &       \multirow{1}{*}{Sensitivity}    &       99.71\% &       98.75\% &       98.91\% &       99.49\% &       99.63\% \\      \cline{2-7}             
                                        &       \multirow{1}{*}{Specificity}    &       99.92\% &       99.87\% &       99.60\% &       99.69\% &       99.92\% \\      \cline{2-7}             
                                        &       \multirow{1}{*}{F-score}        &       99.57\% &       99.19\% &       98.79\% &       99.13\% &       99.61\% \\                      
                \cline{1-7}
        \end{tabular}                                                                                                           
        \end{center}                                                                                                    
}
\parbox{.275\linewidth}{        
        \begin{center}                                                                                                  
        \scriptsize             
        \begin{tabular}{|*{3}{c|}} 
                \multicolumn{1}{c}{}{}\\
                        \multicolumn{1}{c}{}     &      \multicolumn{2}{c}{{\textit{\color{dkgray}Average per-class}}}\\
                        \hline                                                                          
                                \multirow{5}{*}{\centering{\rotatebox[origin=c]{90}{ \scshape{Tree Bagger}}}}                                                        
                                &       \multirow{1}{*}{Accuracy}       &       99.97\% \\      \cline{2-3}
                                &       \multirow{1}{*}{Error rate}     &       \color{prColo}{0.03\%}  \\      \cline{2-3}
                                &       \multirow{1}{*}{Precision}      &       99.92\% \\      \cline{2-3}
                                &       \multirow{1}{*}{Sensitivity}    &       99.93\% \\      \cline{2-3}
                                &       \multirow{1}{*}{F-score}        &       99.93\% \\      
                        \hhline{===}                                                            
                                \multirow{5}{*}{\centering{\rotatebox[origin=c]{90}{ \scshape{Gentle Boost}}}}                                                       
                                &       \multirow{1}{*}{Accuracy}       &       98.97\% \\      \cline{2-3}     
                                &       \multirow{1}{*}{Error rate}     &       \color{prColo}{1.03\%}  \\      \cline{2-3}     
                                &       \multirow{1}{*}{Precision}      &       97.48\% \\      \cline{2-3}     
                                &       \multirow{1}{*}{Sensitivity}    &       97.64\% \\      \cline{2-3}     
                                &       \multirow{1}{*}{F-score}        &       97.56\% \\              
                        \hhline{===}                                                            
                                \multirow{5}{*}{\centering{\rotatebox[origin=c]{90}{ \scshape{SVM (linear)}}}}                                                       
                                &       \multirow{1}{*}{Accuracy}       &       98.93\% \\      \cline{2-3}     
                                &       \multirow{1}{*}{Error rate}     &       \color{prColo}{1.07\%}  \\      \cline{2-3}     
                                &       \multirow{1}{*}{Precision}      &       97.56\% \\      \cline{2-3}     
                                &       \multirow{1}{*}{Sensitivity}    &       97.72\% \\      \cline{2-3}     
                                &       \multirow{1}{*}{F-score}        &       97.64\% \\              
                                \hhline{===}                                                    
                                \multirow{5}{*}{\centering{\rotatebox[origin=c]{90}{ \scshape{SVM (rbf)}}}}                                                  
                                &       \multirow{1}{*}{Accuracy}       &       99.68\% \\      \cline{2-3}     
                                &       \multirow{1}{*}{Error rate}     &       \color{prColo}{0.32\%}  \\      \cline{2-3}     
                                &       \multirow{1}{*}{Precision}      &       99.22\% \\      \cline{2-3}     
                                &       \multirow{1}{*}{Sensitivity}    &       99.30\% \\      \cline{2-3}     
                                &       \multirow{1}{*}{F-score}        &       99.26\% \\              
                \cline{1-3}
        \end{tabular}                                                                                                           
        \end{center}                                                                                                    
}
\end{center}
}
\vspace{-0.7cm}
\caption{\color{dkblue}[Resubstitution prediction]\color{black}- Measures from confusion matrices. \vspace{0.1cm}}
\label{table:cm_clust_resub}    

\centering \rule{14cm}{0.4pt}                                                                                                           
\parbox{1.0\linewidth}{  
\vspace{-0.5cm}  
\begin{center}          
\parbox{.55\linewidth}{                                                                                                 
        \begin{center}                                                                                                  
        \scriptsize             
        \begin{tabular}{|*{7}{c|}} 
                \multicolumn{2}{c}{}     &      \multicolumn{5}{c}{{\textit{\color{dkgray}Measures per class}}}\\                                                                          
                \cline{3-7}                                                                                             
                \multicolumn{2}{c|}{}   &                                                                                                                       
                        \color{trColo}C1        &                                                                                                               
                        \color{trColo}C2        &                                                                                                               
                        \color{trColo}C3        &                                                                                                               
                        \color{trColo}C4        &                                                                                                               
                        \color{trColo}C5        \\                                                                              
                        \hline                                                                                                                                          
                                \multirow{5}{*}{\centering{\rotatebox[origin=c]{90}{ \scshape{Tree Bagger}}}}                
                                        &       \multirow{1}{*}{Accuracy}       &       99.76\% &       99.29\% &       99.19\% &       99.41\% &       99.76\% \\      \cline{2-7}
                                        &       \multirow{1}{*}{Precision}      &       99.71\% &       98.40\% &       98.04\% &       98.61\% &       99.48\% \\      \cline{2-7}
                                        &       \multirow{1}{*}{Sensitivity}    &       98.38\% &       99.02\% &       98.46\% &       98.49\% &       99.03\% \\      \cline{2-7}
                                        &       \multirow{1}{*}{Specificity}    &       99.96\% &       99.39\% &       99.41\% &       99.65\% &       99.90\% \\      \cline{2-7}
                                        &       \multirow{1}{*}{F-score}        &       99.04\% &       98.71\% &       98.25\% &       98.55\% &       99.25\% \\      
                        \hhline{=======}                                                                                                                                
                                \multirow{5}{*}{\centering{\rotatebox[origin=c]{90}{ \scshape{Gentle Boost}}}}                                       
                                        &       \multirow{1}{*}{Accuracy}       &       99.41\% &       97.76\% &       96.59\% &       97.67\% &       99.44\% \\      \cline{2-7}
                                        &       \multirow{1}{*}{Precision}      &       99.22\% &       97.20\% &       92.88\% &       91.83\% &       97.93\% \\      \cline{2-7}
                                        &       \multirow{1}{*}{Sensitivity}    &       96.20\% &       94.55\% &       92.24\% &       97.16\% &       98.66\% \\      \cline{2-7}
                                        &       \multirow{1}{*}{Specificity}    &       99.89\% &       98.97\% &       97.89\% &       97.81\% &       99.59\% \\      \cline{2-7}
                                        &       \multirow{1}{*}{F-score}        &       97.68\% &       95.86\% &       92.56\% &       94.42\% &       98.29\% \\      
                        \hhline{=======}                                                                                                                                
                                \multirow{5}{*}{\centering{\rotatebox[origin=c]{90}{ \scshape{SVM (linear)}}}}                                       
                                        &       \multirow{1}{*}{Accuracy}       &       99.94\% &       97.33\% &       95.50\% &       97.79\% &       99.68\% \\      \cline{2-7}
                                        &       \multirow{1}{*}{Precision}      &       99.62\% &       98.19\% &       90.05\% &       91.22\% &       99.11\% \\      \cline{2-7}
                                        &       \multirow{1}{*}{Sensitivity}    &       99.90\% &       91.96\% &       90.44\% &       98.55\% &       98.96\% \\      \cline{2-7}
                                        &       \multirow{1}{*}{Specificity}    &       99.94\% &       99.36\% &       97.01\% &       97.59\% &       99.82\% \\      \cline{2-7}
                                        &       \multirow{1}{*}{F-score}        &       99.76\% &       94.97\% &       90.24\% &       94.74\% &       99.03\% \\      
                        \hhline{=======}                                                                                                                                
                                \multirow{5}{*}{\centering{\rotatebox[origin=c]{90}{ \scshape{SVM (rbf)}}}}                                                  
                                        &       \multirow{1}{*}{Accuracy}       &       99.71\% &       99.09\% &       99.02\% &       99.56\% &       99.83\% \\      \cline{2-7}
                                        &       \multirow{1}{*}{Precision}      &       99.14\% &       98.57\% &       97.87\% &       98.62\% &       99.26\% \\      \cline{2-7}
                                        &       \multirow{1}{*}{Sensitivity}    &       98.57\% &       98.13\% &       97.87\% &       99.21\% &       99.70\% \\      \cline{2-7}
                                        &       \multirow{1}{*}{Specificity}    &       99.87\% &       99.46\% &       99.36\% &       99.65\% &       99.85\% \\      \cline{2-7}
                                        &       \multirow{1}{*}{F-score}        &       98.86\% &       98.34\% &       97.87\% &       98.92\% &       99.48\% \\      
                \cline{1-7}
        \end{tabular}                                                                                                           
        \end{center}                                                                                                    
}
\parbox{.275\linewidth}{        
        \begin{center}                                                                                                  
        \scriptsize             
        \begin{tabular}{|*{3}{c|}} 
                \multicolumn{1}{c}{}{}\\
                \multicolumn{1}{c}{}     &      \multicolumn{2}{c}{{\textit{\color{dkgray}Average per-class}}}\\
                        \hline                                                                          
                                \multirow{5}{*}{\centering{\rotatebox[origin=c]{90}{ \scshape{Tree Bagger}}}}                                                        
                                &       \multirow{1}{*}{Accuracy}       &       99.48\% \\      \cline{2-3}
                                &       \multirow{1}{*}{Error rate}     &       \color{prColo}{0.52\%}  \\      \cline{2-3}
                                &       \multirow{1}{*}{Precision}      &       98.85\% \\      \cline{2-3}
                                &       \multirow{1}{*}{Sensitivity}    &       98.68\% \\      \cline{2-3}
                                &       \multirow{1}{*}{F-score}        &       98.76\% \\      
                        \hhline{===}                                                            
                                \multirow{5}{*}{\centering{\rotatebox[origin=c]{90}{ \scshape{Gentle Boost}}}}                                                       
                                &       \multirow{1}{*}{Accuracy}       &       98.17\% \\      \cline{2-3}     
                                &       \multirow{1}{*}{Error rate}     &       \color{prColo}{1.83\%}  \\      \cline{2-3}     
                                &       \multirow{1}{*}{Precision}      &       95.81\% \\      \cline{2-3}     
                                &       \multirow{1}{*}{Sensitivity}    &       95.76\% \\      \cline{2-3}     
                                &       \multirow{1}{*}{F-score}        &       95.79\% \\              
                        \hhline{===}                                                            
                                \multirow{5}{*}{\centering{\rotatebox[origin=c]{90}{ \scshape{SVM (linear)}}}}                                                       
                                &       \multirow{1}{*}{Accuracy}       &       98.05\% \\      \cline{2-3}     
                                &       \multirow{1}{*}{Error rate}     &       \color{prColo}{1.95\%}  \\      \cline{2-3}     
                                &       \multirow{1}{*}{Precision}      &       95.64\% \\      \cline{2-3}     
                                &       \multirow{1}{*}{Sensitivity}    &       95.96\% \\      \cline{2-3}     
                                &       \multirow{1}{*}{F-score}        &       95.80\% \\              
                                \hhline{===}                                                    
                                \multirow{5}{*}{\centering{\rotatebox[origin=c]{90}{ \scshape{SVM (rbf)}}}}                                                  
                                &       \multirow{1}{*}{Accuracy}       &       99.44\% \\      \cline{2-3}     
                                &       \multirow{1}{*}{Error rate}     &       \color{prColo}{0.56\%}  \\      \cline{2-3}     
                                &       \multirow{1}{*}{Precision}      &       98.69\% \\      \cline{2-3}     
                                &       \multirow{1}{*}{Sensitivity}    &       98.70\% \\      \cline{2-3}     
                                &       \multirow{1}{*}{F-score}        &       98.69\% \\                      
                \cline{1-3}
        \end{tabular}                                                                                                           
        \end{center}                                                                                                    
}
\end{center}
}
\vspace{-0.7cm}
\caption{\color{dkgreen2}[Test prediction]\color{black}- Measures from confusion matrices. \vspace{0.1cm}}
\label{table:cm_clust_pred}                                                                                                                             
\end{table*}
}

\newpage
\appendix 
{
\renewcommand\thefigure{A-\arabic{figure}} 
\setcounter{figure}{0}

\section{Assigning probabilities} \label{subsec:app0}
{
\noindent In the data, if the noise is assumed Gaussian, additive, and i.i.d., the data model for a single observation $D_k$ is:
\begin{equation}
                \textrm{datum} \hspace{0.1cm} d_k = 
                        \textrm{true} \hspace{0.1cm} x_k 
                        + \textrm{noise} \hspace{0.1cm} n_k;
                \hspace{0.2cm} 
                N_k\sim\mathcal{N}(0; \sigma_k)
        \label{eq:modelLIn}
.\end{equation}
\noindent The variables $d_k$, $n_k$ , and $x_k$ are realizations of the random variables (r.v.) $D_k$, $N_k$ and $X_k$.

\vspace{0.2cm}
\noindent By marginalizing over the r.v. $X_k$ and $N_k$: 
\begin{equation}
        p(D_k | z ,M_t ,I )=\int \int p(D_k, N_k ,X_k | z, M_t ,I )  dX_k  dN_k  \hspace{0.3cm} 
.\end{equation}
Assuming $X_k \indep N_k$: 
\begin{equation}
        \begin{split}
                p(D_k | z ,M_t ,I )= \int \int  p(X_k | z,M_t,I  )\times p(N_k | z,M_t ,I ) \hspace{0.5cm} \\
                        \times \hspace{0.1cm} p(D_k | N_k, X_k ,z ,M_t ,I )  dX_k  dN_k 
        \end{split}
.\end{equation}
From the equation \ref{eq:modelLIn}, the likelihood function is: 
\begin{equation}
                p(D_k | N_k, X_k, z, M_t ,I ) =\delta(d_k- x_k -n_k )= \delta_k
.\end{equation}
         
\noindent The Kronecker function $\delta_k$ implies $n_k = d_k - x_k$  and allows to rewrite $p(D_k | z, M_t,I  )$ as:
\begin{equation}
        \begin{array}{lll}
                p( D_k | z , M_t ,I ) &=& \int dx_k   f_X (x_k )  \int dn_k  f_N (n_k )  \delta_k  \\ 
                                        &=& \int f_X (x_k )  f_N (d_k - x_k )  dx_k\\
        \end{array}
,\end{equation}
where $f_X(x_k)$ and $f_N(n_k)$ are the probability density functions of the random variables $X_k$ and $N_k$.\\

\noindent If the true value $X_k$ is considered as a deterministic variable: 
\begin{equation}
        p(D_k | z, M_t ,I )= f_N (d_k - x_k ) = f_N (n_k ) = p(N_k | z, M_t ,I )
.\end{equation}
Otherwise, the probabilistic model of $x_k$ has to be integrated into the full expression of $p(D_k | z, M_t,I )$.

\vspace{0.2cm}
The likelihood $\mathcal{L}(z,M_t )$ describes the probability of observing the full set of independent observations $D=\{D_k\}_{k\in\Lambda}$ given a redshift $z$ and a template model $M_t$ and any additional information $I$.

\noindent Considering the aforementioned hypotheses on the data model, the likelihood is defined as following:
\begin{equation}
        \begin{array}{lll}
                \mathcal{L}(z,M_t ) 
                        &=& p( D | z , M_t ,I ) = p(D_1, \dots D_n | z, M_t ,I )                                                                            \\
                        &=& p(N_1,\dots N_n | z,M_t ,I ) = \prod_{k\in\Lambda} p(N_k  | z, M_t ,I )                                    \\
                        &=& \prod_{1\leq i\leq n}{{(\sqrt{2\pi}\sigma_i)}^{-1}} \exp \Big({- \frac{1}{2} \chi^2{(z, t)} }  \Big)                \\
                &&\vspace{-.5cm}\\
                \chi^2{(z, t)} &=& \sum_{i=1}^{n} \sigma_i^{-2} [d_i - a_{opt}\hspace{0.05cm}t_{i,z}]^2,                                 \\
        \end{array}
        \label{eq:likeli}
\end{equation}
where 
$\Lambda$ is the wavelength range in use (with $n$ datapoints),
$d_i$ and $\sigma_i$ are the observed flux and noise spectra at pixel $i,$ respectively, 
$t_{i,z}$ is the redshifted template interpolated at pixel $i$, and
$a_{opt}$ is the optimal amplitude obtained from (weighted) Least-Square (LS) estimation.\\

\noindent We would like to point out that the estimation is in reality obtained from marginalizing over nuisance parameters $\theta$, such as the amplitude $A$ (r.v.) in the chi-square expression: 
\begin{equation}
        \begin{array}{lll}
                p(z,M_t | D,I  )        &=& \int p(z,M_t, \theta | D,I ) d\theta                                         \\
        \end{array}
.\end{equation}

\noindent The joint-posterior PDF can be rewritten as:
\begin{equation}
        \begin{split}
                p(z,M_t | D ,I )        
                =  \int \frac{p(\theta , z, M_t | I) \times p(D | z,M_t, \theta,I)}  {p(D)} d\theta                                        \\    
                =  \int \frac{p(z, M_t|I) \times p(\theta | z, M_t,I ) \times p(D | z,M_t, \theta,I)}  {p(D)} d\theta                   \\    
                =  \frac{p(z, M_t | I) \times p(D | z,M_t, \theta_{opt},I )} {p(D)}                                        \hspace{2.65cm}\\
                        \times \color{dkblue}
                                \int p(\theta | z, M_t,I ) \frac{ p(D | z,M_t, \theta,I )} {p(D | z,M_t, \theta_{opt},I )} d\theta  
                        \color{black}           \\      
        \end{split}
,\end{equation}          
where the highlighted integral in blue is usually approximated by a constant, and the computed likelihood in redshift estimation englobes the optimal estimation $a_{opt}$ of the amplitude parameter in Eq. \ref{eq:likeli}.\\

\noindent The amplitude $a_{opt}$ is estimated at each trial $(z, t)$:
\begin{equation}
        \begin{array}{lll}
                a_{opt}         &=& (t_z^\top w \hspace{0.1cm} t_z)^{-1} \hspace{0.1cm} t_z^\top w \hspace{0.1cm} s              \\
                                &=& \Big({ \sum_{i=1}^{n} s_i t_{i,z} \sigma_i^{-2}  }\Big) /   \Big({  \sum_{i=1}^{n} t_{i,z}^2 \sigma_i^{-2}  }\Big),
        \end{array}
\end{equation}
where $w= \textrm{diag}(\sigma_1^{-2},\dots,\sigma_n^{-2})$ is the weight matrix.
}

\section{ECOC for multi-class problems} \label{subsec:app1}
{
\noindent The principle of ECOC ({Error-Correcting-Output-Codes}) is based on the binary reduction of the multi-class problem using a coding matrix 
$\mathcal{M}\in\left\{-1; 0; +1\right\}^{K\times L}$ to design a codeword.
\begin{equation}
 \mathcal{M} = 
        \begin{blockarray}{cccc}
                &l_1 &  & l_L \\
                \begin{block}{c(ccc)}
                  c_1&m_{11}&\cdots&m_{1L}\\
                         & \vdots&\ddots&\vdots\\
                  c_K&m_{K1}&\cdots&m_{KL}\\
                \end{block}
        \end{blockarray}
        \vspace{-0.6cm}
,\end{equation} 
where:
\begin{itemize}
        \item $L$ : number of learners;
        \item $K$ : number of distinct classes.
\end{itemize}
The codewords $\textit{\textbf{m}}_{k}= \it(m_{k,1},\dots, m_{k,L})$ translate the membership information for each class $c_k$ given a binary scheme:
\begin{itemize}
        \item $m_{kj}=-1$ : $c_k$ is the negative class for learner $l_j,$
        \item $m_{kj}= 0$ : All observations associated with $c_k$ are ignored by the learner $l_j,$
        \item $m_{kj}=+1$ : $c_k$ is the positive class for learner $l_j.$
\end{itemize}

Codewords are generated using existing coding strategies such as OVA ({one-versus-all}), OVO ({one-versus-one}) and dense random.
Coding matrices for a example of a four-class problem are shown in Figure~\ref{figure:ecocDesign}.

\begin{figure}[h]
        \centering{ 
        {\includegraphics[width=0.50\textwidth]{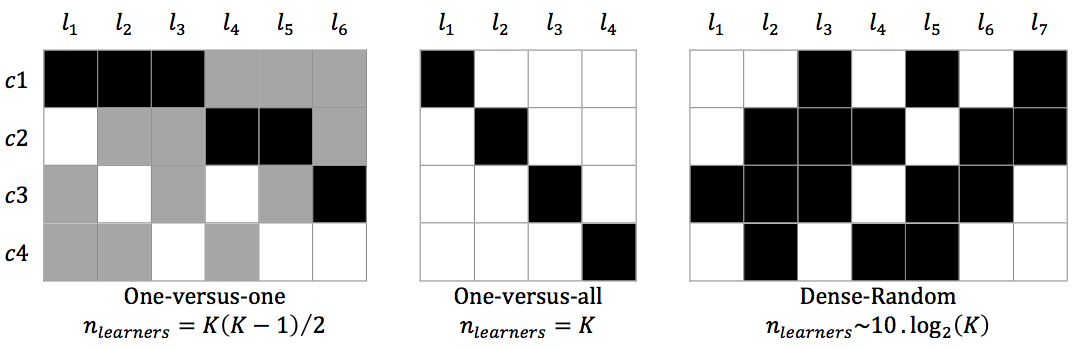}}}
        \caption{\color{dkgray2}[Examples of ECOC coding design] \color{black} The black, white, and gray boxes refer respectively to $m_{k,j}=$ +1, -1 or 0.}
        \label{figure:ecocDesign}
\end{figure}
 
\noindent Each learner $l_j$ is associated with two superclasses, $\left\{S_{+}; S_{-}\right\}$  referring to the positive and the negative classes, respectively, that are used to encode the response vector $\bf Y_{\rm train}$ into a binary vector $[\bf Y_{\rm train}]_{\rm j}$.

\noindent Training the learner $l_j$ with $\left\{\bf X_{\rm train}; [\bf Y_{\rm train}]_{\rm j}\right\}$ is performed with the usual classifiers such as SVM, MLP, and so on.

\noindent A class prediction for an unlabeled spectra $\textit{\textbf{x}}_0$ in $\bf X_{\rm est}$ is achieved in two steps:
\begin{itemize}
        \item[-] \textbf{Step 1: } Each trained learner $l_j$ provides a binary prediction: $(y_0)_j \in\left\{-1; +1\right\}$. \vspace{0.1cm}
        \item[-] \textbf{Step 2: } The bit vector $\textit{\textbf{y}}_0$ for all learners is decoded into the initial K-class by minimizing a distance metric $\Delta$ 
                as the Euclidean distance $\Delta_k={\sum_{j=1}^{L}{(m_{kj}-(y_0)_j)^2}}^{1/2}$ 
                or a binary loss function $\Delta_k=\sum_{j=1}^{L} {| m_{kj} |}\hspace{0.1cm}{g(m_{kj} , s_j)} $, 
                                where $g$ is a binary loss function and $s_j$ the score for learner $j$.\\
                The predicted class $\widehat{c}_k$ for $\textit{\textbf{x}}_0$ is associated with the index $k$ for which the vector $\Delta$ is minimal.
\end{itemize}

}

\section{Description of classification algorithms: SVM - Ensemble methods} \label{subsec:app3}
{
\subsection{Support-Machine Vectors}
{
The SVM method classifies the data by finding the best hyperplane separating the datapoints of one class from those of another category.

\noindent Given a training set of M datapoints $(\textit{\textbf{x}}_i, y_i)_{i\in1\dots M}$, where $\textit{\textbf{x}}_i$ refers to the P-dimensional feature vector  and $y_i$ the associated label that indicates whether the datapoint belongs to the positive class ($y_i=+1$) or the negative class ($y_i=-1$), the objective of SVM is to separate the data into distinct classes using a separating rule in form of a parametrized function $f(\textit{\textbf{x}})$.\\
For linearly separable data, the equation of the hyperplane is:
\begin{equation}
        f(\textit{\textbf{x}}) = f_{\textit{\textbf{w}},b}(\textit{\textbf{x}}) = \textit{\textbf{w}}.\textit{\textbf{x}} + b = 0
        \label{eq:eq_hyper}
,\end{equation} 
where the scalar product $\textit{\textbf{w}}.\textit{\textbf{x}}$ is equivalent to $\textit{\textbf{w}}^{\top}\textit{\textbf{x}}$.

\noindent An infinity  of hyperplanes verify the Eq. \ref{eq:eq_hyper}, but only one hyperplane maximizing the margins between the observations and the hyperplane exists.
This optimal hyperplane verify:
\begin{equation}
        \begin{array} {lll}
        \left\{
                \begin{array}{ll}
                        (\textit{\textbf{w}}.\textit{\textbf{x}} + b) \geq +1      &\textrm{if}\hspace{0.1cm} y_i=+1       \\
                        (\textit{\textbf{w}}.\textit{\textbf{x}} + b) \leq -1      &\textrm{if}\hspace{0.1cm} y_i=-1
                \end{array}
        \right.
        &
        \Leftrightarrow 
        &
        y_i(\textit{\textbf{w}}.\textit{\textbf{x}} + b) \geq +1        \\
        \end{array}
        \label{eq:eq_svm}
.\end{equation} 

\noindent To find the "best" linear hyperplane minimizing the margins $2(\textit{\textbf{w}} . \textit{\textbf{w}}^{\top})^{-1/2}$, the SVM algorithm consists in solving a quadratic problem:
\begin{equation}
         \begin{array}{ll}
                \underset{\textit{\textbf{w}},b}{\text{minimize}}      	&         \frac{1}{2} (\textit{\textbf{w}}  \textit{\textbf{w}}^{\top})   \\
                \textrm{subject to:}                                            		&       y_i (\textit{\textbf{w}}.\textit{\textbf{x}}_i + b) \geq +1.
         \end{array}
         \label{eq:svm_arg}
\end{equation} 

\noindent For non-linearly separable data, the use of a kernel $\varphi$ trick enables to map the distribution of the datapoints $\textit{\textbf{x}}$ into a projected space where $\varphi(\textit{\textbf{x}})$ can be linearly separable, and defines, in the same approach as in Eq. \ref{eq:eq_svm}, a quadratic problem:
\begin{equation}
         \begin{array}{ll}
                \underset{\textit{\textbf{W}},B}{\text{minimize}}       	&         \frac{1}{2} (\textit{\textbf{W}} \textit{\textbf{W}}^{\top})            \\
                \textrm{subject to:}                                            		&       y_i (\textit{\textbf{W}}.\varphi(\textit{\textbf{x}}_i) + B) \geq +1
         \end{array}
,\end{equation} 
where $f(\textit{\textbf{x}}) = f_{\textit{\textbf{W}},B}(\textit{\textbf{x}})=\textit{\textbf{W}}.\varphi(\textit{\textbf{x}}) + B = 0$.\\

\noindent The selection of an adequate kernel $\textbf{K}$ is determined by a list of criteria. By definition, a kernel must be symmetric, definite positive, square integrable and satisfy:
\begin{equation}{
        \left\{
        \begin{matrix}
                        {K}_{i,j} = K(\textit{\textbf{x}}_i, \textit{\textbf{x}}_j) = \varphi(\textit{\textbf{x}}_i). \varphi(\textit{\textbf{x}}_j) \\
                        \exists (\lambda_1 ... \lambda_N )\in \mathbb{R}:   \hspace{0.3cm}   
                                 \sum_{i= 0}^N {\sum_{j= 0}^N {\lambda_i \lambda_j K(\textit{\textbf{x}}_i, \textit{\textbf{x}}_j) \geq0}}. 
        \end{matrix}
        \right.
}\end{equation} 

\noindent Among commonly used kernels:
\begin{itemize} \vspace{-0.2cm}
        \item Linear:           \hspace{1.12cm}  $K(\textit{\textbf{x}}_i, \textit{\textbf{x}}_j) =  \textit{\textbf{x}}_i . \textit{\textbf{x}}_j$
        \item Power:            \hspace{1.15cm}  $K(\textit{\textbf{x}}_i, \textit{\textbf{x}}_j) =  (\textit{\textbf{x}}_i . \textit{\textbf{x}}_j)^m$
        \item Gaussian (rbf):   \hspace{-0.05cm} $K(\textit{\textbf{x}}_i, \textit{\textbf{x}}_j) =  \exp(- \frac{1}{2} | \textit{\textbf{x}}_i - \textit{\textbf{x}}_j |^2 / \sigma^2).$
\end{itemize}
Further details about the SVMs are available in \citealp{vapnik_learn_2000} and \citealp{cristianini_svm_2000}.
}
        
\subsection{Ensemble classifiers} 
{
\noindent The principle of ensemble methodology is to combine a set of predictions from different learners in order to improve the accuracy of a single learner.

\noindent Among the ensemble methods, two distinct approaches are identified:
\begin{enumerate}
        \item   Averaging methods: Bagging/ Random forests...
        \item Boosting methods: AdaBoost/GentleBoost/RSBoost/...
\end{enumerate}

\vspace{0.2cm}

\noindent {\color{dkblue}  - \textit{AdaBoost}} \color{black}

\noindent AdaBoost, known also as "Adaptive Boosting" refers to a specific algorithm of boosted classifier defined as the sum of individual predictions from $T$ weak learners.
The algorithm aims to minimize the (weighted) classification error at each iteration:
\begin{equation}
        \varepsilon_t = \sum_{i=1}^N d_i^{(t)}  \mathbbm{1}(y_i\neq h_t(\textit{\textbf{x}}_i))
,\end{equation}
where:
\begin{itemize}
        \item $x_i$ is the feature vector of the $i$-th observation.
        \item $y_i$ is the true label of the the $i$-th observation.
        \item $h_t$ is the prediction of learner $t$.
        \item $\mathbbm{1}$ is the indicator function.
        \item $d_i^{(t)}$ is the weight of the $i$-th observation at step $t$.
        \item $t$ the iteration step from 1 to $T$.
\end{itemize}
\noindent At the first iteration, the weights $d_i^{(t)}$ are initialized (e.g., $d_i^{(t)}=1/N$) and the weak learner $h_t$ is obtained by minimizing the error $\varepsilon_t$. 
For the next iteration, the weights of the learner ($t+1$) are adjusted according to the performance of the previous one ($t$): whether increase $d^{(t+1)}_i$ for misclassified observations by learner $t$, or reduce the weights otherwise. The learner $h_{t+1}$ is trained using the updated weights $d_i^{(t)}$ in the error $\varepsilon_{t+1}$.

After training, the prediction for a new data point, $\textit{\textbf{x}}$, is obtained by combining the individual predictions of all weak learners:
\begin{equation}
        f(\textit{\textbf{x}}) = \sum_{t=1}^{\top} \alpha_t h_t(\textit{\textbf{x}}) ;       \hspace{0.5cm}
        \alpha_t = \frac{1}{2} \log\Big(\frac{1-\varepsilon_t }{\varepsilon_t}\Big) 
.\end{equation}

\noindent The AdaBoost algorithm can also be viewed as a minimization of an exponential loss function:
\begin{equation}
        \sum_{i=1}^N w_t \exp(-y_i f(\textit{\textbf{x}}_i))    
,\end{equation}
where $w_t$ are normalized observational weights.

\vspace{0.45cm}

\noindent {\color{dkblue}  - \textit{LogitBoost}} \color{black}

\noindent Following a similar approach to AdaBoost, the LogitBoost consists in training learners sequentially by minimizing an error function $\varepsilon_t$; the only difference being the minimization of the error function with respect to a fitted regression model $\widetilde{y}$ instead of ${y}$:
\begin{equation}
                \varepsilon_t =\sum_{i=1}^N d_i^{(t)} (\widetilde{y}_i  - h_t(\textit{\textbf{x}}_i)  )^2         ; \hspace{0.5cm}
                \widetilde{y}_i = \frac{ y_i^* - p_t(\textit{\textbf{x}}_i)}{p_t(\textit{\textbf{x}}_i) (1 -  p_t(\textit{\textbf{x}}_i) )}
,\end{equation}
where:
\begin{itemize}
        \item   $y_i^*$ are modified labels: $y_i^*=0$ if $y_i=-1$; and  $y_i^*=1$ otherwise.
        \item   $p_t(\textit{\textbf{x}}_i)$ is the predicted class probability for the $i$-th observation to be in the positive class "+1" given by the learner $t$.
\end{itemize}

\vspace{0.45cm}

\noindent {\color{dkblue}  - \textit{GentleBoost}} \color{black}

\noindent Also called Gentle AdaBoost, this algorithm combines the methodology of AdaBoost and LogitBoost.
An exponential loss function is minimized with a different optimization strategy to AdaBoost. Further, similarly to LogiBoost, weak learners fit a regression model $\widetilde{y}$ to the response variables $y$. 

\vspace{0.45cm}

\noindent {\color{dkblue}  - \textit{Bagging}} \color{black}

\noindent Bagging, referring to "bootstrap aggregation", consists in generating $m$ new training sets $P_j$, each of size $N'$, by uniformly sampling with replacement from the initial training set $I = (\textit{\textbf{x}}_i, y_i)_{i\in1\dots N}$.

\noindent The $m$ models are trained separately 
and the class prediction of an unlabeled data $\textit{\textbf{x}}$ is obtained by combining the individual predictions of the $m$ models : '{averaging}' if regression, or '{voting}' if classification.

\vspace{0.2cm}
Further details on the ensemble algorithms can be found in \citealp{dietterich_ensemble_2000}.
}
}

\section{Measures for multi-class classification}\label{subsec:app5}
{
For a binary classification, the confusion matrix represents the fraction of predicted labels versus the true classes.
Four quantities are directly measured:
\begin{itemize}
        \item $TP$ : True Positives
        \item $TN$ : True Negatives
        \item $FP$ : False Positives
        \item $FN$ : False Negatives
\end{itemize}

\begin{table}[ht]
        \begin{center}                                                                                                  
        \mynewformat
        \begin{tabular}{ l|c|c|c|c }                                                                                                    
                \multicolumn{2}{c}{}     &      \multicolumn{2}{c}{{\scshape{\color{trColo}True}}}&\\                                                                           
                \cline{3-4}                                                                                             
                \multicolumn{2}{c|}{}&                                                                                          
                        \color{trColo} \textit{pos}&                                                                                    
                        \color{trColo} \textit{neg}&                                                                                    
                         \textit{\color{dkgray}Total}   \\                                                                              
                \hhline{~-|--|-}                                                                                                
                \multirow{1}{*}{\centering{\rotatebox[origin=c]{90}{\hspace{0.5cm}\scshape{\color{prColo}Predicted}}}}                                                                          
                &\color{prColo}\textit{pos}             &       \cellcolor{lightGreen} TP      &       FP                                      &       \color{dkgray}TP+FP             \\
                \hhline{~-|--|}                                                                                                 
                &\color{prColo}\textit{neg}     &       FN                                      &       \cellcolor{lightGreen}  TN      &       \color{dkgray}FN+TN     \\
                \cline{2-5}                                                                                             
                \multicolumn{1}{c}{} &                                                                                          
                 \multicolumn{1}{c|}{\textit{\color{dkgray}Total}} &                                                                                            
                  \multicolumn{1}{c}{\color{dkgray}     TP+FN   } &                                                                             
                  \multicolumn{1}{c}{\color{dkgray}     TN+FP   } &                                                                             
                  \multicolumn{1}{|c}{\color{dkgray}    TP+FP+TN+FN     } \\                                                                              
        \end{tabular}   
        \end{center}
\end{table}

\noindent For multi-class classification, the approach consists in estimating these measures for each class.
For example:
\begin{table}[ht]                                                                                                       
        \begin{center}                                                                                                  
        \mynewformat
        \begin{tabular}{ l|c|c|c|c|c }                                                                                                  
                \multicolumn{2}{c}{}     &      \multicolumn{3}{c}{{\scshape{\color{trColo}True}}}&\\                                                                           
                \cline{3-5}                                                                                             
                \multicolumn{2}{c|}{}&                                                                                          
                        \color{trColo}"  0 "&                                                                                   
                        \color{trColo}"+1"&                                                                                     
                        \color{trColo}"+2"&                                                                                     
                         \textit{\color{dkgray}Total}   \\                                                                              
                \hhline{~-|---|-}                                                                                               
                \multirow{4}{*}                                                                                         
                {\centering{\rotatebox[origin=c]{90}{\hspace{0.5cm}\scshape{\color{prColo}Predicted}}}}                                                                                         
                &       \color{prColo}" 0 "     &\cellcolor{lightGreen} 4503    &       2       &       1       &       \color{dkgray}  4506    \\      
                \hhline{~-|---|}                                                                                                
                &       \color{prColo}"+1"       &      9       &\cellcolor{lightGreen} 3581    &       203     &       \color{dkgray}  3793    \\      
                \hhline{~-|---|}                                                                                                
                &       \color{prColo}"+2"       &      0       &       0       &\cellcolor{lightGreen} 8047    &       \color{dkgray}  8047    \\      
                \cline{2-6}                                                                                             
                \multicolumn{1}{c}{} &                                                                                          
                 \multicolumn{1}{c|}{\textit{\color{dkgray}Total}} &                                                                                            
                  \multicolumn{1}{c}{\color{dkgray}     4512    } &                                                                             
                  \multicolumn{1}{c}{\color{dkgray}     3583    } &                                                                             
                  \multicolumn{1}{c}{\color{dkgray}     8251    } &                                                                             
                   \multicolumn{1}{|c}{\color{dkgray}   16346   } \\                                                                            
         \end{tabular}                                                                                          
        \end{center}    
        \begin{center}                                                                                                  
        \mynewformat
        \begin{tabular}{|*{4}{c|}} 
                \multicolumn{1}{c}{}     &      \multicolumn{3}{c}{{\textit{\color{dkgray}Measures per class}}}\\                                                                          
                \cline{2-4}                                                                                             
                \multicolumn{1}{c|}{}           &                                                                                               
                        \color{trColo}"  0 "    &                                                                                       
                        \color{trColo}"+1"      &                                                                                       
                        \color{trColo}"+2"      \\                                                                              
                \hline 
                        \multirow{1}{*}{TP}             &       {4503}          &       {3581}          &       {8047}                  \\      \hline 
                        \multirow{1}{*}{FP}             &       {3}                     &       {212}           &       {0}                             \\      \hline 
                        \multirow{1}{*}{FN}             &       {9}                     &       {2}                     &       {204}                   \\      \hline 
                        \multirow{1}{*}{TN}             &       11831            &       12551           &       8095                            \\
                \cline{1-4}
        \end{tabular}                                                                                                           
        \end{center}
\end{table}

\noindent From the confusion matrix, the overall performances of the classification are quantified with the following measures:
\setlength\extrarowheight{5pt}
\renewcommand{\arraystretch}{1.2}
\begin{table}[ht]
        \begin{center}                                                                                                  
        \mynewformat
        \begin{tabular}{|*{3}{c|}} 
                \multicolumn{1}{c}{}     &      \multicolumn{2}{c}{{\textit{\color{dkgray} multi-class classification $(C_k)_{ k\in\{1\dots K\}}$  }}}\\
                \hline 
                        \multirow{6}{*}{\centering{\rotatebox[origin=c]{90}{ \scshape{Measures per-class}}}} 
                                &       \scshape{Accuracy($C_k$)}       &       $\frac{TP_k + TN_k}{TP_k + FN_k + TN_k + FP_k}$             \\[4pt] \cline{2-3}
                                &       \scshape{Precision($C_k$)}      &       $\frac{TP_k}{TP_k + FP_k}$                                                \\[4pt] \cline{2-3}
                                &       \scshape{Sensitivity($C_k$)}    &       $\frac{TP_k}{TP_k + FN_k}$                                                \\[4pt] \cline{2-3}
                                &       \scshape{F-score($C_k$)}                &       $\frac{2*TP_k}{2*TP_k + FN_k + FP_k}$                                 \\[4pt] \cline{2-3}
                                &       \scshape{Specificity($C_k$)}    &       $\frac{TN_k}{FP_k + TN_k}$                                                \\[4pt] 
                \hhline{---}     
                \multicolumn{3}{c}{} \\[-10pt]
                \hhline{---}  
                        \multirow{6}{*}{\centering{\rotatebox[origin=c]{90}{ \scshape{Average per-class}}}} 
                                &       \scshape{Accuracy}      &       $\frac{1}{K} \sum_k Accuracy(C_k)$                                                   \\[4pt] \cline{2-3}
                                &       \scshape{Error rate}    &       $\frac{1}{K} \sum_k Error(C_k)$                                                      \\[4pt] \cline{2-3}
                                &       \scshape{Precision}     &       $\frac{1}{K} \sum_k Precision(C_k)$                                                  \\[4pt] \cline{2-3}
                                &       \scshape{Sensitivity}   &       $\frac{1}{K} \sum_k Sensitivity(C_k)$                                                        \\[4pt] \cline{2-3}
                                &       \scshape{F-score}       &       $\frac{1}{K} \sum_k Fscore(C_k)$                                                     \\[4pt]  
                \cline{1-3}
        \end{tabular}                                                                                   
        \end{center}    
\end{table}

Further details are provided in \citealp{fawcett_introduction_2006}.
}

\section{Description of the FCM clustering algorithm} \label{subsec:app4}
{
Similar to the {k-means} algorithm that aims to minimize the intraclass variance, the FCM ({Fuzzy C-Means}) algorithm exploits additional information about membership of the data to multiple clusters.\\
To partition a dataset $\textit{\textbf{X}} = (\textit{\textbf{x}}_{1} \dots \textit{\textbf{x}}_{M})^\top$ of P-dimensional vectors into $K$ clusters, the algorithm aims to solve a quadratic problem in order to determine the optimal solution (\textbf{U},\textbf{G}), where $\textbf{G} = (\textbf{g}_{1} \dots \textbf{g}_{M})$ refers to the centroids of the final $K$ clusters and $\textbf{U} = ({\mu}_{ij})_ {\left\{\substack{1\leq i\leq K \\  1\leq j\leq M}\right. }$ is a coefficient matrix of class memberships for each element.
\begin{equation}
         \begin{array}{lll}
                \underset{\textbf{U,G}}{\text{minimize}}        &       J(\textit{\textbf{X}}; \textbf{U,G})           &       \\
                \text{subject to:}                                      &       \sum_{i=1}^{K} \mu_{ij} = 1 ,  &\; j = 1, \ldots, M.
         \end{array}
\end{equation}
The algorithm proceeds iteratively and converges when the estimated coefficient matrix at the iteration $t$ is not very different from its previous estimation:
 \begin{equation}
        \begin{matrix}
                \| \textbf{U}^{(t)} -  \textbf{U}^{(t-1)}    \|  < \varepsilon_{0}              &         \text{, where $\varepsilon_{0}$ is fixed by the user}
        \end{matrix}    
 .\end{equation}
\noindent Further, the elements $(\textit{\textbf{x}}_j)_{ j = 1, \ldots, M}$ are said to belong to the class $(c_i)_{i= 1, \ldots, K}$ for which the final coefficient $\mu_{ij}$ is maximal.\\
The matrix $\bf U$ can be further exploited to assess the membership level of the element $\textit{\textbf{x}}_j$ to its predicted class.
 
\noindent The cost function to minimize is:
\begin{equation}
        J(\textit{\textbf{X}}; \textbf{U,G}) = \sum_{i=1}^{K} \sum_{j=1}^{M}  \mu_{ij}^{m} D_{ijA}^2
,\end{equation}
where the distance $D_{ijA}$, the coefficient $\mu_{ij}$ and the centroid $\textbf{g}_i$ are defined as following:
\begin{equation}
        \begin{array}{ll}
                D_{ijA}^2               &= \| \textit{\textbf{x}}_j - \textbf{g}_i \|^{2}_{A}                                                                              
                                          = (\textit{\textbf{x}}_j - \textbf{g}_i)^\top A (\textit{\textbf{x}}_j - \textbf{g}_i)                
        \end{array}
        \label{eq:norm}
,\end{equation}  
\begin{equation}
        \begin{array}{ll}
                \mu_{ij}                &= \Big(  \sum_{k=1}^{K} {D_{ijA}^2}/{D_{kjA}^2}  \Big)^{-{2}/{(m-1)}}                           \\
        \end{array}
,\end{equation}  
\begin{equation}
        \begin{array}{ll}
                \textbf{g}_i    &=  \big( { \sum_{j=1}^{M} \mu_{ij}^m \textit{\textbf{x}}_j  }\big) / \big({ \sum_{j=1}^{M} \mu_{ij}^m } \big)
        \end{array}
.\end{equation}

In the FCM, the fuzzifier parameter $m\geq1$ is used to determine the level of fuzziness: if $m=1$, the coefficient matrix is binary, which is equivalent to a hard partitioning.
Usually, in the absence of prior information about the datamodel, the value $m=2$ is used.\\
For the norm $ \| . \|^{2}_{A} $in Eq \ref{eq:norm}, a common choice for the matrix $A$ is the identity matrix, but it can be designed to incorporate individual variances of the data as $A= \textrm{diag} (\sigma_1^{-2} \dots \sigma_M^{-2})$ or the inverse of the covariance matrix.
}
}


\end{document}